# Leveraging Data Mining, Active Learning, and Domain Adaptation in a Multi-Stage, Machine Learning-Driven Approach for the Efficient Discovery of Advanced Acidic Oxygen Evolution Electrocatalysts

Rui Ding[a, b, c †], Jianguo Liu[d †], Kang Hua[d], Xuebin Wang[e], Xiaoben Zhang[a, c], Minhua Shao[f,g], Yuxin Chen[h*], and Junhong Chen[a,c*]

[a] Pritzker School of Molecular Engineering, University of Chicago, 5640 S Ellis Ave., Chicago, IL 60637, United States

[b] Data Science Institute, University of Chicago, 5801 S Ellis Ave., Chicago, IL 60637, United States

[c] Chemical Sciences and Engineering Division, Physical Sciences and Engineering Directorate, Argonne National Laboratory, 9700 S Cass Ave., Lemont, IL 60439, United States

[d] Institute of Energy Power Innovation, North China Electric Power University, 2 Beinong Road, Beijing 102206, P. R. China

[e] National Laboratory of Solid State Microstructures, College of Engineering and Applied Sciences, Nanjing University, 22 Hankou Road, Nanjing 210093, P. R. China

[f] Department of Chemical and Biological Engineering, The Hong Kong University of Science and Technology, Clear Water Bay, Kowloon, Hong Kong 999077, China

[g] Energy Institute, The Hong Kong University of Science and Technology, Clear Water Bay, Kowloon, Hong Kong 999077, China

[h] Department of Computer Science, University of Chicago, 5730 S Ellis Ave., Chicago, IL 60637, United States

[*]Corresponding Authors: junhongchen@uchicago.edu, chenyuxin@uchicago.edu

[†]Co-first authors

## Abstract

Developing advanced catalysts for acidic oxygen evolution reaction (OER) is crucial for sustainable hydrogen production. This study introduces a novel, multi-stage machine learning (ML) approach to streamline the discovery and optimization of complex multi-metallic catalysts. Our method integrates data mining, active learning, and domain adaptation throughout the materials discovery process. Unlike traditional trial-and-error methods, this approach systematically narrows the exploration space using domain knowledge with minimized reliance on subjective intuition. Then the active learning module efficiently refines element composition and synthesis conditions through iterative experimental feedback. The process culminated in the discovery of a promising Ru-Mn-Ca-Pr oxide catalyst. Our workflow also enhances theoretical simulations with domain adaptation strategy, providing deeper mechanistic insights aligned with experimental findings. By leveraging diverse data sources and multiple ML strategies, we establish an efficient pathway for electrocatalyst discovery and optimization. This comprehensive, data-driven approach represents a paradigm shift and potentially new benchmark in electrocatalysts research.

## 1. Introduction

The development of advanced electrocatalysts for the acidic oxygen evolution reaction (OER) in proton exchange membrane (PEM) water electrolysis is imperative for enabling sustainable hydrogen production and achieving carbon neutrality targets[1]. Despite their promise due to their electronic properties[2,3], $RuO_2$ and $IrO_2$-based materials still face inherent trade-offs between activity and stability under harsh acidic conditions[4]. Prior efforts in the field via nearly exhaustive, brute-force search (**Figure S1a**) approaches have extensively explored a wide range of doping[5-10] and morphological[11-13] strategies, yet a comprehensive understanding of the optimal electrocatalytic systems remains elusive. The allure of multi-metallic or "high-entropy" alloy/oxide materials, with their multi-principal element composition, lies in their potential for synergistic catalytic effects[14-16]. Recent advances have proven such solutions as promising to break the limitations of traditional catalysts[17-19]. However, the exploration of such complex multi-metallic systems for acidic OER remains in its infancy, grappling with the immense inherent complexity and the formidable task of navigating the vast compositional and parametric spaces inherent in these systems.

Traditionally, the development of electrocatalysts has focused on empirical, trial-and-error methodologies, heavily dependent on limited prior knowledge and heuristic exploration[20]. The Edisonian approaches, while grounded in chemical intuition, are insufficient for the nuanced optimization required for our targeted systems, given their expansive array of constituent elements and synthesis parameters (**Figure S1b**). Moreover, the intrinsic limitations of trial-and-error approaches stem from an over-reliance on subjective intuition, leading to a narrow and potentially suboptimal exploration of the materials landscape.

To address these challenges and unlock the untapped potential of multi-metallic oxides for the acidic OER, we pioneered a transformative, multi-stage ML-driven approach. Our workflow embodied the synergistic integration of data mining, active learning, and domain adaptation at different discovery stages. By harnessing the collective power of these ML techniques, our methodology has minimized subjective biases and maximized data-driven decision-making in a rationally hierarchical way. As illustrated in **Figure 1**, we began by harnessing the breadth of available domain knowledge and conducting data mining. This step successfully distilled key parameters, established foundational pattern understandings, and systematically narrowed the initial exploration space. Subsequently, an active learning strategy was employed and synergistically coupled with high-throughput experimental feedback. This iterative active learning-driven process navigated an efficient and refined search within the vast parameter space of quaternary element compositions and synthesis conditions. With this approach, called "DASH", the overpotentials at 10 mA $cm^{-2}$ ($\eta_{10}$) observed in the best samples from each of the five experimental batches systematically decreased from 209 mV to 154 mV, reflecting the dynamic and continuous improvement characteristic of active learning. This dynamic optimization led towards the discovery of a promising Ru-Mn-Ca-Pr oxide catalyst out of an enormous candidate chemical and engineering parameter space. Like directly "dashing" to the endpoint in a winding maze, this is likely unachievable through subjective expertise and intuition.

In the final stage after material characterization, domain adaptation employed within density functional theory (DFT) simulations effectively narrowed the theoretical design space to configurations that are potentially valid. Through enhanced ML surrogate modeling, this approach allowed us to allocate limited resources to critical DFT simulations, enabling a broader and in-depth investigation and thus providing valuable atomic scale insights. Our multi-faceted, integrated ML-powered methodology transcends the mere acceleration of optimal candidate discovery, heralding a transformative shift in electrocatalysis research. By weaving various ML techniques with diverse knowledge sources throughout the materials discovery process, this work eclipses the current stage-specific, single-expert-system ML applications in OER electrocatalysts[21-25]. Demonstrating the efficacy of this comprehensive ML-driven strategy in acidic OER, we have charted an unprecedented blueprint for the future of electrocatalyst research.

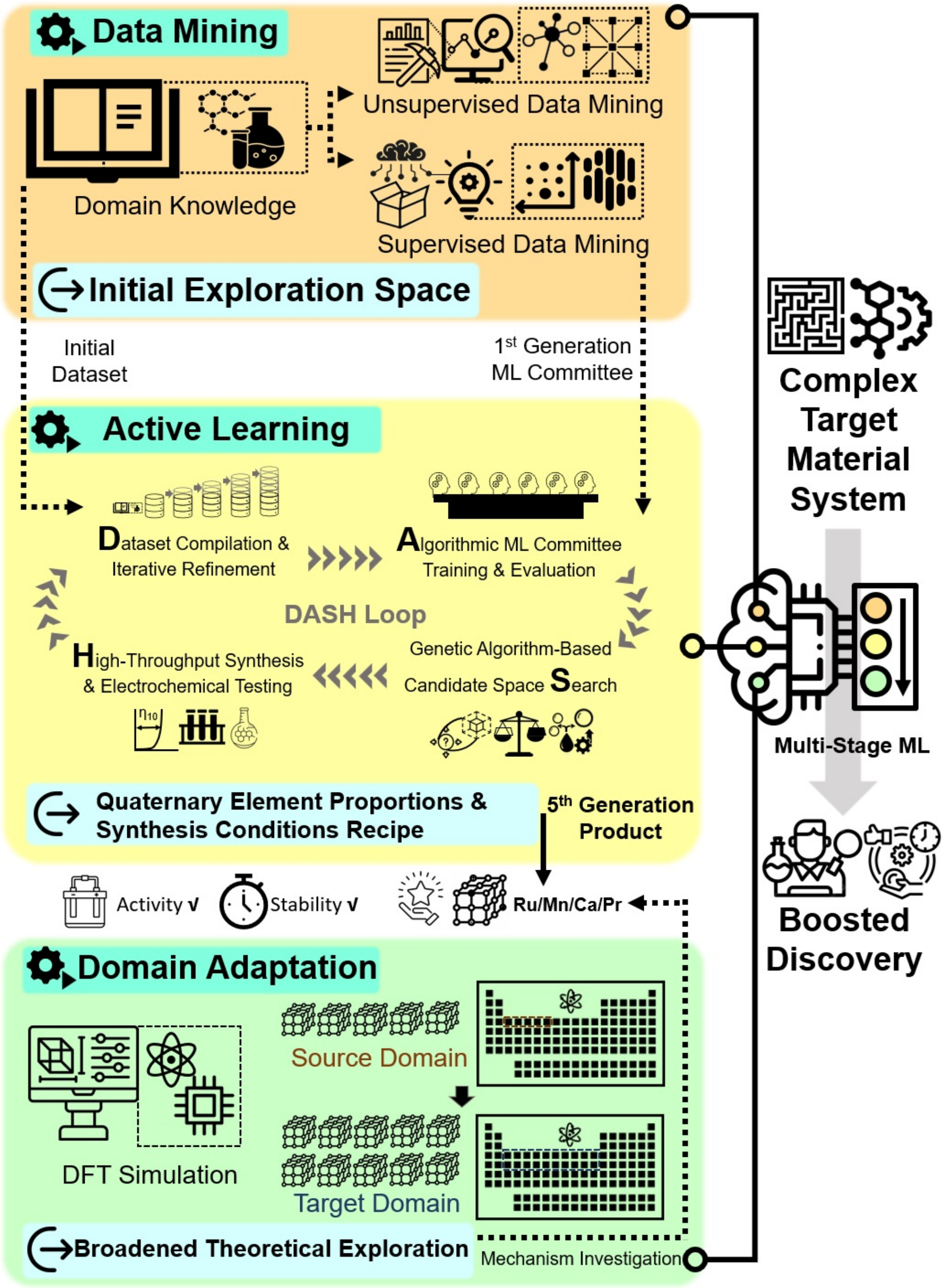

**Figure 1** Schematic illustration of the comprehensive, multi-stage ML workflow (DASH) that guides the entire discovery process of OER electrocatalyst materials. This workflow incorporates different specialized ML expert modules that are derived from various models and data sources.

## 2. Data Mining on Domain Knowledge for Initial Exploration Space

Facing the astronomical number of possible candidate configurations, we need to rationally decide on the initial exploration space. In the past, this was often accomplished by reading numerous domain publications in combination with researchers' own subjective experience to decide on the initial trial schemes. Serving a similar but more rational role, we have undertaken a comprehensive data mining approach centered around the digitalization and distillation of domain knowledge as a starting point of the data science workflow. This endeavor culminates in the digitalization of an extensive literature body (**Table S1** & **Table S2**): 534 high-quality publications dating from November 2009 to January 2023. Four complementary datasets were extracted for subsequent data mining, segmented based on (1) different publication qualities: an initial full dataset and a high-quality subset filtered by criteria such as citation counts, and (2) different fitting targets: $\eta_{10}$ and decay rate, representing OER activity and stability, respectively (**Figure S2**, **Supplementary Note 1**).

### 2.1. Unsupervised Data Mining

The initial phase of our research involved straightforward yet effective unsupervised techniques to identify fundamental patterns in the datasets extracted from the literature. Corresponding results are provided in **Figure 2** for activity data and **Figure S3** for stability data. This included the Bibliometric Interconnected Network Graph[26,27] and Apriori associate rule mining[28,29]. We use the former (**Figure 2 a-b, Figure S3 a-b**) to visualize the frequency of the occurrence of common metal elements alone and in pairs in the aforementioned datasets, as well as their relationship as itemset with $\eta_{10}$ and decay rate. The latter (**Figure 2 c-d, Figure S3 c-d**) further extends the association rules from element types to more empirical synthesis parameters, resulting in statistical suggestions based on domain-wide knowledge. With the support of unsupervised data mining of the extensive domain expertise (**Supplementary Note 2**), we could derive key rational objective decisions on defining the initial exploration.

From the element network analysis, we found that among the knowledge base, the most frequently used elements are Ir, Ru, Sr, Zn, Mn, Ni, and Co. And among them, Ru is the most competitive element. Ru shows promising activity, with qualification rates significantly better than Ir and other elements, thus aligning with domain consensus[30]. Furthermore, opposite of consensus[31], Ru-based candidates demonstrate evenly matched stability expectations compared to Ir. Such advantages also existed with Ru's coexistence with other dopants. Therefore, we believe that **(1) Ru vs. Ir:** Ru is an indispensable element to explore with other elements as dopants because it might have better potential than Ir to achieve both good activity and stability through synergistic effects. (For consistency in terminology, all metal elements in the precursor are referred to as "dopants", even though the first "dopant" is the primary metal. These are arranged in descending order of their proportions, from the first to the fourth.) Such a conclusion could be drawn by the distribution of percentages highlighted in dark grey in **Figure 2** and **Figure S3** for qualification rates (percentage that reached <250 mV $\eta_{10}$/ <1 mV $h^{-1}$ decay rate, color bar) lower than 20.0%. Delving deeper into the Apriori associate rule mining, we obtained several other key insights that help determine the initial exploration space, especially for empirical parameters**. (2) Hydrothermal:** For the hydrothermal process (which broadly includes precursor mixing), room temperature is sufficient from an activity perspective, and the corresponding time range suggested was 12-24 hours. However, for stability, a temperature range of 50-100°C was preferred, and the corresponding time range was 6-12 hours. To balance these factors, we hence define the hydrothermal parameter candidate space for temperature and time to be 25-60°C and 6-24 hours, respectively. **(3) Annealing:** For the annealing process, the activity side suggests a temperature range of 300-400°C, while the stability side prefers 500-600°C. For the annealing time, the activity side requests three to six hours, while the stability side requests over six hours. **(4) Dopants:** The Apriori analysis also revealed that Sr, Zn, and Mn are recognized as good secondary elements, potentially serving as major dopants. They have high lift values (ratio of observed to expected co-occurrence), indicating a strong association with good activity. Moreover, the stability-focused itemsets suggested that a third type of metal should be present, with a proportion of 5%~15% in the precursor. This emphasizes the potential of multi-metal synergy for improving catalyst stability.

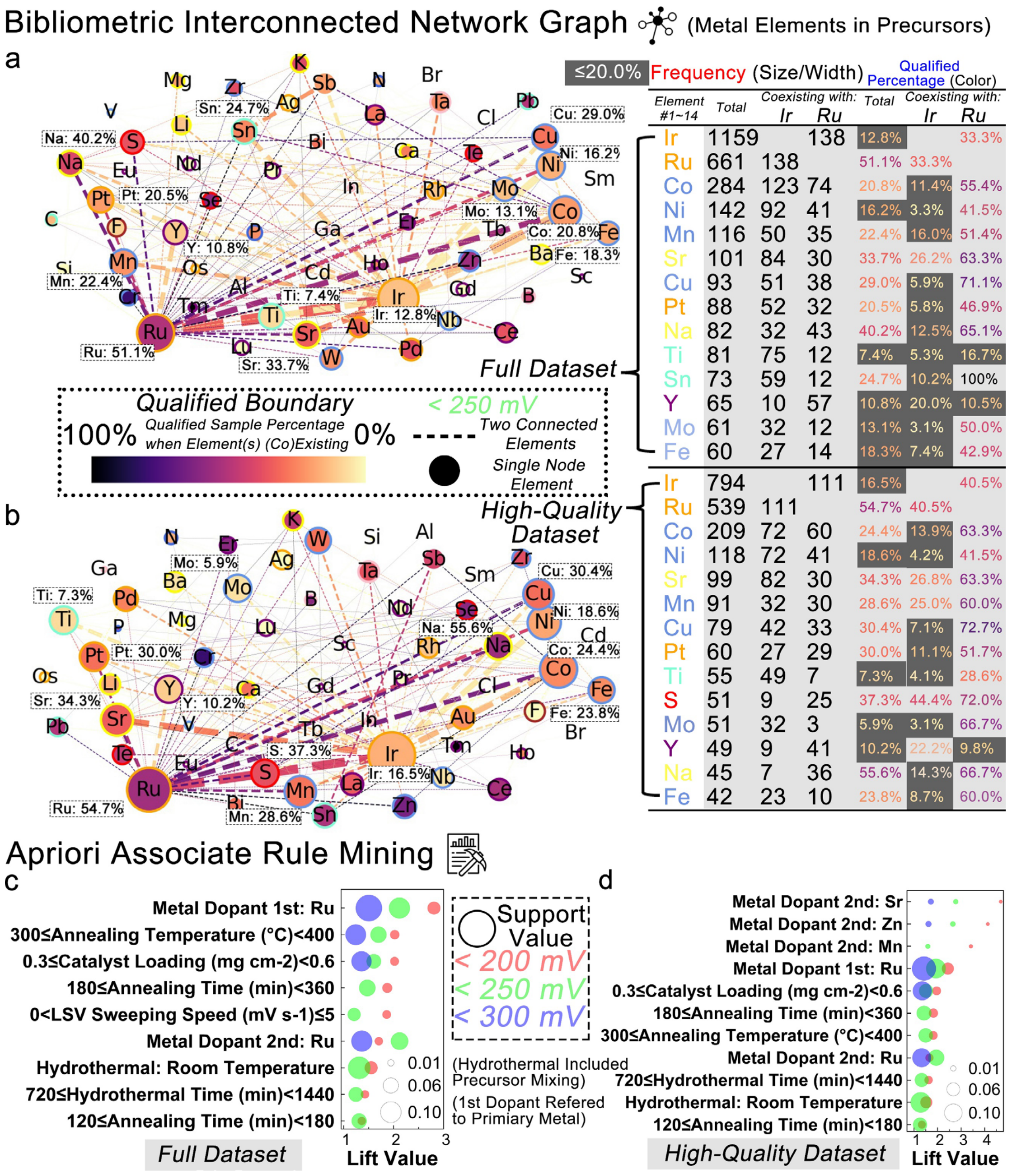

**Figure 2** Key results derived from unsupervised data mining focused on OER activity. A) Network graph based on the full dataset of chemical elements with a qualified overpotential boundary set at 250 mV. B) Network graph based only on the high-quality dataset of chemical elements with a qualified overpotential boundary set at 250 mV (for network graphs herein, different colors at the edges of nodes represent different groups of elements. Aquamarine: C group; royal blue: N group; red: O group; pink: B group; brown: halogen; cornflower blue: earth abundant–transition metal; purple: rare earth metal; yellow: first/second group metal; and orange: noble metal.). c) Results of high lift values in Apriori associate rule mining based on the full dataset with a frequent itemset length of two for activity-related insights; d) Results of high lift values in Apriori associate rule mining based only on the high-quality dataset with the frequent itemset length of two for activity-related insights.

## 2.2. Supervised Data Mining

Drawing on the insights from unsupervised data mining, we crafted from scratch a set of essential design rules to locate our initial exploration spaces to ensure balance between activity and stability. However, through dimensionality reduction techniques such as principal component analysis (PCA)[32] and t-distributed stochastic neighbor embedding (t-SNE)[33] (**Supplementary Note 3**), we delineated the intricate, high-dimensional nature of the datasets. This further revealed the limitations of unsupervised methods and underscored the need for more advanced ML strategies to accurately parse and interpret the complex patterns within our data. By leveraging ML models to discern data patterns, we can uncover more accurate qualitative patterns and laws governing parameter influence, enabling us to precisely define the most promising regions for continued investigation.

Based on the same datasets, we adopted a committee ensemble approach by training and evaluating (**Supplementary Discussion S1**) various ML algorithms. The optimized models were then integrated into a committee, and the influence of each member ML algorithm was determined by its regression performance. Specifically, models capable of generating more accurate predictions quantified by $R^2$ yield greater influence on the overall decision-making process of the committee, whereas those with negligible or negative $R^2$ values have minimal or no impact on the overall decision-making process. The committee-based query method used here is a common strategy in batch active learning[34]. Unlike ensemble algorithms such as Random Forest, which uses decision trees as basic learners, our approach aims to integrate models with completely different architectures and complexities on a higher scale, ranging from support vector regression (SVR), gradient boosting, LightGBM, CatBoost, k-nearest neighbors (KNN), AdaBoost, and Decision Tree, to eXtreme gradient boosting (XGBoost) to artificial neural networks (ANN), in addition to using Random Forest as one of the committee members. These different algorithmic architectures are capable of learning via different paths on the same dataset, enabling a more comprehensive and robust evaluation of the corresponding uncertainty. **Figure S4-Figure S11** illustrate the regression metrics of the four committees by two datasets and two prediction targets: full dataset-activity, full dataset-stability, high-quality-dataset-activity, and high-quality-dataset-stability. A low mean average error (MAE) and a high $R^2$ value notably highlight some of the committee members, especially the promising prediction accuracies of the Boosting-based algorithms. The gradient boosting regressor could achieve an $R^2$ value of 0.84 and a MAE value of 29.76 mV for $\eta_{10}$ on a full dataset (**Figure S6**). For a high-quality dataset (**Figure S7**), XGBoost stood out with the lowest MAE of 27.21 mV. Based on **Figure S2b**, we could infer that the best members of the two activity-predicting committees could limit their quantitative prediction error to less than 10% of the recognized level, illustrating its microscopic insights. Similarly, for the two stability-predicting committees, we could observe that a high $R^2$ value such as 0.86 could be achieved by XGBoost on a full dataset. And on a high-quality dataset, CatBoost could reach an impressive $R^2$ value of 0.89, demonstrating its excellent forecast consistency and macroscopic grasp of long-term stability and complexity. These best members of the two stability-predicting committees have acceptably overcome the inherent selection bias and data noise caused by the selective reporting and missing details in the literature.

Hence, we transitioned to an in-depth analysis of the models' decision-making processes, systematically leveraging and integrating the interpretative tools Shapley Additive Explanations (SHAP)[35], assisted by Friedman's H statistics[36], and Partial Dependence Plot (PDP)[37]. Based on the SHAP matrices assembled by the top performers in the committees, as illustrated in **Figures 3a-b**, we can discern patterns that illustrate how features influence the activity and stability of acidic OER catalysts. The pipeline employed is depicted on top of **Figure 3**, and unfolds as follows: initially, we identify the most impactful input features through the weighted average SHAP values, which are then coupled with H statistics to pinpoint the most nonlinearly correlated pairs of informative features for two-dimensional dependence plots. By thoroughly analyzing the results (**Supplementary Note 4**), we vividly capture and present the key insights from this data-mining phase in **Figure 3** (two committees trained on the high-quality datasets) and **Figure S12** (two committees trained on the full datasets).

We begin with **(1) Activity vs. Stability Factors:** Feature importance analyses are provided via cohort bar plots. For activity, element-related features, primarily atomic properties, were particularly influential. Both full and high-quality activity datasets consistently pointed to the atomic radius of the primary metal element as the most critical factor (**Figure 3c**, **Figure S12c**). In contrast, for stability, parameters related to synthesis thermodynamics and kinetics, such as temperature and time duration, prevailed (**Figure 3d**, **Figure S12d**). Interestingly, while the choice of the primary metal element appears crucial for activity, the atomic radius of the second and third elements ranked highly for stability despite their lower proportions. This suggests from the domain level that generally the primary metal essentially dictates activity, but the selection of secondary dopant metals, like the second or third, is equally pivotal for stability. **(2) Primary and Secondary Element Choices:** Further examination of the cohort plots unveiled a critical threshold at 127.5 pm for the atomic radius of the primary metal, a value autonomously identified by SHAP. Beyond this threshold, elements with a larger atomic radius exhibit a significant increase in importance. Dependence plots in **Figure 3e** and **Figure S12e** further indicate positive SHAP values under 127.5 pm as undesirable for decreasing $\eta_{10}$. A cluster of negative SHAP values around 138 pm identifies preferred elements such as Rh, Pd, Re, Os, Ir, and Ru, with Ru standing out as the most favorable, corresponding to the peak. For stability, **Figure 3f** and **Figure S12f** indicate a preference for a higher proportion of the second metal (greater than ~35%) and the selection from a broad spectrum of larger atomic radius elements (greater than ~136 pm), predominantly alkali/alkaline or heavy metals. Hence, it is inferred from the domain perspective that while Ru is essential as the primary metal for ensuring activity, extensive doping with a diverse range of elements is beneficial for stability.

Other broad domain-level insights include: **(3) Testing Parameters:** Despite its high-importance ranking, the nearly always reported stability testing time was included in the ML model to bolster data integrity, as it serves as the denominator when calculating time-averaged decay. Intriguingly, the cohort plots identified approximately 19 hours as a pivotal point, suggesting that annealing parameters assume greater importance over extended durations. Similarly, catalyst loading plays a notable role, shedding light on macro-scale marginal effects. As depicted in **Figure S12g**, beyond a preferred loading of 0.5 mg $cm^{-2}$, the reduction in SHAP values becomes less marked. Additionally, supporting materials like $TiO_x$ or carbon particles did not exhibit a focused distribution in the analysis, indicating their negligible impact and thus leading to our decision to exclude them from further consideration. **(4) Synthesis Parameters:** The thermodynamic and kinetic parameters (temperature and time) in the sample synthesis, such as annealing and hydrothermal conditions, are suggested to be nearly as crucial as the choice of metal elements for both activity and stability. Although they could be optimized directly like the proportion of metal elements, the exploratory costs in experimental synthesis differ, especially since the annealing step is time-consuming, and fine-tuning the optimal temperature can lead to significant energy consumption. Thus, a prudent approach is to confirm them as constants through domain-knowledge data mining. **Figure 3g-h** and **Figure S12h** visualize the pattern distributions of annealing temperatures in different committees. **Figure 3g** reveals a notable negative peak in the wavy SHAP value distribution trend around 200~500°C. Intriguingly, **Figures 3h** and **Figure S12h** from stability perspective display a similar yet inverse peak, with the intersection point with the SHAP value zero line around 450°C. Given the initial coarse-grained results from unsupervised data mining, we believe setting the annealing temperature directly to 400°C and 500°C for exploration will most likely balance activity and stability. Finally, the interaction features of the annealing and hydrothermal time did not exhibit a concentrated color distribution. Hence, considering the need to control costs, we fixed the annealing time to six hours based on unsupervised data-mining outcomes, while the hydrothermal time is included as one of the variables for optimization in the subsequent module.

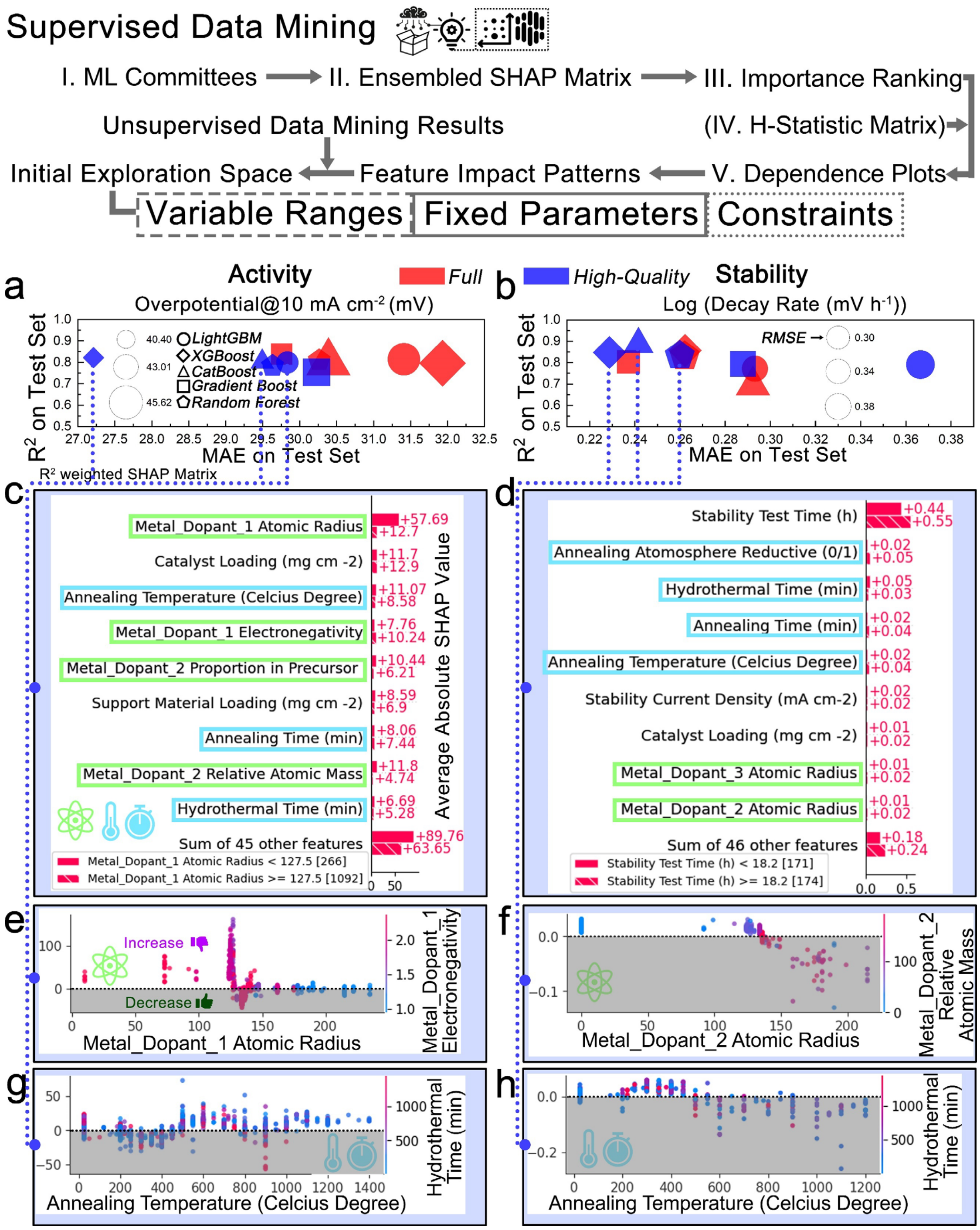

**Figure 3** Top: workflow schematic of supervised data mining module. Key results derived from supervised data mining based on two committees trained on the high-quality datasets. a) and b): The top-five ML models, as identified from committees trained on various datasets, are evaluated based on $R^2$ and MAE metrics on the test set. c)-d) SHAP cohort bar plots that highlight the important features with light green and blue frames highlighting the element-related features: atomic properties and synthesis condition parameters, respectively. e)-h) Selected SHAP two-dimensional interaction plots that feature an interaction of the primary studied feature on the X-axis with a second feature, which is indicated by the color bars. The second features, also chosen from the top features with a similar type, are those with a high degree of interaction in the Friedman's H-statistic interaction matrix. Dashed lines

at y=0 in each dependence plots split-grey areas that indicate the preferred value ranges.

# 3. Active Learning-Guided Experimental Exploration

Our data mining results provide qualitative ranges rather than precise guidance for optimizing complex multi-element systems, especially concerning the selection and proportions of various metal elements. Nevertheless, we have narrowed the exploration space based on domain-knowledge data mining, preparing for the next phase of our learning-based adaptive experimental design pipeline. This module uses an active learning strategy to address optimization in nuanced recipes, including the optimal numerical values for elements and proportions, and for undetermined synthesis conditions. **Figure S13** both displays our experimental process for synthesizing multi-metallic oxide samples (**Supplementary Discussion S2**) and lists the specific parameter determined by the ML committee in each round by a balanced query strategy. **Table S3** further records the set ranges, constants, and constraints (we required Ru to be one of the four elements in the selection group) for the corresponding experimental variables, namely the initial exploration space defined through previous data mining on domain knowledge. For cost efficiency, we focused on observing $\eta_{10}$ in synthesized samples, refining through activity-focused ML committees and experimental feedback.

Our comprehensive OER experimental outcomes, conducted in 0.5 M $H_2SO_4$, are summarized in **Figure 4**. **Figure 4a** (referred to as “batch 0” in the dataset) displays the Linear Sweep Voltammetry (LSV) polarization curves with Ru as the sole metal precursor. **Figures 4b-f** present the top samples and their corresponding curves from each experimental batch. Additionally, **Figure 4g** employs violin plots to showcase the distribution of the total 258 samples in five iterations, exclusively guided by the ML committee without human decision interference. A discernible trend emerges across iterations, with the best samples exhibiting progressively lower $\eta_{10}$ values, underscoring the active learning workflow’s efficacy in refining this complex material system. Initially, in the first batch our domain knowledge-based ML committee showed negligible enhancements compared to the baseline $RuO_2$-400 and $RuO_2$-500 in **Figure 4a**. However, by the fifth iteration, as depicted in **Figure 4f**, the final samples demonstrated significant improvements. These top-performing samples, labeled A to D, with sample B achieving an exceptionally low $\eta_{10}$ of 154 mV, surpassed 99.5% of the samples in our domain-knowledge datasets. Notably, our synthesis approach, characterized by its simplicity, holds potential for scalability.

While $\eta_{10}$ is an important metric, it is not the sole criterion for practical applications. Examining the Tafel slopes of the samples in **Figures 4a-f**, as presented in **Figure S14**, reveals that the top-ranked samples do not always excel in kinetic performance. Specifically, a reversed ranking from $\eta_{10}$ could be found in the final batch. Sample B, despite having the lowest $\eta_{10}$, exhibits the highest Tafel slope (80.9 mV $dec^{-1}$), indicative of its slower electrochemical kinetics compared with sample C (69.0 mV $dec^{-1}$). Further evaluation of the samples' electrochemically-active surface area (ECSA) is presented in **Figure S15**, where the ECSA values were calculated based on the measured double-layer capacitance ($C_{dl}$). B exhibits the largest ECSA (2357 $cm^2$) and C has the lowest ECSA (923.3 $cm^2$). However, normalizing the OER current to ECSA at $\eta$ = 200 mV, C demonstrates the highest specific current density (0.0186 mA $cm^{-2}$ ECSA). We also conducted nitrogen adsorption-desorption isotherms of A~D (**Figure S16**). The fitting results indicate a consistent trend: sample B exhibits the highest Brunauer–Emmett–Teller (BET) surface area of 94.35 $m^2$ $g^{-1}$, greater than that of A at 88.88 $m^2$ $g^{-1}$, D at 74.47 $m^2$ $g^{-1}$, and C at 39.92 $m^2$ $g^{-1}$. Based on these results, we conclude that A/B/D might have advantages in surface area, which could enhance exposure and result in a higher density of active sites and thus lower their overpotentials under low-current density half-cell tests. However, C exhibits a better reaction kinetic rate, supported by its best Tafel slope (smallest) and specific current density (highest) normalized by ECSA.

This insight led to additional testing in real-world scenarios, specifically focusing on single-cell PEM electrolyzers to assess performance at higher current densities, reaching levels of 1 A $cm^{-2}$. As depicted in **Figure 4h**, sample C, despite having a comparatively higher $\eta_{10}$ and superior kinetics among the final four, displayed remarkable performance under high-current density conditions in the single cell. A commercially purchased $IrO_x$ sample with a 0.5 $mg_{Ir}$ $cm^{-2}$ loading was outperformed by sample C. While sample C achieved a significant current density of 3 A cm$^{-2}$ at 2V, samples A/B/D were deemed subpar. Electrochemical impedance spectroscopy (EIS) conducted on the corresponding MEA samples (**Figure S17**), verified that sample C, as part of the MEA component, exhibited comparatively lower contact, charge transfer, and mass transfer resistances at both lower (0.1 A cm$^{-2}$) and higher (1 A cm$^{-2}$) current densities. This outcome suggests that despite its smaller surface area, sample C demonstrates superior comprehensive behavior under high-current density conditions. Conversely, samples A/B/D experienced higher impedance in practical applications despite their initial advantage in $\eta_{10}$. Therefore, to provide a more comprehensive evaluation, we further conducted accelerated stability tests under different scenarios. Initially, we performed half-cell constant current tests, which are quicker due to harsher conditions. Consistent with predictions by the stability ML committee in the supervised data-mining module that sample C is the most competitive in long-term performance (**Figure S18**), real experimental results consistently revealed that sample C indeed outperformed all others (**Figure S19**). In practical-level electrolyzer cell tests, MEAs loaded with sample C demonstrated considerable stability, as illustrated in **Figure 4i**. Over a prolonged testing duration of 125 hours, the average decay rates were only 0.1728 and 0.1964 mV h$^{-1}$ at 10 and 20 mA cm$^{-2}$, respectively.

In summary, while the ML committee's proposed element recipes for samples A/B/D (all containing Ru, Ca, Sr, and Nd) had maximized $\eta_{10}$ reduction after five iterations in half-cell tests, they underperformed in practical single-cell tests due to kinetic or stability issues. In contrast, sample C showed enhanced kinetics, charge/mass transfer characteristics, and electrochemical stability, which are crucial for device-level applications. Hence, while recipes for A/B/D may result in a higher surface area, C demonstrates superior performance overall and the potential for practical PEM electrolyzer applications. This outcome highlights the necessity of assessing electrocatalyst performance across diverse test conditions and metrics. It also emphasizes that expert input and thorough analysis are essential to complement ML-driven results for the multi-objective optimization that is crucial for practical scenarios, thus ensuring the continuous refinement and applicability of ML strategies.

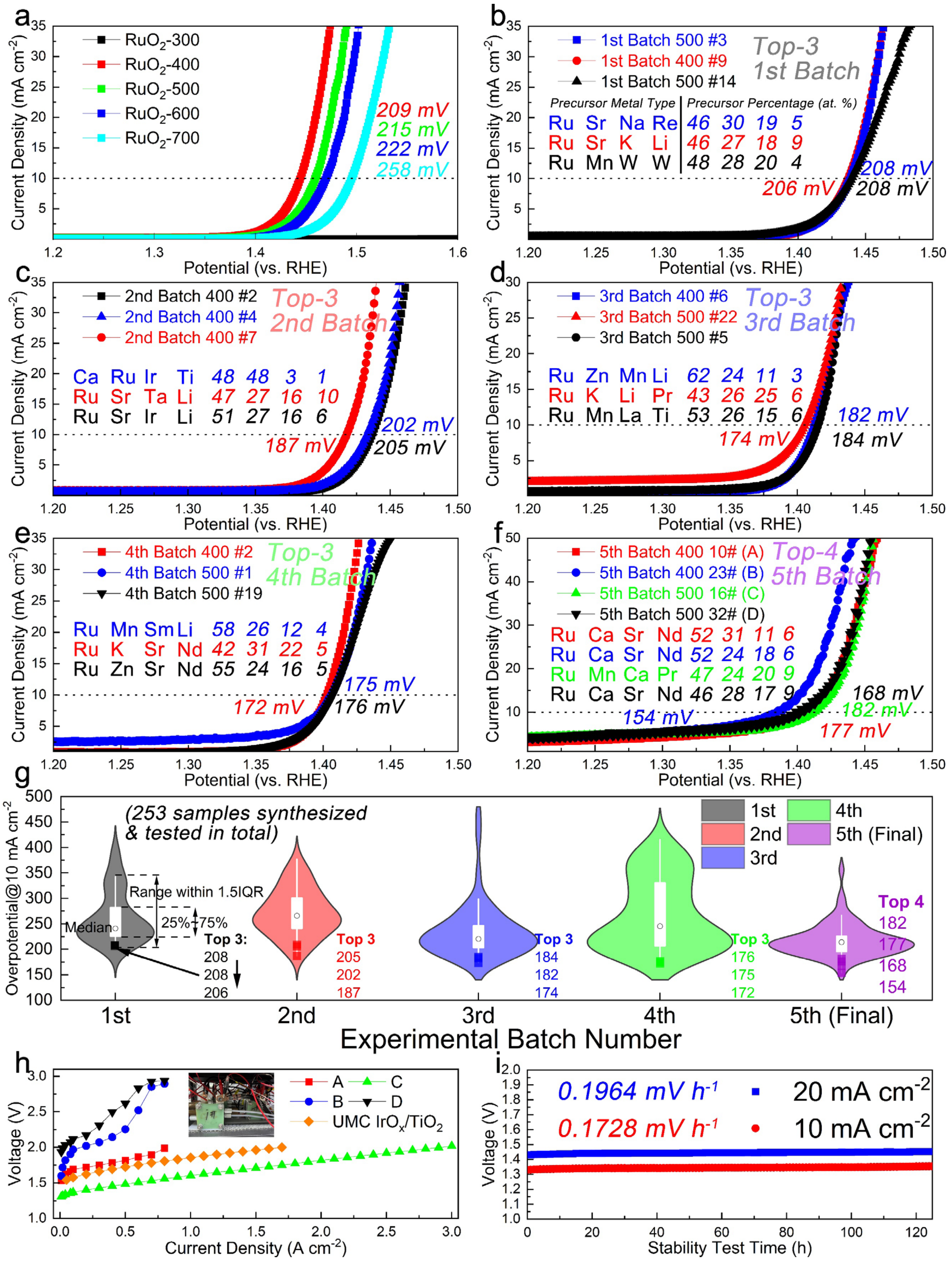

**Figure 4** a) LSV curves of batch 0, Ru-only samples obtained by different temperatures. All LSV curves are represented using line-scatter plots with legends. Due to inferior activity, the curve of $RuO_2$-300 nearly overlaps with the X-axis. b)-f) LSV curves of the best-performing top samples in the first to the fifth (final) ML committee-guided experiment batch. g) The violin plots showing the distribution pattern of $\eta_{10}$ values of the samples tested and synthesized under the guidance given by the DASH workflow. h) The polarization curves in a PEM electrolyzer single cell with MEA samples loaded with different samples of A-D obtained in the final batch shown in **Figure 4f**.

i) The V-t curves of sample C loaded on MEA in a PEM electrolyzer single cell with 10/20 mA $cm^{-2}$ constant current density.

## 4. Material Characterization

After identifying the optimal sample C through the DASH loop, we sought to gain deeper insights behind the enhanced performance. A comprehensive suite of characterization techniques from a materials science perspective was conducted (**Supplementary Note 5**, results of 11 samples other than C). A total of 12 samples were included: samples A~D, $RuO_2$-400/500, and their respective states before the acid wash (identified with the suffix "Pre"). High-resolution transmission electron microscopy (TEM) results of the 12 samples are provided in **Figure 5a-b** and **Figures S20-S22**. In general, the samples obtained consist of small, single-crystalline $RuO_2$ particles, approximately 5 to 10 nm in diameter, interconnected to form a polycrystalline structure. As indicated by **Figure 5c** and **Figure S22a**, sample C's lattice spacing of the (110) lattice plane expanded to 3.25 Å when compared to the standard lattice spacing of the pure Ru samples both before ($RuO_2$-400-Pre, 3.12 Å) and after ($RuO_2$-400, 3.09 Å) the acid wash process. Subsequent EDX analysis via TEM in **Figure 5d** further strengthened our belief that the second to fourth metal elements are uniformly distributed with non-aggregated signals detected, suggesting their existence as dopants in the major $RuO_2$ lattice. To further validate our hypothesis, we examined X-ray diffraction (XRD) spectra for more crystallographic information. Having been acid-washed, sample C seemed to be dominated by the $RuO_2$ phase (**Figure 5e**) with the impurities in C-Pre well removed; and no traces of $CaRuO_3$ and $SrRuO_3$ could be found like that in A/B/D. Such perovskites are known by consensus to dissolute cations and cause the irreversible collapse of polymetallic oxide structures[38-41], thus possibly contributing to the inferior stability behavior as observed in experiments. Additionally, we observed noticeable leftward shifts in the XRD peaks for the main phases, signifying slight expansions in the lattice parameters in all A~D samples, especially for (110) and (101) facets compared with $RuO_2$-400/500[42]. This is further supported by the comparison of the average interplanar spacing from TEM images across 12 samples (**Figure S23**).

To further prove the doping of metal elements into $RuO_2$ lattice, we obtained X-ray photoelectron spectroscopy (XPS) spectra for the surface chemical states, with a focus on sample C. Combining the XPS semi- and inductively coupled plasma (ICP) quantitative analysis summarized in **Figure S24**, we confirmed that Mn was the major dopant. Hence, we first examined the XPS Mn-2p spectra in **Figure 5f**. Compared with that before acid wash, the $Mn^{2+}$ 3/2 peak of sample C shifts toward a lower binding energy, which indicates Mn cation's difference with that in Mn oxides and as a dopant into the Ru oxide lattice as reported[43]. This was further supported by electron paramagnetic resonance (EPR) comparisons of magnetic properties (**Figure S25**). We further investigated the Ru 3d and 3p XPS spectra in **Figure 5g**. Different from A/B/D, the Ru 3d 3/2 peak of sample C exhibited a slight shift towards a higher binding energy compared with $RuO_2$, consistent with previous reports indicating Mn's doping effect on Ru's local electronic structures[44,45]. As for Ru's 3p peaks, no obvious shift in peaks could be found, but the results are hidden in peak-fitting. Based on previous reports, we calculated the $Ru^{>4+}$ species ratio, which has been proven to optimize the OER reaction pathway and enhance the charge transport by both experimental and theoretical approaches[43]. As shown in **Figure 5h**, we could find that sample C has the highest ratio (35.24%) among A~D. The result again supports C's superiority in terms of intrinsic activity and charge transfer characteristic, which could overcome its disadvantage in surface area and become the most practical sample.

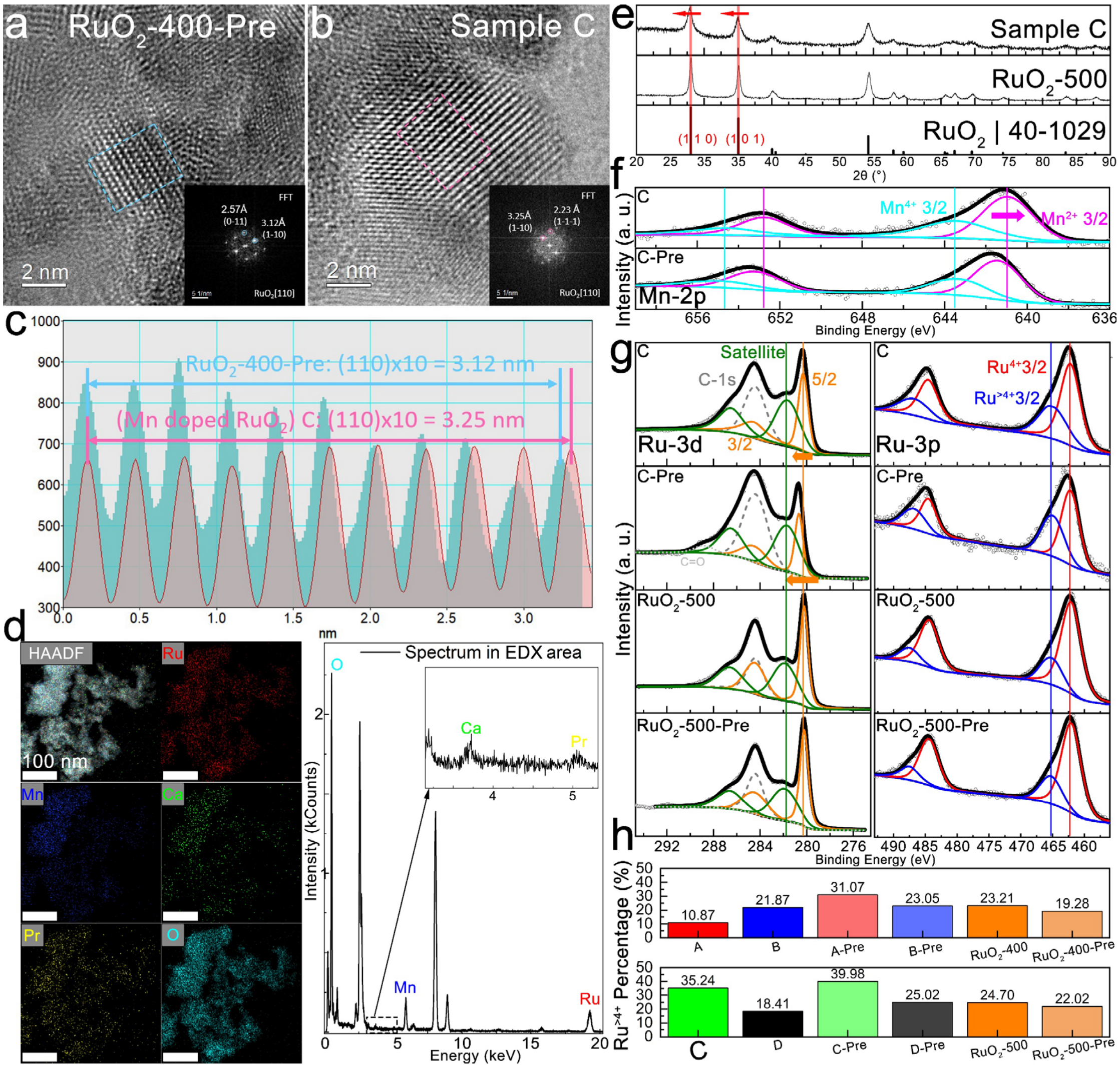

**Figure 5** a) and b) are high-resolution TEM images of $RuO_2$-400-Pre and sample C, respectively, with cyan and pink frames selected for performing Fast Fourier Transform (FFT) to analyze the diffraction pattern. c) Corresponding statistical analysis of the average interplanar spacing; d) EDX elemental distribution mapping spectrum for various elements in sample C. e) XRD spectra of sample C compared with $RuO_2$-500. f) XPS Mn-2p spectra of sample C and its pre-acid wash state: "C-Pre". g) XPS Ru-3d and 3p spectra of sample C, $RuO_2$-500, and their corresponding states before acid wash. h) Statistical analysis of the XPS-derived $Ru^{>4+}$ species percentage in the samples.

## 5. Domain Adaptation-Assisted DFT Theoretical Simulation

To further understand the origin of the superiority of C's activity and stability, DFT simulation in the next stage is performed. However, since numerous elements are involved with varying doping ratios, we need to perform a large number of expensive DFT relaxations to find the stable structure of each sample before conducting surface catalysis simulations. Therefore, in this last module, a novel domain-adaptation strategy based on commonly reported DFT ML surrogate modeling[46-48] was illustrated.

The characterization results suggested that the C is $RuO_2$ doped with multiple metal elements, and the same for A/B/D's surfaces after Sr and Ca were dissolved during acid wash. Therefore, in order to simplify and unify the qualitative discussion, we selected $RuO_2$ (110) as the subject of multi-metal doping research. **Figure 6a** presents the schematic workflow for the domain adaptation strategy. We aimed to achieve effective prediction of stable configurations by the ML committee when dopant contents were provided. As we are interested in handling a wide range of both the types and amounts of dopants, similar to previous experimental modules, the DFT relaxation calculations for sampling this vast candidate space from scratch would be prohibitively expensive for ML surrogate modeling. Hence, we divided the entire set of metal elements of interest into two domains: the source (S) domain candidates include common earth-abundant metals like Fe, Co, Ni, and the relatively rare elements such as La and Ce are categorized into the target (T) domain. We provided abundant data entries in the S domain (3,973) through high-throughput DFT relaxation calculations (**Supplementary Note 6**), while only about one-fifth (848) of the data entries were provided with the T domain concerned. Therefore, in **Figure 6b**, Committee S achieved satisfactory predictions for slab energy values across its ML committee members that had $R^2$ values close to or over 0.99. In comparison, Committee T exhibited inferior performance due to its broader candidate space to be sampled and fewer data entries available as expected. Next, instead of directly expanding the dataset size, we employed domain adaptation to avoid potential exponential growth in the supplemental DFT calculations needed to cover additional elements in the T domain. This involved fine-tuning the members of Committee S on Dataset T for a limited number of epochs. **Figure 6b** along with **Figure S26** demonstrates the corresponding results, showing that the fine-tuned Committee S-T significantly improved its regression metrics on Dataset T. Compared to Committee T, Committee S-T was able to use the same small-volume dataset T to adaptively fine-tune the general pattern learned from the large-volume dataset S, further including situations when elements in the T domain were added. Moreover, similar patterns (**Figures S27-S29**) could be observed when OER-related intermediate species, such as O, OH, and OOH, are present on the slab surface, further supporting the broad effectiveness of this strategy. In **Supplementary Note 7**, we further address the potential "forgetting" issues[49] of Committee S-T, underscoring the substantial domain differences between candidates S and T through cross-domain evaluation, yet affirming its cost-effectiveness for multi-metallic system exploration. Consequently, Committee S-T can effectively serve as the surrogate model with significantly lower costs in the far vaster candidate space that we are interested in exploring.

Next, we provided the quantitative composition results to Committee S-T and identified stable doping configurations, specifically the lowest slab energy, using the method described in DASH as illustrated in **Figure 6c**. Without extensive calculations to determine the locations of the dopant atoms, four representative structures for samples A~D were selected using our ML surrogate based on domain adaptation. Subsequently, we conducted OER simulations on these slabs. Initially, we comprehensively investigated the theoretical OER activity by simulating reaction pathways on the surface covered by various species[25,50]. **Figure 6d** indicates that the theoretical OER overpotentials for sample C, computed from the energy barriers of the rate-determining steps (**Figure S30**), were overall the best compared to the other samples. Although slightly higher than sample B and undoped $RuO_2$ in the O-covered scenario, the OER overpotentials for C were the lowest in the other two scenarios at 403 mV and 376 mV, with the latter considered the commonly studied descriptor of OER activity[30]. Further examination of the density of states in **Figure S31** revealed that C's Ru 4d and O 2p band centers were closer to the Fermi energy level, contributing to its optimized reaction pathway as reported[51-53]. This might explain our experimental observation that C possesses the best electrochemical kinetics among A~D. It has optimal intrinsic single-site activity due to its refined electronic structures. We also assessed the theoretical stability (shown in **Figure S32** and summarized in **Figure 6e**). Although not the best (-1.20 eV), C and all other samples demonstrated favorable water adsorption energies ($\triangle G^{*}_{H2O}$), which can be regarded as a straightforward yet effective descriptor of oxide's stability in water[54-56]. Consistent with experimental observations, they are stable and unlikely to dissolve. Further calculations of the surface Ru dissolution potentials ($U_{diss}$)[46,57,58] and surface O vacancy formation energy ($\triangle G_{VO}$) indicated that C was the best, with values of 3.34 V and 3.84 eV, respectively. The former suggests that C is the most electrochemically stable. Importantly, the latter, which is commonly examined in OER-related studies[59-61], indicates the stability of Ru-based materials in acidic electrolytes, as O vacancies could lead to over-oxidation and the generation of soluble Ru species. In summary, our DFT simulations of both theoretical activity and stability for samples A~D correlate well with the experimental results. Finally, for readers interested in further explorations based on Committee S-T, **Supplementary Note 8** provides additional insights into broader candidate spaces and innovative dopants.

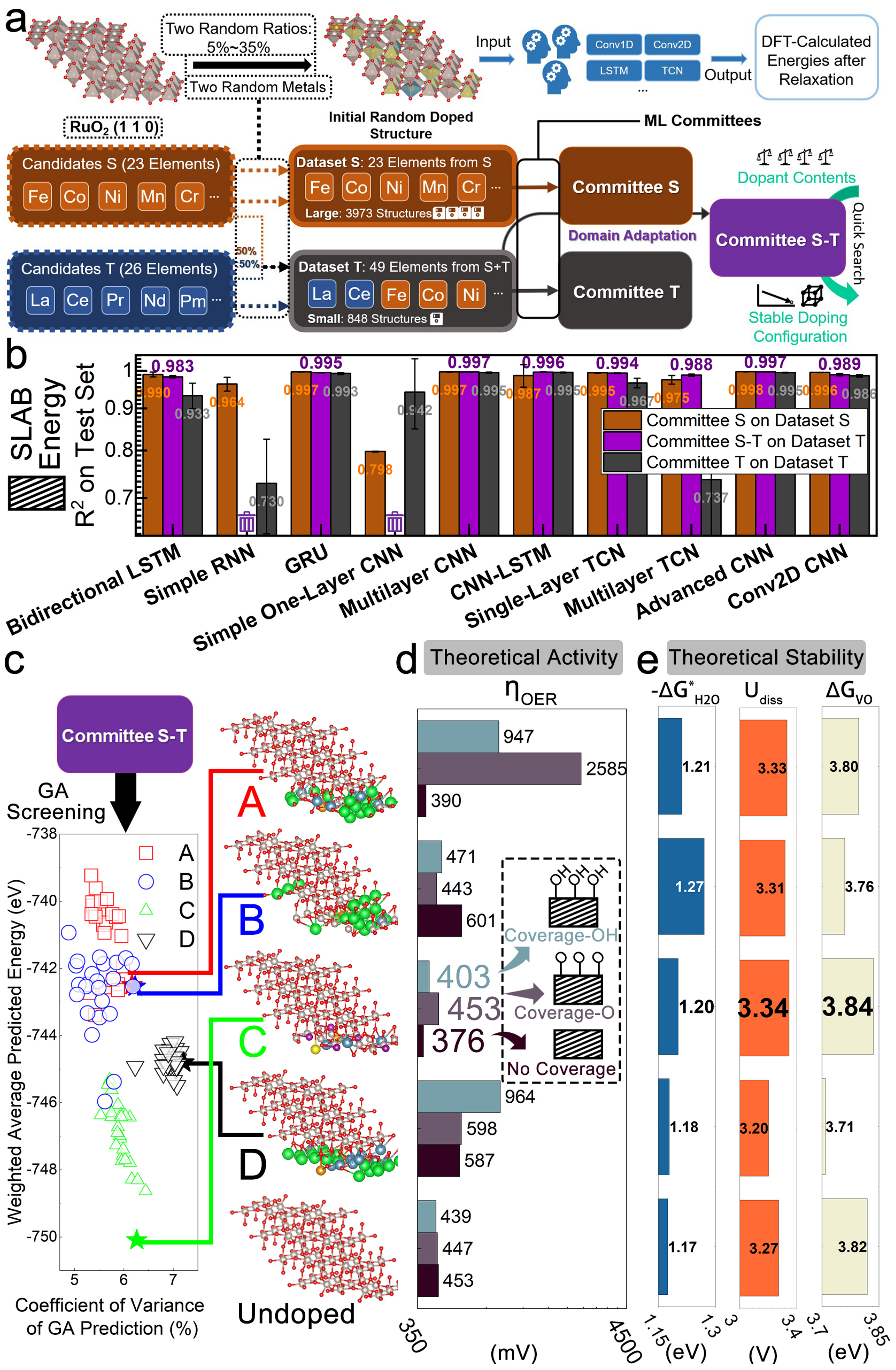

**Figure 6** a) Schematic of the domain adaptation workflow to obtain Committee S-T. b) A summary of the committees' $R^2$ values on the test sets. The trash bin icons were placed to indicate that for Committee S-T, corresponding members from Committee S and Committee T were deprecated due to inferior performances. c) Results of the GA screening using committee models S-T for identifying the lowest slab energy of compositions corresponding to experimentally obtained samples A~D, and the corresponding determined doping configuration structures with the lowest slab energy. d) Summary of the theoretical OER overpotential of the samples under different coverage scenarios. e) Summary of the theoretical stability descriptors of the samples.

# 6. Conclusion

This comprehensive study represents a paradigm shift in the development of advanced electrocatalysts for acidic OER. Our research unfolds a multi-stage, machine learning-driven approach that embodies the organic fusion of data mining, active learning, and domain adaptation, leveraging these methodologies synergistically across different stages of materials discovery based on varied data sources. This integration not only streamlines the journey from conceptualization to experimental validation and theoretical investigation but also ensures a flexible and reliable data-driven exploration of complex multi-metallic systems. Unlike traditional methods that rely on chemical intuition and trial-and-error, our approach applies a rational, hierarchical, data-driven decision-making process across an immense compositional space, using different ML modules at various stages of the pipeline. Furthermore, while current ML studies in the field often remain confined to single stages of research based on a single modality of data, our work extends the application of ML to encompass the entire process of materials discovery, thus addressing challenges that single-stage models cannot. Through this innovative framework, we have successfully identified an Ru-Mn-Ca-Pr catalyst that exhibits exceptional activity and stability and has been validated extensively through experimental and theoretical approaches. This methodology establishes a new benchmark for the field of electrocatalysis, demonstrating the potential to revolutionize the discovery and optimization of OER catalysts through a comprehensive, data-driven strategy.

.

# Method

## Domain-Knowledge Dataset

A systematic literature review was conducted to compile a dataset of experimental studies on metal-oxide type electrocatalysts for OER in acidic conditions. Studies involving theoretical DFT or molecular dynamics simulations were excluded due to the variability in outcome metrics compared to experimental results. The literature search was performed using the Web of Science, resulting in a comprehensive collection of relevant publications. The curated dataset includes input features such as transition metal elements, proportions of precursors, conditions of hydrothermal mixing (precursor mixing), annealing process conditions, post-treatment conditions, and testing conditions. Output fitting targets for catalyst activity and stability, specifically $\eta_{10}$ and time-averaged voltage decay rate, were recorded in the data entries. For minor missing values in the dataset, the median value from the dataset was used for imputation, a method validated by previous studies [62] in the field of ML[63-65]. Criteria were established to differentiate a "high-quality dataset" from the initial dataset, focusing on records from high-impact journals with high citation counts and more recent publication dates to ensure data reliability. The high-quality dataset post-screening included 1,358 entries (**Figure S2b**) related to electrocatalytic activity and 345 entries addressing stability under constant current density tests (1,847/453 in the full dataset). Four versions of datasets capturing domain knowledge were thus established: high-quality-activity, full-activity, high-quality-stability, and full-stability.

In both unsupervised and supervised data mining, insights from the high-quality datasets and the full initial datasets were integrated. This strategy was also applied in the active learning module called DASH flow. Two ML committees, one initiated from the high-quality dataset and the other from the full dataset, were updated iteratively with the data supplements derived from experimental observations. Candidate observations were determined based on recommendations from both committees, maintained at a 1:1 ratio in all genetic algorithm (GA) search result batches, to ensure diversity and reduce potential biases in the learning process. Details regarding reproducibility are documented in **Supplementary Note 1**.

**Unsupervised Data Mining**

The Bibliometric Interconnected Network Graph and the Apriori Associate Rule Mining are applied, building upon the foundational work referenced in previous studies[27]. These methods aim to explore relationships among itemsets categorized as "qualified" or "disqualified". Thresholds for OER activity have been established at 200, 250, and 300 mV for overpotential; for OER stability, thresholds are set at -1, 0, +1 log (mV $h^{-1}$) for decay rate, consistent with those used in the level set estimation.

A technique similar to that employed by VOSviewer bibliometric software[26] facilitates visualization of the distribution of key elements within the dataset via an interconnected network graph. In such graphs, nodes symbolize different elements, where the color of the node edges differentiates element groups and node size denotes occurrence frequency. The intensity of a node's inner color indicates the proportion of high-quality samples associated with that element (e.g., $\eta_{10}$ of OER < 250 mV, as depicted in **Figure 2 a-b**); here, its values are referred to as "quantification rates". The width of the lines between the nodes represents the frequency of element pairs co-occurring, and the depth of line color reflects the likelihood of these pairs being part of high-quality entries. Qualification rates, defined as the probability that the presence of an element or a pair of elements in the system contributes to reducing overpotentials or decay rates, are determined. The colors of the value numbers in the summary columns of **Figure 2** match the corresponding node/edge colors. Network graphs are produced using the Python package, NetworkX.

Similarly, Apriori Associate Rule Mining is applied to extend the analysis to additional features for pattern discovery. The same thresholds previously mentioned are used for itemset comparison across different datasets. This method incorporates continuous variables such as synthesis and testing parameters into the analysis. For interpreting results, particularly in bubble figures, two key metrics are emphasized: "Lift" and "Support". A high "Lift" value, calculated as the ratio of the observed support to the expected support if the items were independent, indicates a strong association with a target itemset linked to favorable performance. Conversely, a high "Support" value, representing the frequency of occurrence of an itemset within the dataset, shows its significance in the analysis. In visualized figures, itemsets with the highest lift values are displayed, indicating the parameters most likely associated with high performance. Details on reproducibility are provided in **Supplementary Note 2**.

**Supervised Data Mining**

A committee ensemble strategy was used in the supervised data mining module to enhance the robustness of the ML workflow and minimize potential biases or errors. This ensemble, comprising 11 models including SVR, K-Neighbors Regressor (KNR), Light Gradient-Boosting Machine (LightGBM), CatBoost, XGBoost, Gradient Boosting, Random Forest, Decision Tree, AdaBoost, and Multi-Layer Perceptron (MLP) with one or two hidden layers, integrates insights from diverse architectures. It is noted that LightGBM, CatBoost, XGBoost, Gradient Boosting, Random Forest, and AdaBoost are generally recognized as "ensemble models" in computer science, indicating their composition from base learners such as single decision trees. In this study, each of these "ensemble models" is considered a member of a higher-level ensemble, the "committee," alongside other models like SVR or KNR. Each committee member was subject to five-fold cross-validation to fine-tune hyperparameters, aiming to optimize performance.

Interpretation of the internal decision-making processes of these models was primarily performed using SHAP analysis, supplemented by partial dependence plot (PDP) and Friedman's H statistics. For a broader perspective, the concept of insight assembly was expanded, as indicated by dashed lines in **Figure 3**. Rather than relying solely on the SHAP value matrix of the best-performing committee member model, the weighted average SHAP values of the top-three performing models (e.g., LightGBM, XGBoost, and CatBoost for the committee trained on the high-quality dataset) based on $R^2$ values were used. This approach enhances the SHAP matrix to support parameter insights in SHAP analysis plots. Additionally, due to high input dimensionality in supervised ML models, the SHAP results were segmented by predefined feature categories into two domains: "Element" and "Synthesis & Testing" for clearer analysis. This segmentation facilitates intuitive inspection and clearer insights. Alongside this, Friedman's H statistics were employed to identify high-order, nonlinear correlations within influential features. Target feature pairs, exhibiting both high importance in SHAP and elevated second-order H statistics, indicative of significant nonlinear dependencies, were analyzed. The use of PDPs together with SHAP dependence plots for these feature pairs allows a comprehensive understanding of the patterns and relationships that influence model predictions. Details regarding reproducibility can be found in **Supplementary Note 4**.

### Active Learning-Guided Experimental Exploration

The DASH loop was developed to synthesize Ru-based multi-metallic oxides with up to four metal elements, using an iterative active learning cycle driven by ML committees for predicting activity. For each iteration, the two ML committees, trained on the current datasets (high-quality and full datasets, with supplement from each iteration), estimate the expected values and uncertainties for unexplored points within the exploration space, as indicated in **Table S3**, derived from earlier domain-knowledge data mining. The robustness and fairness of these estimations are ensured through a weighted and averaged approach based on the $R^2$ value of each model in the committee from the test set, consistent with methods used in prior data-mining modules. For inferior models with $R^2$ values lower than 0, its corresponding weight coefficient would be set to 0 to nullify its contribution in the committee decision. In the first iteration, the two ML committees based on high-quality and full datasets for activity in previous supervised data-mining modules are directly applied for a batch parameter suggestion GA search. Similarly, after the first iteration's experimental observations, the corresponding data point records would be supplemented with the existing datasets for the next iteration.

The GA search identifies a batch of preferred observation points within the exploration space. Parameters determined during the GA search include types of metal elements, corresponding precursor proportions, and hydrothermal conditions, as detailed in **Table S3**. A balanced batch query strategy accounts for both the expected $R^2$ weighted low average (overpotential) and high variance values in predictions (prediction discrepancies among committee members). The GA search is conducted at random, generating a batch of suggestions and maintaining a 1:1 ratio between suggestions with lower weighted averages and those with higher variances. Furthermore, the pyrolysis temperature settings alternate between 400 and 500 °C, also in a 1:1 ratio. The ML committee alternates between models based on the full dataset and high-quality-only dataset, in a 1:1 ratio, leading to 2×2×2=8 types of parameter suggestions for each batch for experimental synthesis and validation, all in equal proportion to minimize bias. This method incorporates diversity in batch candidate generation, in line with the community consensus regarding the integration of various decision-making systems and expert systems within the active learning framework, as supported by recent studies[66,67]. The hyperparameters for the GA search are set to a population number of 3,000, a maximum of 50 iterations, and a variation probability of 0.01. Each parameter suggestion type is repeated eight times, resulting in a batch of 64 suggestions for experiments. Following the synthesis of the recommended formulations indicated by the 13-dimensional vector from the GA search, the corresponding experimental synthesis and sample evaluations are undertaken. These observations are then digitized and incorporated into the dataset to inform updates. Subsequently, the next iteration of the ML committee is trained on this enriched dataset to commence the next round.

**Domain Adaptation-Assisted DFT Theoretical Simulation**

Domain adaptation was applied in the DFT section to align the predictive capabilities of machine-learning models between two datasets: Dataset S, which includes transition metals commonly used as dopants, and Dataset T, which encompasses a wider array of rare earth elements. To begin, separate ML committees—Committee S and Committee T—were formed and trained rigorously on their respective datasets to achieve robust predictive accuracy for each specific domain. Training included a variety of deep-learning architectures such as bidirectional long short-term memory (LSTM), gated recurrent unit (GRU), convolutional neural networks (CNNs), and temporal convolutional networks (TCNs), which underwent systematic, ten-fold cross-validation to ensure model robustness.

Domain adaptation was carried out by fine-tuning the best-performing models from Committee

S using Dataset T, adapting their learned features to the unique characteristics of the rare elements in Dataset T. Fine-tuning was carefully managed to limit the number of training epochs, and employed a protocol starting from the pretrained weights and biases of Committee S to minimize the risk of catastrophic forgetting. During this adaptation, simpler models that showed poor performance or instability—namely the simple RNN and one-layer CNN—were omitted from Committee S-T to ensure the reliability and stability of the adapted models. In the pre-processing stages, advanced normalization techniques were used to standardize input and output data. Different scaling techniques, including MinMaxScaler and RobustScaler, were applied to various parts of the data to maintain realistic conditions and variability in training. The performance of the adapted models in Committee S-T was then evaluated to confirm their improved predictive accuracy on Dataset T.

Following the establishment of Committee S-T, GA was again used as in the final step of the DASH loop to determine the doping configuration, specifically the position of dopant atoms leading to the lowest slab energy. This process was iterated 25 times to generate a batch of calculation results. The weighted (by $R^2$) average and variance of the committee members' predictions in Committee S-T were evaluated. The optimal structure was selected based on the optimistic estimation strategy[68], specifically choosing the point in the repeated results that has the most negative value of committee expectation (y-axis value) * [1 + uncertainty percentage (x-axis value)].

Refined simulations were then performed on screened slabs to assess the theoretical OER activity and stability of the samples. DFT calculations assessed the slab energies and adsorption properties of OER intermediates (O, OH, and OOH) on doped $RuO_2$ surfaces under various surface-coverage scenarios based on a standard adsorption evolution mechanism. The assessment of stability involved calculations of Gibbs free energy changes for water adsorption on the surfaces ($\triangle G^*_{H2O}$), the dissolution potentials of surface Ru ($U_{diss}$), and the formation energies of O vacancies( $\triangle G_{VO}$). Secondary reproducibility details are provided in **Supplementary Note 6**.

**Data Availability**

In line with the principles of open access and knowledge-sharing in the ML community, all ML training and data mining scripts used in this study, datasets extracted from the literature for data mining and initial ML committee training, high-throughput experimental/DFT computational data, characterization results of the samples, and other supplementary data mining results and discussion are publicly accessible at https://github.com/ruiding-uchicago/DASH for interested readers to review in detail.

**Supplementary Materials**

The secondary details are available in the **Supplementary Materials** document, including **Figures S1-S32**, **Tables S1-S3**, and **Supplementary Discussion S1-S2**. More trivial reproducibility details are available in **Supplementary Notes 1~8** (Available on https://github.com/ruiding-uchicago/DASH).

# Acknowledgements

This project is supported by the Eric and Wendy Schmidt AI in Science Postdoctoral Fellowship, a Schmidt Sciences program, at the University of Chicago. This work is also in early stages, partially supported by the National Key Research and Development Program of China (2022YFB4002303), the National Natural Science Foundation of China (U23B2075, 52272039, 51972168), and the Research Grant Council of Hong Kong Special Region (16308420). Part of the computational resources used by this work are provided by the University of Chicago's Research Computing Center and the High-Performance Computing Center of Nanjing University, China.

# Author contributions

**Rui Ding:** Conceptualization, Experiments, Characterization, Data Analysis, ML Modeling, DFT Simulation, Drafting.

**Kang Hua**: Experiments, Characterization

**Jianguo Liu**: Resources, Manuscript Reviewing

**Xuebin Wang**: Resources, Manuscript Reviewing

**Xiaoben Zhang**: Data Analysis, Manuscript Reviewing

**Minhua Shao**: Resources, Manuscript Reviewing

**Yuxin Chen**: Supervision, Resources, ML Methods, Manuscript Reviewing

**Junhong Chen**: Supervision, Resources, Editing, Manuscript Reviewing

**Supplementary Materials** for

# Leveraging Data Mining, Active Learning, and Domain Adaptation in a Multi-Stage, Machine Learning-Driven Approach for the Efficient Discovery of Advanced Acidic Oxygen Evolution Electrocatalysts

Rui Ding[a, b, c †], Jianguo Liu[d †], Kang Hua[e], Xuebin Wang[e], Xiaoben Zhang[a, c], Minhua Shao[f], Yuxin Chen[g*], and Junhong Chen[a,c*]

[a] Pritzker School of Molecular Engineering, University of Chicago, 5640 S Ellis Ave., Chicago, IL 60637, United States

[b] Data Science Institute, University of Chicago, 5801 S Ellis Ave., Chicago, IL 60637, United States

[c] Chemical Sciences and Engineering Division, Physical Sciences and Engineering Directorate, Argonne National Laboratory, 9700 S. Cass Avenue Lemont, IL, United States

[d] Institute of Energy Power Innovation, North China Electric Power University, 2 Beinong Road, Beijing 102206, P. R. China

[e] National Laboratory of Solid State Microstructures, College of Engineering and Applied Sciences, Nanjing University, 22 Hankou Road, Nanjing 210093, P. R. China

[f] Department of Chemical and Biological Engineering, The Hong Kong University of Science and Technology, Clear Water Bay, Kowloon, Hong Kong 999077, P. R. China

[g] Department of Computer Science, University of Chicago, 5730 S Ellis Ave., Chicago, IL 60637, United States

[*]Corresponding Authors

[†]Co-first authors

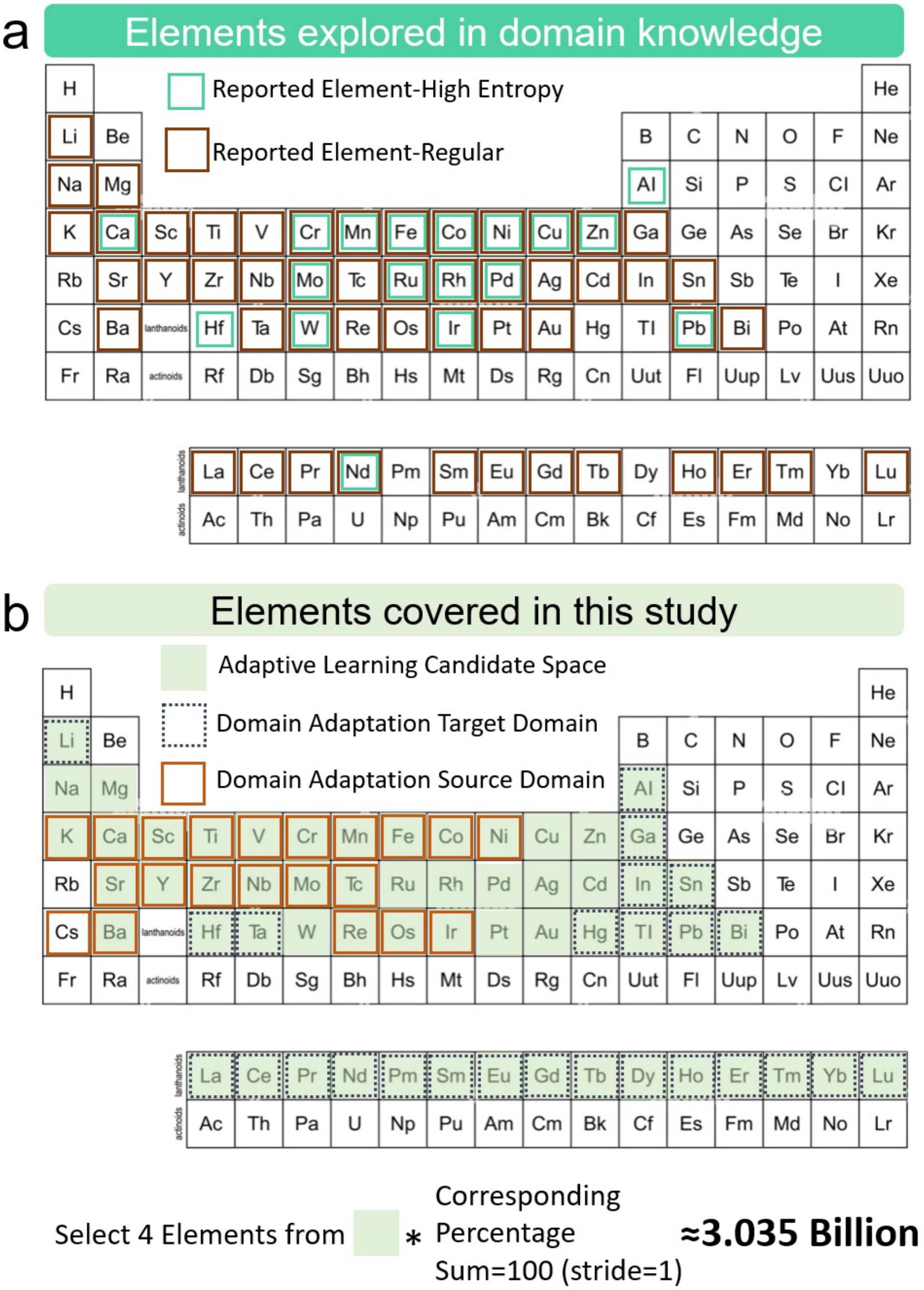

**Figure S1** a) The light blue highlighted elements represent the metals involved in the precursor synthesis for acidic OER electrocatalysts in high-entropy or multi-metallic systems, as explored in recent publications [1-12] within the last two years. In these studies, the number of metal elements in the precursors would equal or exceed four. These studies typically do not investigate the elemental ratios, often defaulting to equimolar proportions in high-entropy systems. In contrast, the brown highlighted elements denote metals explored over the past 20 years in conventional research for acidic OER catalyst systems, used for doping or as base materials, as identified in the domain knowledge dataset used during the data-mining phase of this work (see **Figure S2**). In these studies, the metal element numbers in the precursors are predominantly lower than four, usually less than three. b) The candidate metal element space explored in this study is highlighted in light green, representing the candidate elements involved in the DASH loop search within the active-learning module, expanded from the elements covered in the current domain as shown in a) (including the remaining lanthanides and additions of Hg, Tl). The orange solid box highlights the candidate doping elements in the source domain for the domain adaptation module related to DFT theoretical simulations, while the blue dashed box outlines the corresponding target domain elements.

Note: To estimate the total number of combinations for designing a quaternary oxide with four elements selected from 58 possible elements and having their proportions sum up to 100% (with each proportion being an integer), the following steps were performed:

(a) Element Selection: Four elements are chosen from a pool of 58 candidates. (b)Percentage Assignment: For each selected element, a percentage value is assigned between 1 and 100, inclusive, with a step size of 1. (3) Constraint Application: The sum of the four percentage values must be equal to 100. This introduces a constraint on the percentage combinations, ensuring that the sum of the assigned percentages for the four elements equals 100. (4) Duplication Removal: Duplicate combinations are removed to retain only unique combinations. This step ensures that permutations of the same combination are counted as one unique combination. The result shows that the total possibilities would reach a huge number of 3.035 billion (3,034,803,310). This calculation encompasses both the distribution of percentages to the elements and the selection of the elements themselves. Corresponding script is available at the online repository's "./ML Databases and Scripts/Number of Possibilities Calculation" directory.

In conclusion, considering all possible combinations of the candidate metal elements highlighted in light green, assuming the sum of proportions equals 100% with a stride of 1%, the total possible formulations approach over six billion, even with only four precursor metals. This calculation excludes other synthesis parameters, such as annealing duration and temperature, post-treatment steps, and electrochemical factors, thus underscoring the "expansive array of constituent elements and synthesis parameters." This vast candidate space highlights the complexity of optimizing such a system, necessitating a multi-stage strategy that employs various ML techniques for decision-making at different phases.

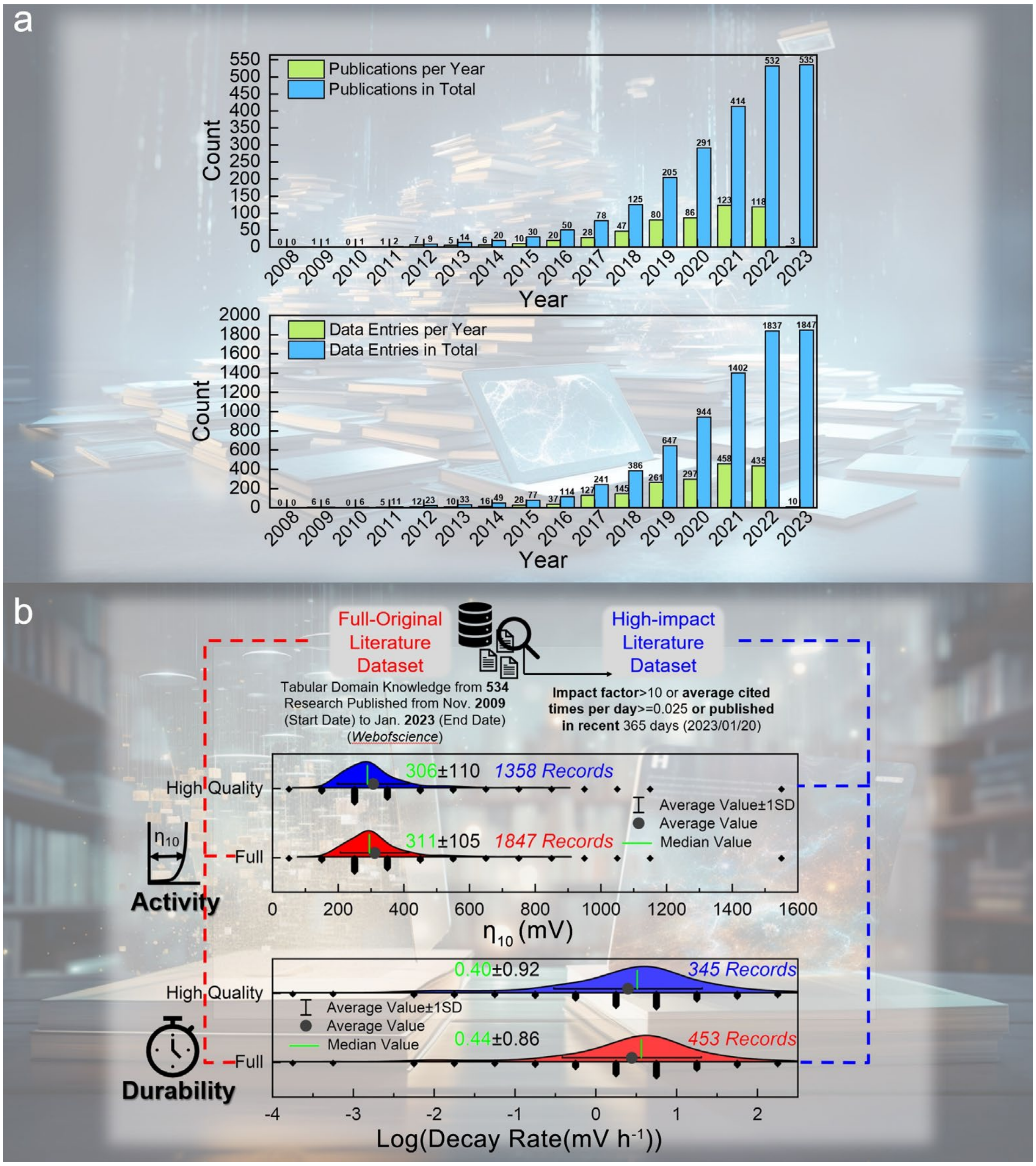

**Figure S2** a) Publication and data entries (number per year vs. total) in the full initial domain knowledge dataset. b) Half-violin plots that reflect the distribution patterns of the domain knowledge datasets.

**Table S1** The features in the domain knowledge dataset (full dataset, 1,847 entries) for catalytic activity and corresponding variable range.

| Feature Name (Unit) | Variable Range |
|---|---|
| Metal_Dopant_1[a] （1~4 represent the proportion in precursor from high to low） | 49 different metal elements[a,b] |
| Metal_Dopant_2 | 55 different metal elements or none |
| Metal_Dopant_3 | 39 different metal elements or none |
| Metal_Dopant_4 | 7 different metal elements or none |
| Metal_Dopant_1 Proportion in Precursor （at. %; refers to that in total of four types of metal） | 28.57~100 |
| Metal_Dopant_2 Proportion in Precursor （at. %） | 0~50 |
| Metal_Dopant_3 Proportion in Precursor （at. %） | 0~33.33 |
| Metal_Dopant_4 Proportion in Precursor （at. %） | 0~17.67 |
| Hydrothermal Temperature (°C) (or precursor mixing) | -196~320 |
| Hydrothermal Time (min) (or precursor mixing) | 0 ~86,400 |
| Hydrothermal Still/Stirring (0/1) (or precursor mixing) | 0: still incubation; 1: stirring or sonication |
| Hydrothermal Strong Reductant in Liquid (0/1) (or precursor mixing) | 0: False; 1: True |
| Hydrothermal Weak Reductant in Liquid (0/1) (or precursor mixing) | 0: False; 1: True |
| Mixed in Solid or Liquid (0/1) | 0: False (liquid mixing or hydrothermal); 1: True (ball milling) |
| Annealing Temperature (°C) | 25~1,400 |
| Annealing Time (min) | 0~20,160 |
| Annealing Still/Stirring (0/1) | 0: False; 1: True |
| Annealing Atmosphere Inert (0/1) | 0: False; 1: True |
| Annealing Atmosphere Reducing (0/1) | 0: False; 1: True |
| Post-processing Acid Wash, etc. (after annealing; 0/1) | 0: False; 1: True |
| Catalyst Loading (mg $cm^{-2}$) | 0.000714~490 |
| Support Material Loading (mg $cm^{-2}$) | 0~18 |
| Support Material is Not Carbon (support material, $TiO_x$, etc.; 0/1) | 0: False; 1: True |
| Electrode Type_Glassy Carbon/Carbon Paper or Ti Mesh (0/1) | 0: Glassy Carbon;1: Carbon Paper or Ti Mesh |
| LSV Scanning Speed (mV $s^{-1}$) | 0.1~100 |
| Electrolyte Proton Concentration (M) | 0.01~6 |

Note:

a) For clarity and consistency in terminology, all metal elements in the precursor are referred to as "dopants", even though the first "dopant" is actually the primary metal. These are arranged in descending order of their proportions, from the first to the fourth.

b) When conducting non-ML method data mining, the dataset directly treats elements as frequent items. However, for ML modeling and fitting, the elemental information is digitized and represented by properties such as relative atomic mass, atomic number, period, group, ionization energy, electronegativity, number of outermost d electrons, and atomic radius, as we have illustrated in detail in **Supplementary Note 1**, **Figure SN 1-1**.

c) Additionally, in some studies, such as those using TiN or TiC as catalyst supports, the elements N and C do not exist in the final mixed oxide product in the same way as they do in the precursor organic compounds. Therefore, despite the term "metal elements" used in this paper, a few records in the dataset will include N and C. This is in consideration of their stable presence during the catalytic process and to account for their potential synergistic effects in catalysis.

**Table S2** The features in the domain knowledge dataset (full dataset, 453 entries) for stability and corresponding variable range.

| Feature Name (Unit) | Variable Range |
|---|---|
| Metal_Dopant_1 (1~4 represent the proportion in precursor from high to low) | 41 different metal elements |
| Metal_Dopant_2 | 53 different metal elements or none |
| Metal_Dopant_3 | 32 different metal elements or none |
| Metal_Dopant_4 | 3 different metal elements or none |
| Metal_Dopant_1 Proportion in Precursor (at. %; refers to that in total four types of metal) | 28.57~100 |
| Metal_Dopant_2 Proportion in Precursor (at. %) | 0~50 |
| Metal_Dopant_3 Proportion in Precursor (at. %) | 0~32.34 |
| Metal_Dopant_4 Proportion in Precursor (at. %) | 0~2.06 |
| Hydrothermal Temperature (°C) (or precursor mixing) | -77~320 |
| Hydrothermal Time (min) (or precursor mixing) | 0 ~10,080 |
| Hydrothermal Still/Stirring (0/1) (or precursor mixing) | 0: still incubation; 1: stirring or sonication |
| Hydrothermal Strong Reductant in Liquid (0/1) (or precursor mixing) | 0: False; 1: True |
| Hydrothermal Weak Reductant in Liquid (0/1) (or precursor mixing) | 0: False; 1: True |
| Mixed in Solid or Liquid (0/1) | 0: False (liquid mixing or hydrothermal); 1: True (ball milling) |
| Annealing Temperature (°C) | 25~1,200 |
| Annealing Time (min) | 0~20160 |
| Annealing Still/Stirring (0/1) | 0: False; 1: True |
| Annealing Atmosphere Inert (0/1) | 0: False; 1: True |
| Annealing Atmosphere Reducing (0/1) | 0: False; 1: True |
| Post-processing Acid Wash, etc. (after annealing; 0/1) | 0: False; 1: True |
| Catalyst Loading (mg $cm^{-2}$) | 0.000714~490 |
| Support Material Loading (mg $cm^{-2}$) | 0~5.58 |
| Support Material is Not Carbon (support material, $TiO_x$ etc.; 0/1) | 0: False; 1: True |
| Electrode Type_Glassy Carbon/Carbon Paper or Ti Mesh (0/1) | 0: Glassy Carbon;1: Carbon Paper or Ti Mesh |
| Electrolyte Proton Concentration (M) | 0.01~6 |
| Stability Constant Current Density (mA $cm^{-2}$) | 0.1~1,000 |
| Stability Test Time (h) | 0.28~8,000 |

Note:

After integrating the fundamental intrinsic atomic properties information into the dataset, we further examined the correlation matrices for the initial dataset using the Kendall, Spearman, and Pearson methods. Interested readers can check the corresponding results in the online GitHub repository (https://github.com/ruiding-uchicago/DASH) ("/Online Repository Figures/"): **Figure OR1~OR12**. While there is a high degree of inter-correlation among features that inevitably express fundamental intrinsic atomic properties information of elements, such as relative atomic mass, atomic number, period, group, ionization potential, electronegativity, number of d electrons, and atomic radius, the correlations are significantly lower among other features. Specifically, this lower correlation is observed both within features related to material synthesis and testing and between these features and the fundamental intrinsic atomic properties of elements (except for a reasonable correlation between annealing time and temperature). This finding is particularly encouraging for several reasons. First, the lower degree of correlation among these sets of features suggests a diverse and rich dataset, which is crucial for the robustness of ML models. This correlation indicates that our dataset encompasses a wide range of independent variables, thus enhancing the potential for our models to capture and learn from a broad spectrum of material behaviors and properties. Second, the low correlation between material synthesis/testing features and elemental fundamental intrinsic atomic properties implies that our dataset is not dominated by any single type of information. This diversity ensures that our ML models are not biased towards fundamental intrinsic atomic properties features alone but are also informed by practical, experimental data. In summary, this characteristic of our dataset is advantageous for developing nuanced and comprehensive ML models, as it allows for the exploration of complex interactions within materials, potentially leading to novel insights and breakthroughs in the field of electrocatalysis.

The corresponding unprocessed domain knowledge dataset .csv file (that is readable in Excel software), Python script files, and generated pkl dataset files (for supervised learning and training first-iteration ML committees based on the previously mentioned domain knowledge dataset) are stored and publicly available in the: "/ML Databases and Scripts/Domain Knowledge Database Preprocessing" directory of the DASH online repository.

More secondary implementation details and discussions are available in **Supplementary Note 1.**

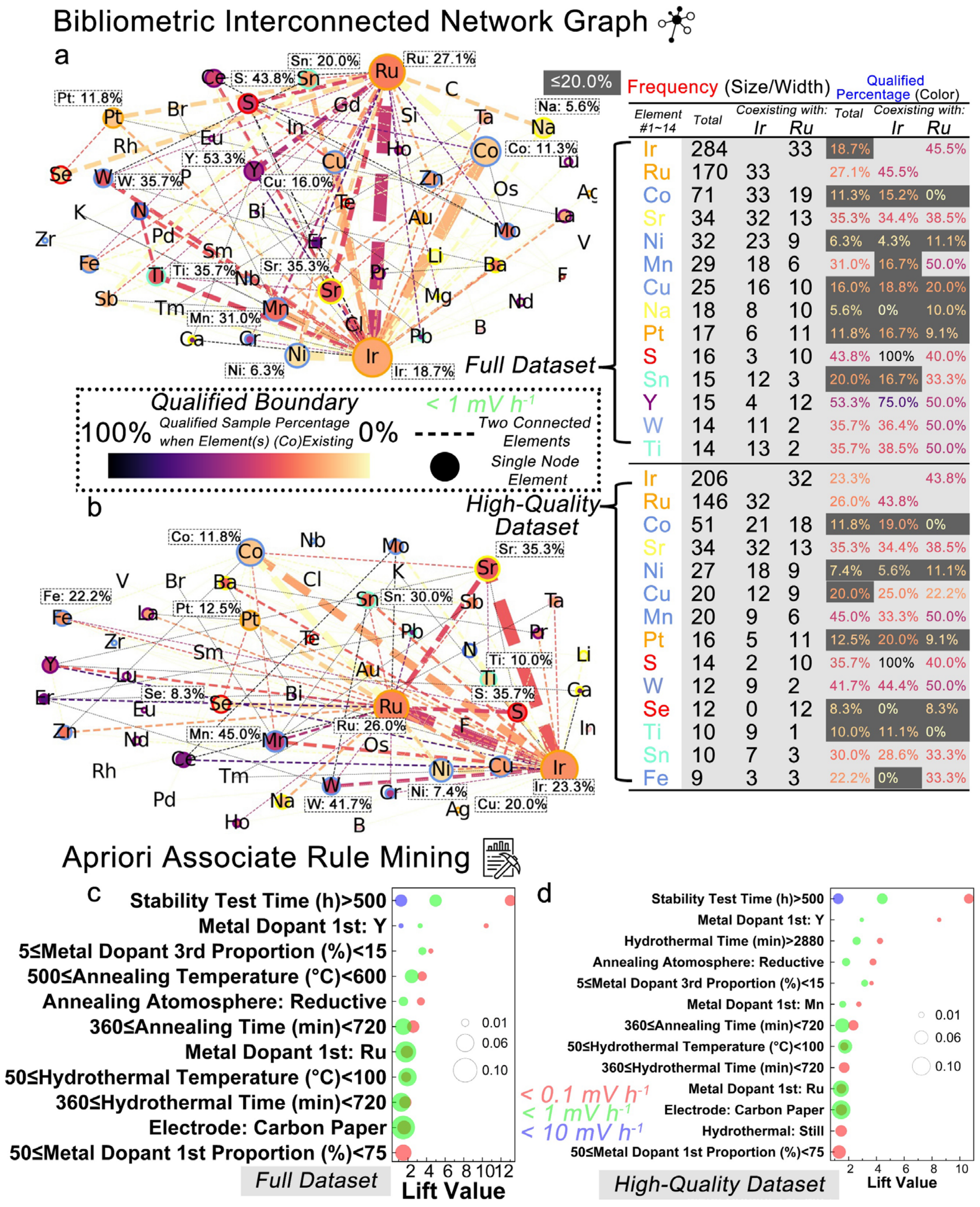

**Figure S3** Key results derived from unsupervised data mining focused on OER stability. a) Network graph based on the full domain knowledge datasets of chemical elements with a qualified decay rate boundary set at 1 mV $h^{-1}$; b) Network graph based on a high-quality domain knowledge dataset of chemical elements with a qualified decay rate boundary set at 1 mV $h^{-1}$ (For network graphs herein, different colors at the edges of nodes represent different groups of elements. Aquamarine: C group; royal blue: N group; red: O group; pink: B group; brown: halogen; cornflower blue: earth abundant–transition metal; purple: rare earth metal;

yellow: first/second group metal; and orange: noble metal); c) Results of high lift values in Apriori associate rule mining based on the full domain knowledge dataset with a frequent itemset length of two for stability-related insights; d) Results of high lift values in Apriori associate rule mining based on a high-quality domain knowledge dataset with the frequent itemset length of two for stability-related insights.

**Supplementary Discussion S1**

In the supervised data-mining process, we adopt a robust and comprehensive approach to hyperparameter optimization in our regression ML models in the committees, a process that is instrumental in maximizing model performance and ensuring the validity of our results.

**Hyperparameter Optimization Strategy**

Our strategy involves a meticulous application of five-fold cross-validation (CV) exclusively on the training data set. This process segregates the training data into five distinct subsets. In each iteration of the CV, four subsets are used for training the model, while the remaining subset serves as the validation set. This cycle is repeated five times, ensuring that each subset is used for validation once. This approach allows each model to be trained and validated on different data segments, thus effectively honing its hyperparameters. By confining this process within the training data, we eliminate the risk of data leakage and ensure that the test data remains an unbiased arbitrator of model performance.

The absence of test data involvement in this phase is crucial for maintaining the integrity of our evaluation process. It guarantees that the model's hyperparameters are optimized without any foreknowledge of the test data, thus ensuring that the final evaluation on the test set is conducted under fair and unbiased conditions. This method not only enhances the reliability of our results, but also ensures that the models are evaluated based on their ability to generalize to new, unseen data.

**Performance Metrics**

To comprehensively assess the performance of our ML models, we employ a suite of statistical metrics, each providing unique insights into the model's accuracy and efficacy.

1. Mean Absolute Error (MAE): MAE measures the average magnitude of errors in a set of predictions without considering their direction. It is calculated using the formula:

$$\mathrm{MAE} = \frac{1}{n}\sum_{i=1}^{n}\left|y_i - \hat{y}_i\right|$$

where $\hat{y}_i$ represents the actual value and $y_i$ represents the predicted value. MAE is particularly useful because it provides a straightforward interpretation of the average error magnitude.

2. Mean Squared Error (MSE): MSE provides a measure of the quality of an estimator—it is always non-negative, and values closer to zero are better. It is given by:

$$\mathrm{MSE} = \frac{1}{n}\sum_{i=1}^{n}(y_i - \hat{y}_i)^2$$

MSE penalizes larger errors more severely than smaller ones, making it sensitive to outliers in the data.

3. Root Mean Squared Error (RMSE): RMSE is the square root of the MSE and is used to measure the difference between values predicted by a model and the values observed. The formula is:

$$\mathrm{RMSE} = \sqrt{\frac{1}{n}\sum_{i=1}^{n}(y_i - \hat{y}_i)^2}$$

RMSE is in the same units as the response variable and is particularly useful for understanding the error magnitude in the context of the measured data.

4. Coefficient of Determination ($R^2$): $R^2$ quantifies how well future outcomes are likely to be predicted by the model. It is defined as:

$$R^2 = 1 - \frac{\sum_{i=1}^{n}(y_i - \hat{y}_i)^2}{\sum_{i=1}^{n}(y_i - \bar{y})^2}$$

where $\bar{y}$ is the mean of the observed data. $R^2$ provides a scale for model comparison, indicating the proportion of variance in the dependent variable that can be explained by the independent variables in the model.

**Regression Performance Discussion**

As we briefly discussed in the main text, most of the regression models in the four committees, either based on full or high-quality datasets or focused on activity or stability, have shown comparatively good predicting abilities. In particular, the best-performing algorithms among the 11 models in our committee in terms of the $R^2$ values were identified as ensemble models, such as CatBoost, Random Forest, and Gradient Boosting, thus highlighting their effectiveness and the value of their SHAP matrices as representative for deeper materials science insights. Our domain knowledge dataset, rooted in experimental research literature, inevitably contains significant experimental detail variations such as disparities in synthesis preparation and testing characterization methods among different laboratories and researchers. These variations may introduce unavoidable but non-negligible noise in the dataset. However, as demonstrated through the statistical analysis of four key metrics for different regression models in **Figures S6**, **S7**, **S10**, **S11**, it is evident that the top-performing ensemble algorithms often achieve $R^2$ values close to or exceeding 0.8 on the test set. The $R^2$ value, commonly regarded as highly significant due to its ability to quantify the proportion of variance in the dependent variable that is predictable from the independent variable, underscores the reliability of our domain knowledge dataset and the robust predictive power of the ML models derived from it.

These models exhibit substantial robustness against outliers in the dataset, effectively learning the overarching trends within the studied material system. Additionally, an intriguing observation from our study is that despite the high-quality dataset and the full dataset appearing to share a similar distribution pattern, as shown in **Figure S1b**, ML models trained on these datasets exhibited noteworthy differences in performance. As summarized in **Figure 3a**, the top-performing algorithms demonstrated a lower MAE when using the high-quality dataset. This trend was further corroborated by the MSE and RMSE metrics. This finding aligns with our initial assumption that data entries sourced from varying levels of reliability may differentially impact model performance due to their respective outlier and noise levels. Scripts in this part for activity or stability regression models can be retrieved in the directory "Domain Knowledge-Based Initial ML Committee and Blackbox Interpretation/ Regression/Activity/Initial ML Committee Training/" (for stability regression, find the folder named "Stability", which is a same-level directory of "Activity").

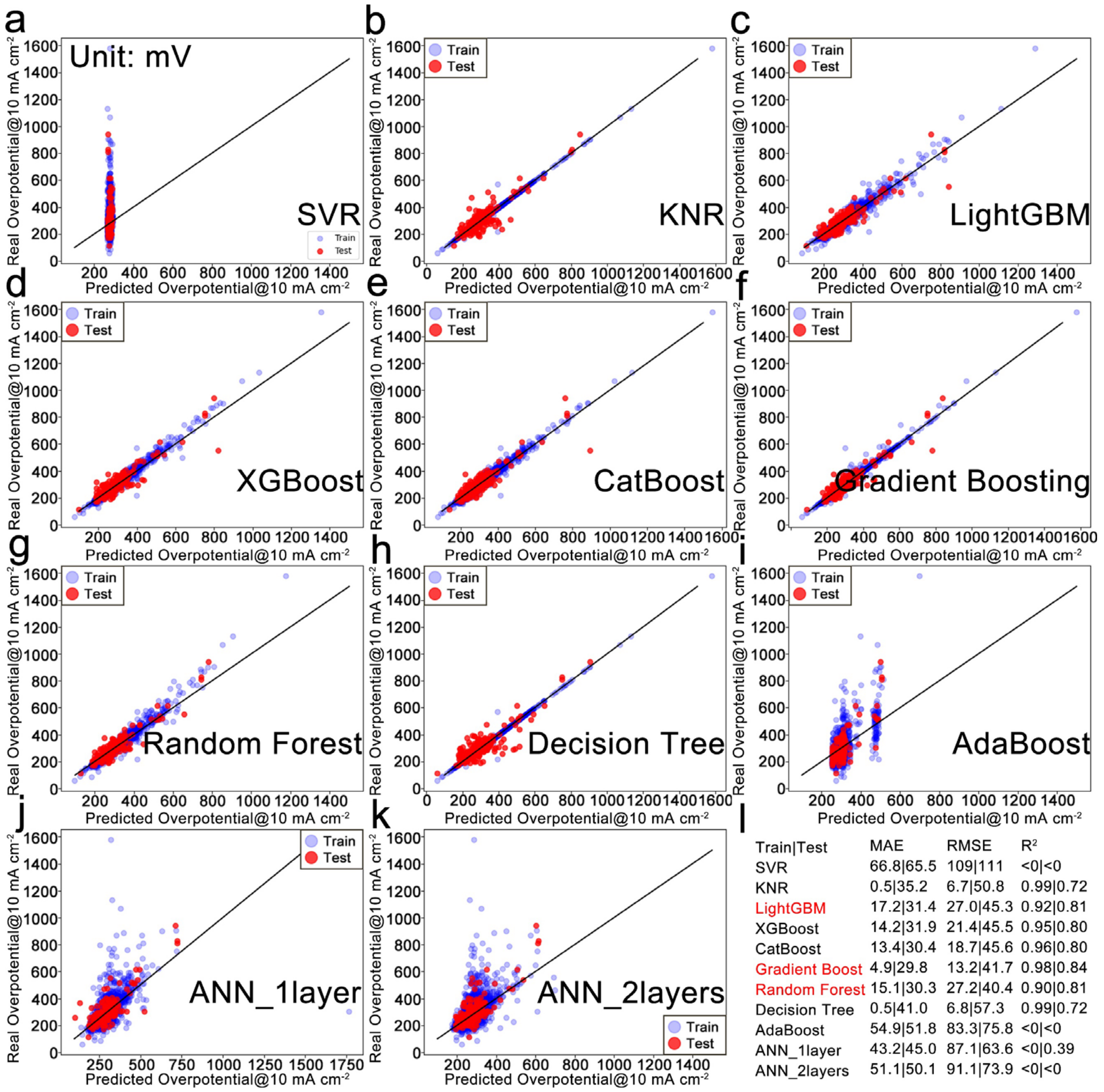

**Figure S4** Scatter diagrams of the ML model committee members, namely different algorithms trained to predict $\eta_{10}$ based on the full domain knowledge dataset. The red point's X-axis value indicates the predicted outcome from the machine-learning model for a specific sample in the test set, whereas its Y-axis value reflects the actual recorded value in the dataset. Similarly, the blue points represent the results in the training set. The black line, represented by y=x, functions as a benchmark: the proximity of these red points to the y=x line is indicative of the model's prediction accuracy. Diagrams corresponding to the different hyperparameter-optimized algorithms: a) SVR, b) KNR, c) LightGBM, d) XGBoost, e) CatBoost, f) Gradient Boost, g) Random Forest, h) Decision Tree, i) AdaBoost, j) ANN with one hidden layer, k) ANN with two hidden layers, and l) summary of the key performance metrics for regression, with the top three in terms of $R^2$ highlighted in red font.

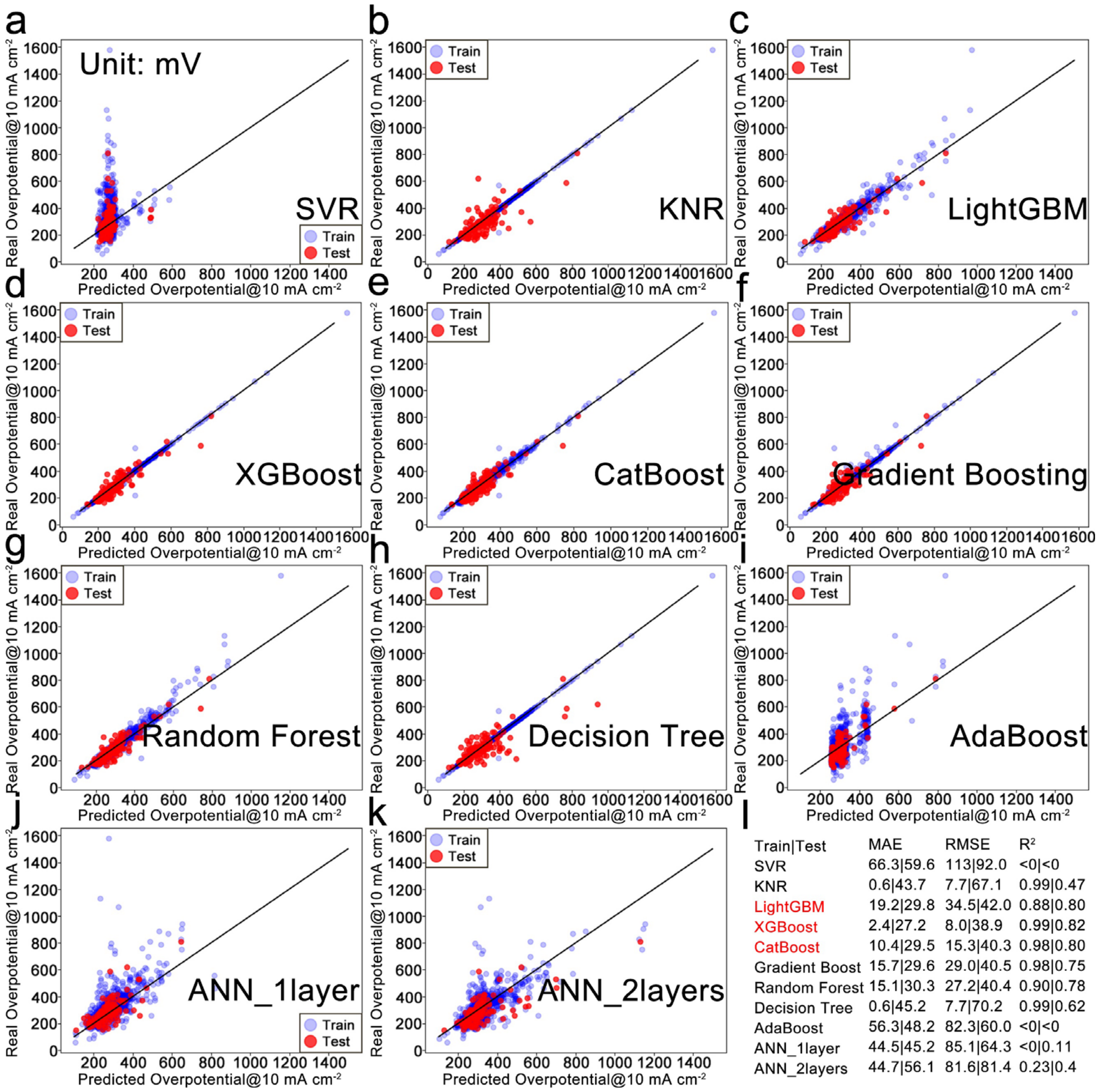

| Train\|Test | MAE | RMSE | $R^2$ |
|---|---|---|---|
| SVR | 66.3\|59.6 | 113\|92.0 | <0\|<0 |
| KNR | 0.6\|43.7 | 7.7\|67.1 | 0.99\|0.47 |
| LightGBM | 19.2\|29.8 | 34.5\|42.0 | 0.88\|0.80 |
| XGBoost | 2.4\|27.2 | 8.0\|38.9 | 0.99\|0.82 |
| CatBoost | 10.4\|29.5 | 15.3\|40.3 | 0.98\|0.80 |
| Gradient Boost | 15.7\|29.6 | 29.0\|40.5 | 0.98\|0.75 |
| Random Forest | 15.1\|30.3 | 27.2\|40.4 | 0.90\|0.78 |
| Decision Tree | 0.6\|45.2 | 7.7\|70.2 | 0.99\|0.62 |
| AdaBoost | 56.3\|48.2 | 82.3\|60.0 | <0\|<0 |
| ANN_1layer | 44.5\|45.2 | 85.1\|64.3 | <0\|0.11 |
| ANN_2layers | 44.7\|56.1 | 81.6\|81.4 | 0.23\|0.4 |

**Figure S5** Similar to **Figure S4**, scatter diagrams of the ML model committee members, namely different algorithms trained to predict $\eta_{10}$ based on the high-quality domain knowledge dataset. Diagrams corresponding to the different hyperparameter optimized algorithms: a) SVR, b) KNR, c) LightGBM, d) XGBoost, e) CatBoost, f) Gradient Boost, g) Random Forest, h) Decision Tree, i) AdaBoost, j) ANN with one hidden layer, k) ANN with two hidden layers, and l) summary of the key performance metrics for regression, with the top three in terms of $R^2$ highlighted in red font.

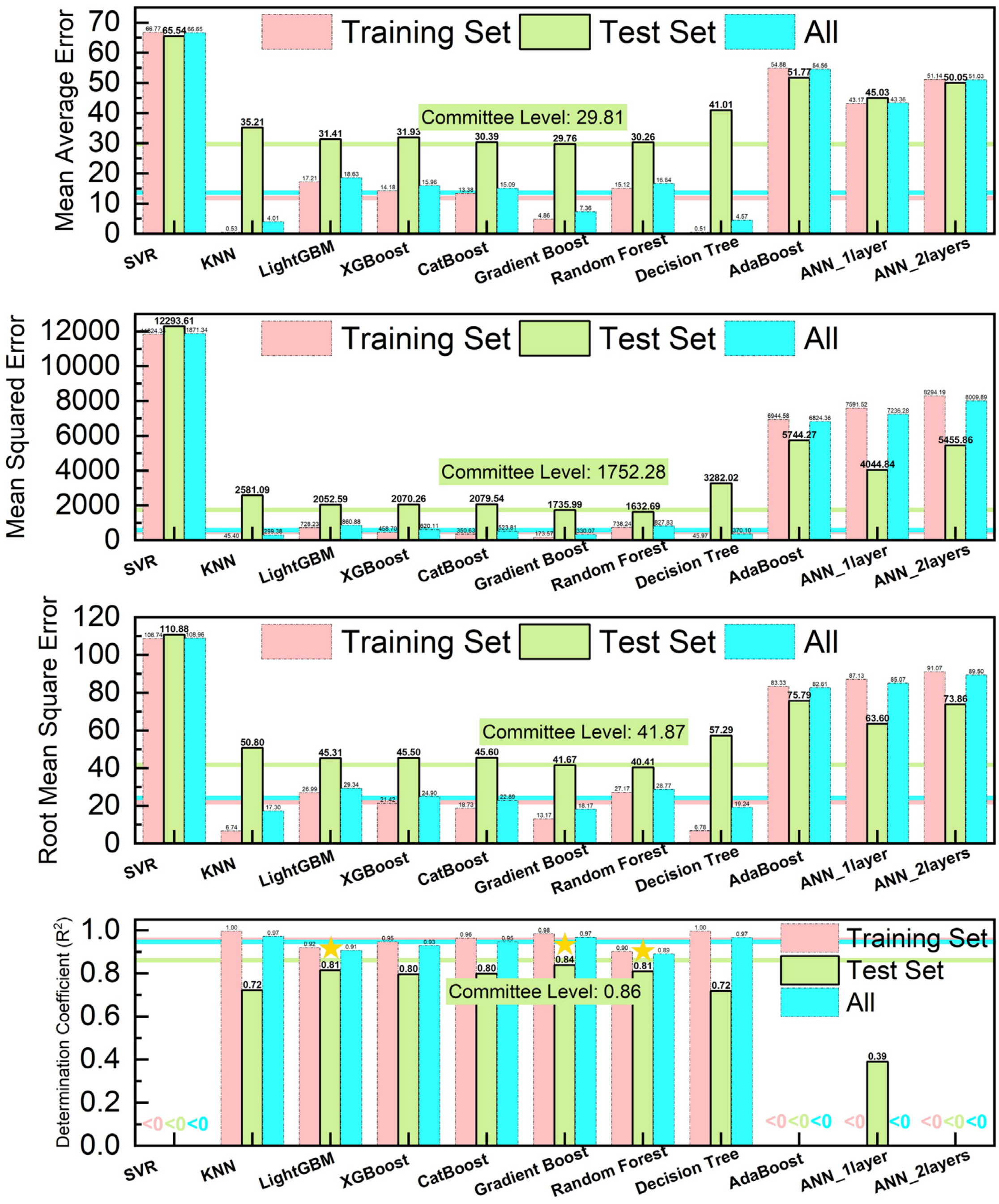

**Figure S6** Summary of the regression performance metrics on ML models illustrated in **Figure S4**.

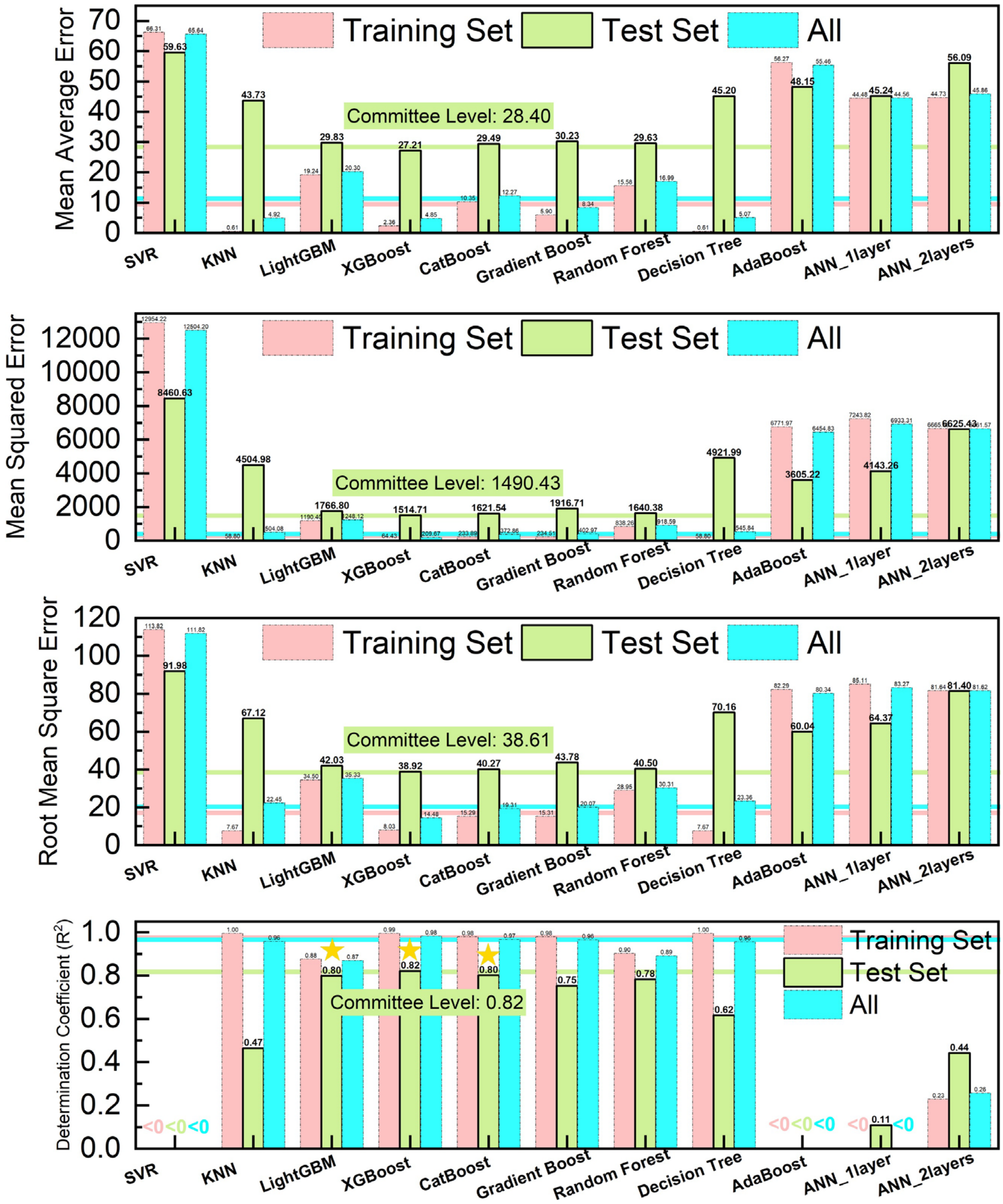

**Figure S7** Summary of the regression performance metrics on ML models illustrated in **Figure S5**.

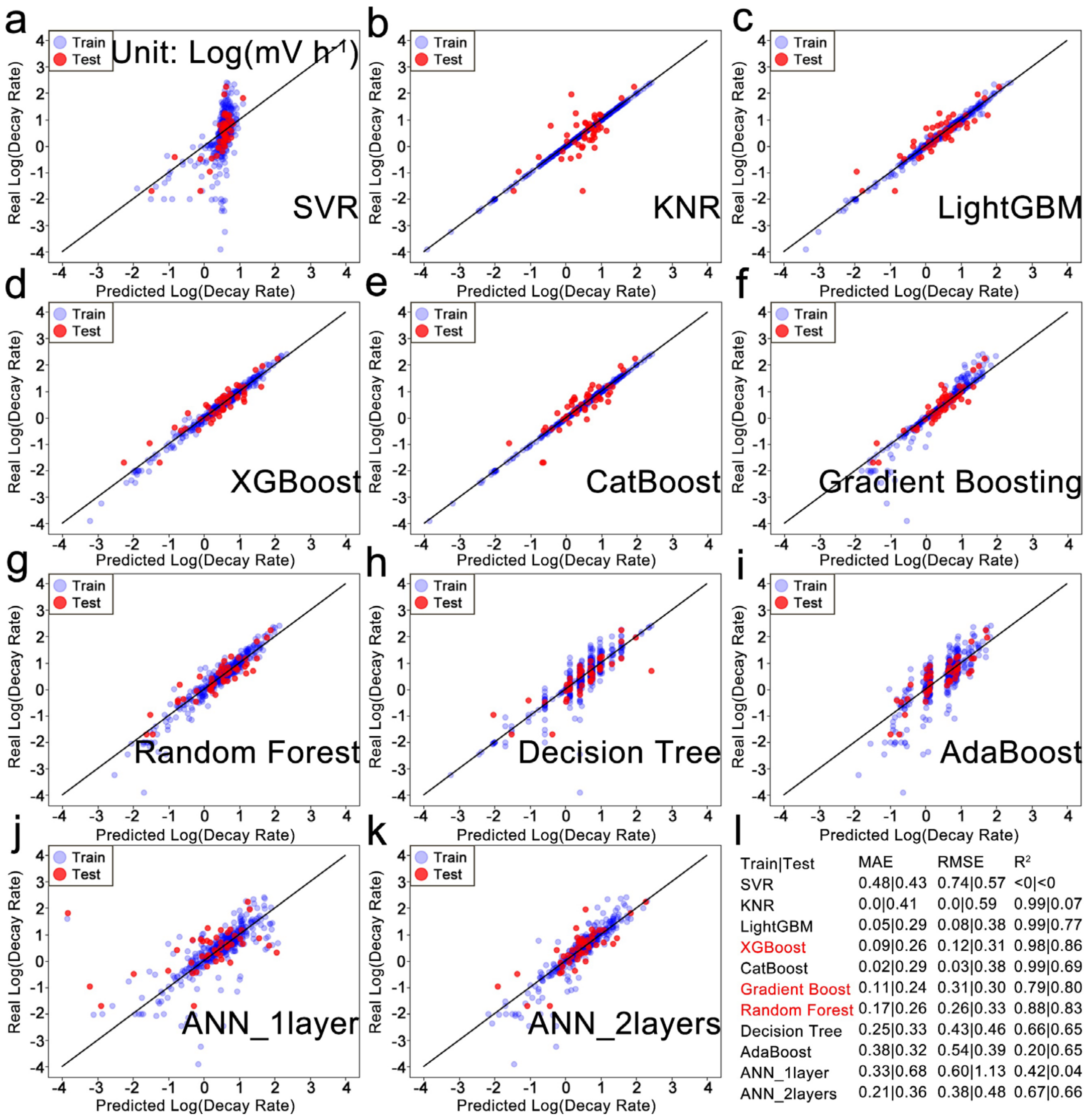

**Figure S8** Scatter diagrams of the ML model committee members, namely different algorithms trained to predict the decay rate based on the full domain knowledge dataset. Diagrams corresponding to the different hyperparameter optimized algorithms: a) SVR, b) KNR, c) LightGBM, d) XGBoost, e) CatBoost, f) Gradient Boost, g) Random Forest, h) Decision Tree, i) AdaBoost, j) ANN with one hidden layer, k) ANN with two hidden layers, and l) summary of the key performance metrics for regression, with the top three in terms of $R^2$ highlighted in red font.

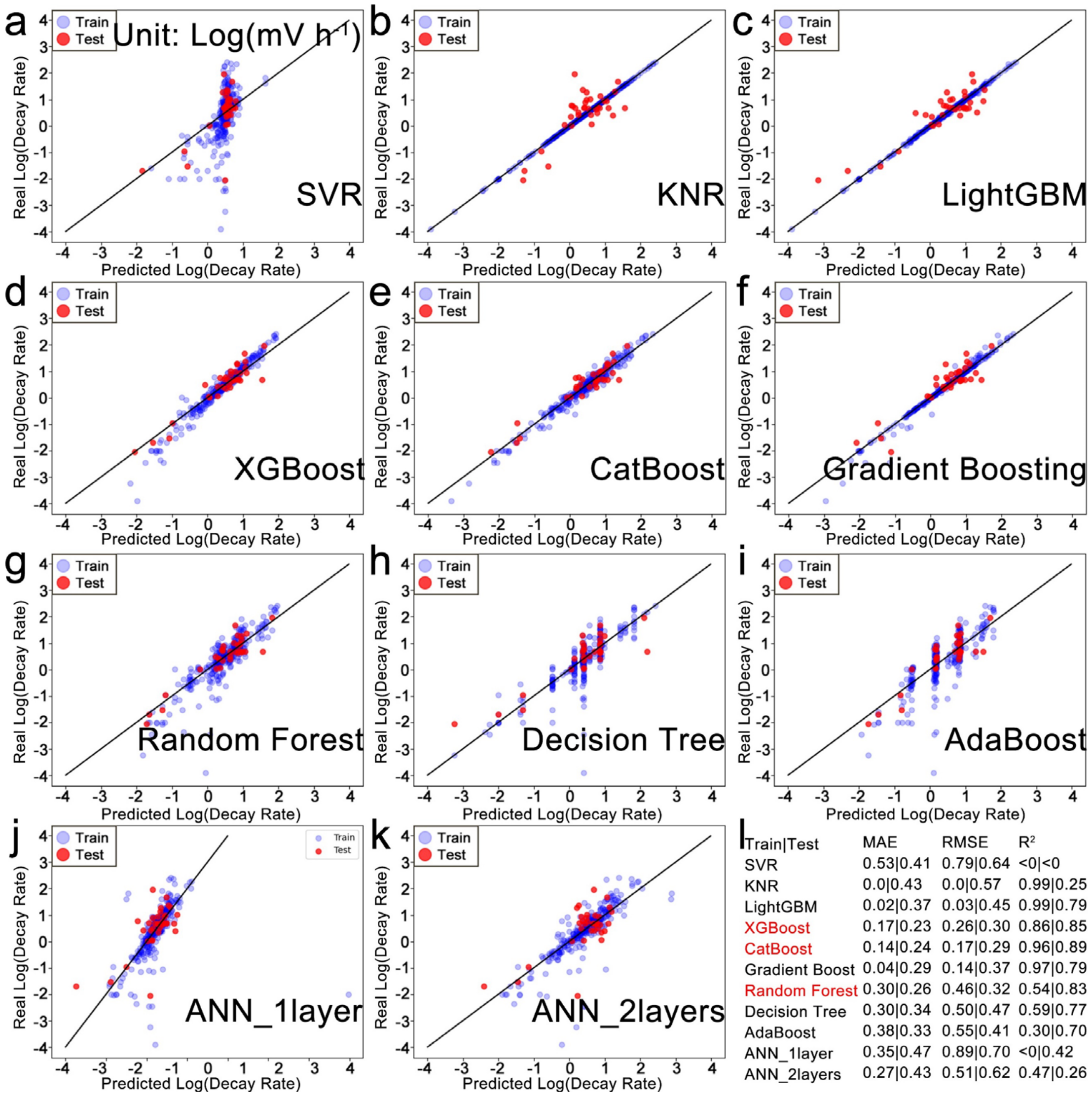

| Train\|Test | MAE | RMSE | $R^2$ |
|---|---|---|---|
| SVR | 0.53\|0.41 | 0.79\|0.64 | <0\|<0 |
| KNR | 0.0\|0.43 | 0.0\|0.57 | 0.99\|0.25 |
| LightGBM | 0.02\|0.37 | 0.03\|0.45 | 0.99\|0.79 |
| XGBoost | 0.17\|0.23 | 0.26\|0.30 | 0.86\|0.85 |
| CatBoost | 0.14\|0.24 | 0.17\|0.29 | 0.96\|0.89 |
| Gradient Boost | 0.04\|0.29 | 0.14\|0.37 | 0.97\|0.79 |
| Random Forest | 0.30\|0.26 | 0.46\|0.32 | 0.54\|0.83 |
| Decision Tree | 0.30\|0.34 | 0.50\|0.47 | 0.59\|0.77 |
| AdaBoost | 0.38\|0.33 | 0.55\|0.41 | 0.30\|0.70 |
| ANN_1layer | 0.35\|0.47 | 0.89\|0.70 | <0\|0.42 |
| ANN_2layers | 0.27\|0.43 | 0.51\|0.62 | 0.47\|0.26 |

**Figure S9** Scatter diagrams of the ML model committee members, namely different algorithms trained to predict the decay rate based on the high-quality domain knowledge dataset. Diagrams corresponding to the different hyperparameter optimized algorithms: a) SVR, b) KNR, c) LightGBM, d) XGBoost, e) CatBoost, f) Gradient Boost, g) Random Forest, h) Decision Tree, i) AdaBoost, j) ANN with one hidden layer, k) ANN with two hidden layers, and l) summary of the key performance metrics for regression, with the top three in terms of $R^2$ highlighted in red font.

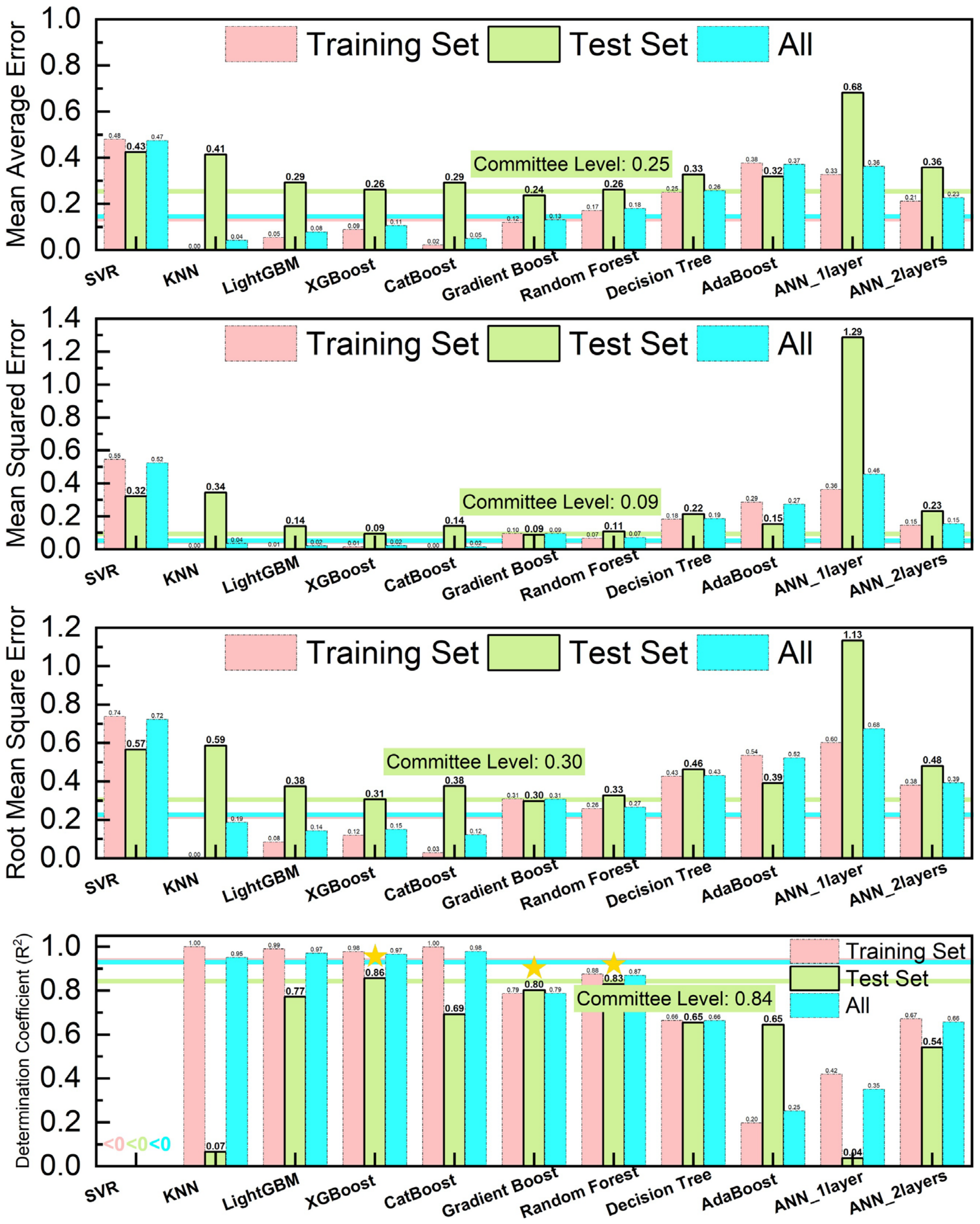

**Figure S10** Summary of the regression performance metrics on the ML models illustrated in **Figure S8**.

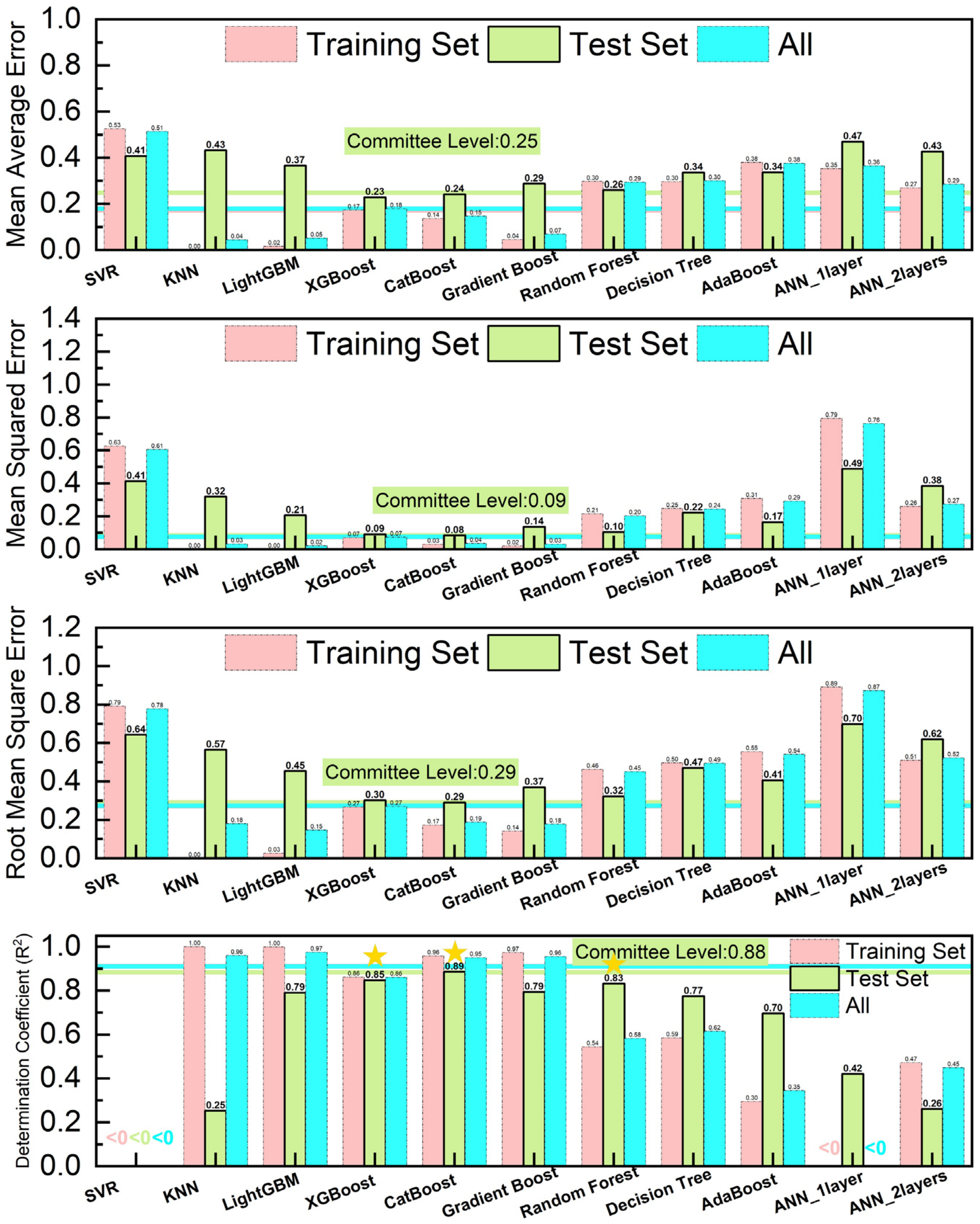

**Figure S11** Summary of the regression performance metrics on the ML models illustrated in **Figure S9**.

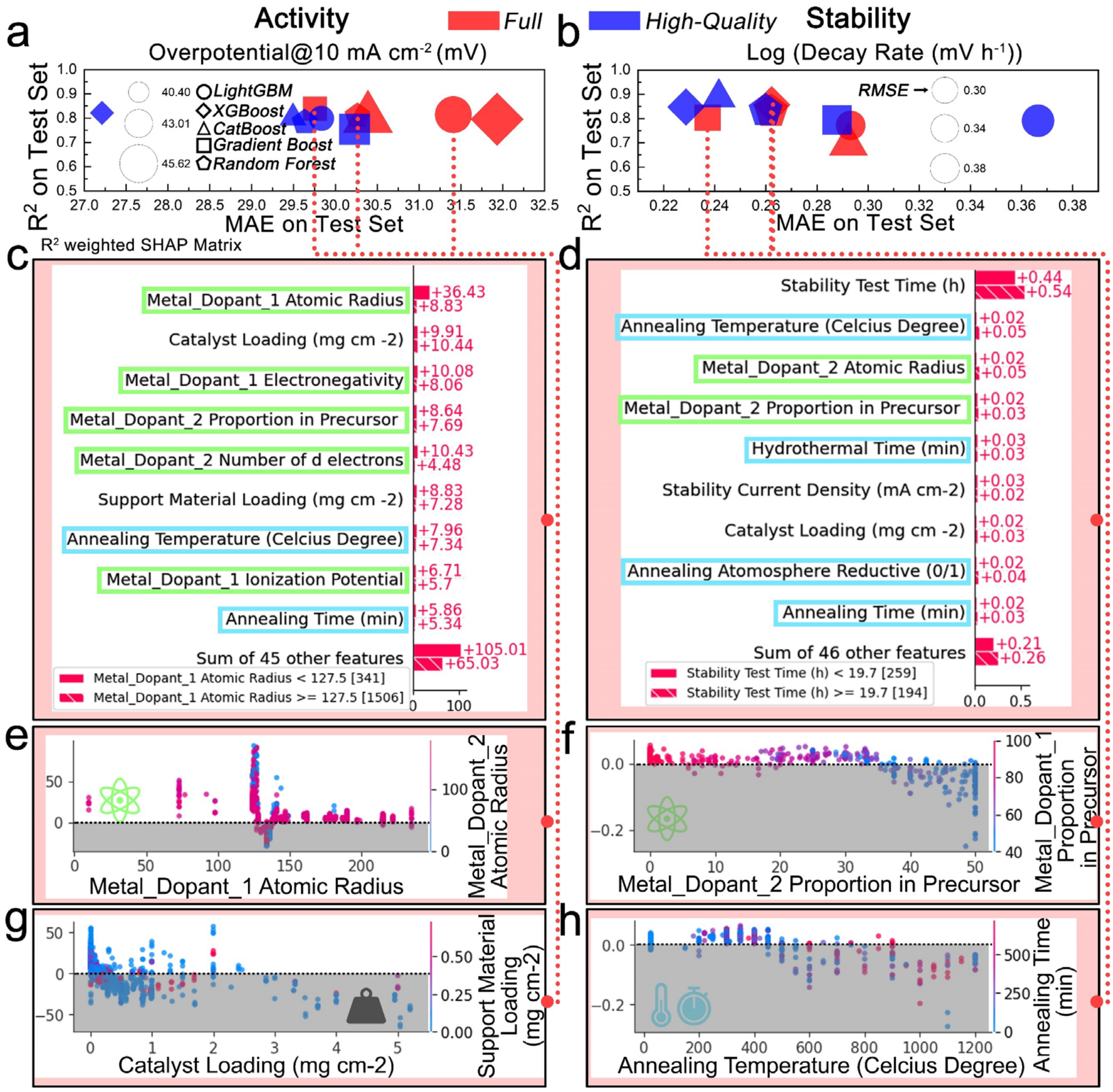

**Figure S12** Similar to **Figure 3**, key results derived from supervised data mining based on two committees trained from the full datasets. a) and b): The top five ML models, as identified from committees trained on various domain knowledge datasets, are evaluated based on $R^2$ and MAE metrics on the test set. c)-d) SHAP cohort bar plots that highlight the important features with light green and blue frames highlighting the element-related features: atomic properties and synthesis condition parameters, respectively. e)-h) Selected SHAP two-dimensional interaction plots that feature an interaction of the primary studied feature on the X-axis with a second feature, which is indicated by the color bars. The second features, also chosen from the top features with a similar type, are those with a high degree of interaction in the Friedman's H-statistic interaction matrix. Dashed lines at y=0 in each dependence plots split-grey areas that indicate the preferred value ranges.

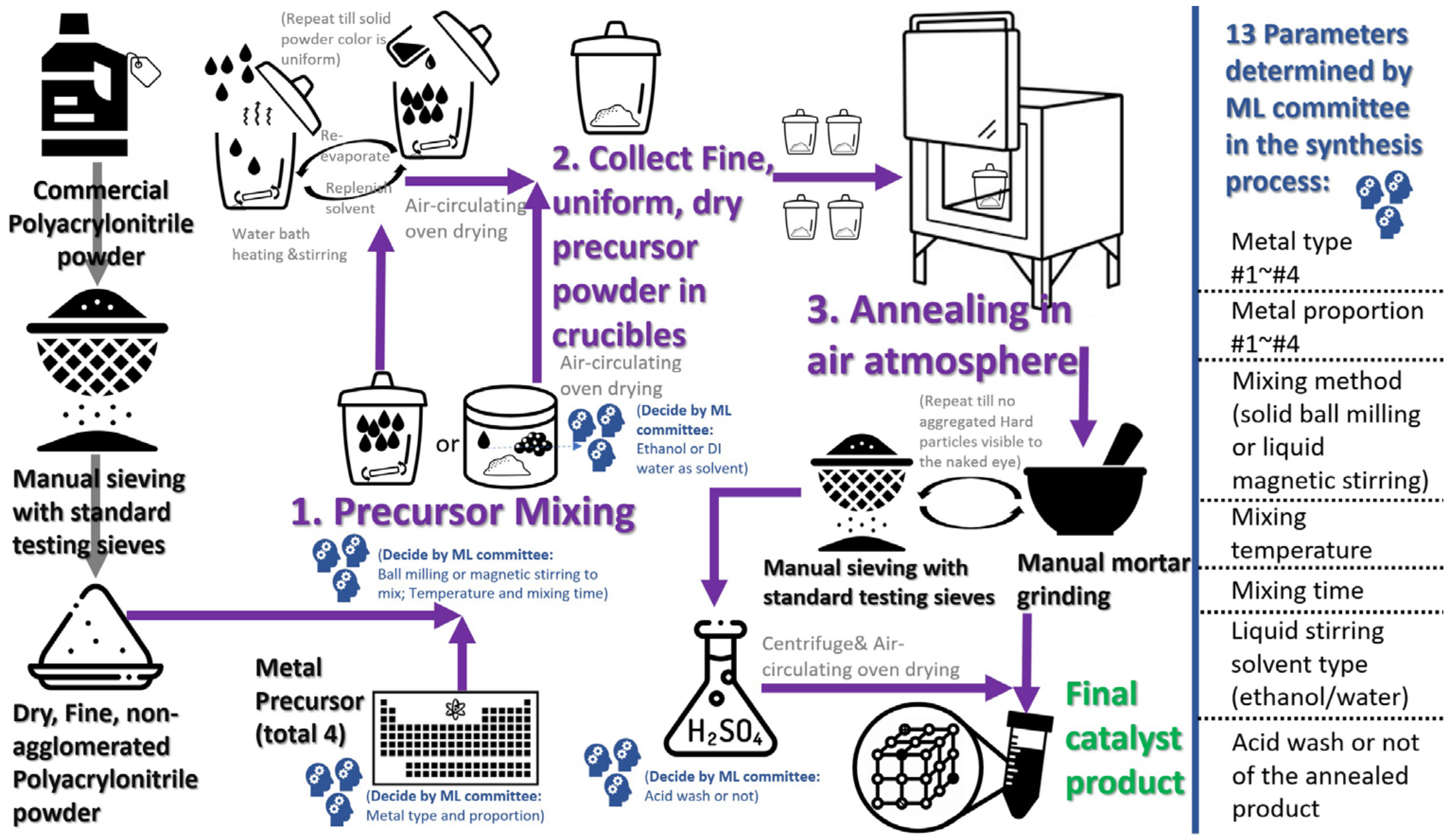

**Figure S13** Visualized synthesis process of the OER electrocatalyst samples via experiments in this work, with vivid icons and details.

**Supplementary Discussion S2**

**Details of the Experimental Synthesis and Electrochemical Evaluation Processes**

**Synthesis of Oxide Samples**

The preparation of polyacrylonitrile powder began with commercial polyacrylonitrile, which underwent manual sieving using standard testing sieves to yield a dry, fine, non-agglomerated product weighing 150 mg. Concurrently, a solution of metal precursors, comprising a total of four types, was prepared in a water: ethanol (1:1) mixed solvent. These metal salt precursors were all dissolved in the 1:1 water: ethanol mixture for future use, with the specific molar amount determined based on the equivalent molar quantity to 50 mg of hydrated ruthenium chloride, per the ML committee's prediction in each iteration. This involved drawing the metal solution with a pipette to extract a certain amount.

The mixing method, either ball milling or magnetic stirring, was decided by the ML committee. For ball milling, constraints were applied during the ML algorithm prediction: when ball milling was selected, all solvents and reducing agents were set to zero in the GA computing processes, although the temperature was not constrained. The milling jar was preheated in an air-circulating oven to a specific temperature (if above room temperature) as predicted by the ML model before beginning the milling process. In the case of magnetic stirring, polyacrylonitrile powder and various metal solution precursors were dissolved in 20 mL of a specified solvent (ethanol or water). Since polyacrylonitrile powder has low solubility, the product was slurry-like. The mixture was then placed in a 30 mL crucible under water bath heating and stirring conditions at 90 °C. Magnetic stirring continued until the solvent completely evaporated, leaving a dry, non-clumpy powder. This solvent addition and evaporation process may be repeated multiple times as necessary.

Following this, the precursor powder from the previous step was transferred into a 30 mL crucible (the same one if the precursor was already in a liquid phase) and calcined in a muffle furnace at either 400 °C or 500 °C under air. The heating rate was maintained at 1.67 °C $min^{-1}$ for six hours to obtain oxide samples. The resulting black mixed polymetallic oxide powder was then manually sieved using standard testing sieves and ground in a mortar until no aggregated hard particles were visible. This sieving and grinding process may be repeated as necessary. An optional acid washing step may be followed, as dictated by ML model predictions. If required, the product underwent acid washing in 0.5M $H_2SO_4$ at room temperature for eight hours. The acid-washed product was then collected by centrifugation, rinsed with DI water, and dried in an air-circulating oven to yield a fine powder. Ultimately, the final product, whether directly obtained from sieving and grinding or after undergoing acid washing, was ready for use as a catalyst for further OER electrochemical testing purposes.

**Electrochemical Test in Half Cell**

Half-cell tests for OER were conducted using a Rotating Disk Electrode (RDE). The experimental setup involved a typical standard three-electrode system, with a 5.00 mm-diameter RDE serving as the working electrode. A Reversible Hydrogen Electrode (RHE) was employed as the reference electrode after calibration, and a graphite rod was introduced as the counter electrode. The catalyst ink for the OER derived from oxide nanoparticles had a concentration of 11.78 mg $mL^{-1}$. The solvent for the ink was a mixture of isopropanol and Nafion perfluorinated resin solution (5 wt. % in isopropanol), with a volume ratio of 1:0.05. To prepare the ideal half-cell ink, the solid catalyst was pre-ground, as mentioned in **Synthesis of polymetallic oxide samples**, followed by ice-bath ultrasonication of the solid-liquid mixture until a uniform black suspension was obtained.

For electrochemical measurements, a specific amount of the ink was pipetted onto the glassy carbon of the RDE and naturally air-dried to achieve a smooth catalyst layer with a loading of 0.50 mg $cm^{-2}$. The evaluation of the OER catalytic performance was carried out in a 0.50 M $H_2SO_4$ aqueous solution (pH~0.3)[13] using Linear Sweep Voltammetry (LSV). The test temperature was maintained at 25 °C using a circulating water bath. During the LSV tests, the rotation speed was set at 1,600 rpm, with a scan rate of 10 mV $s^{-1}$.

The electrochemically-active surface area (ECSA) of the catalyst samples is calculated from the double-layer capacitance according to the equation[14]:

$$ECSA = C_{dl}/C_s$$

Where $C_s$ is the specific capacitance of the sample. Hence, we use general specific capacitances of $C_s = 0.03\ mF/cm^2$ based on typical reported value. The double-layer capacitance ($C_{dl}$) can be measured via cyclic voltammetric scan (CVs), a potential range in which no apparent Faradaic processes occurred. The range for CVs is 0.85-0.95 V. Scan rates are 20, 40, 60, 80, 100, and 120 mV/s, respectively.

ECSA-normalized activity at a given overpotential is another metric for evaluating the activity of a catalyst. The ECSA-normalized activity definition is based on the specific current density per ECSA ($J_{ECSA-200}$), which is calculated using the current density ($J_{200}$) at overpotential 200 mV and according to the equation:

$$J_{ECSA-200} = J_{200}/ECSA$$

We conducted the half-cell stability test in the same standard three-electrode system, but used sintered titanium fiber felts (Bekipor, Bekaert) with a thickness of 250 μm (cleaned in an ultrasonic bath with ethanol and water) as the electrode with a size of 1 cm*1 cm, and the catalyst loading is unchanged. Without rotating the electrode, we placed a magnetic stirrer to ensure the mass transfer of the electrolyte. The constant current density was kept at 10 mA $cm^{-2}$ for over 36 hours. All the half-cell tests were performed on electrochemical workstations CHI650e and CHI750e. Readers can retrieve all the tested data in the online repository in the directory “Experimental Records and Raw Data”.

**Membrane Electrode Assembly Test in a Single Cell**

The preparation of the Membrane Electrode Assembly (MEA) begins with the formulation of the ink. The procedure is as follows: 100 mg of 60 wt. % Pt/C catalyst is weighed and dispersed in ultrapure water. Acetone and Nafion D520 solutions are then added to the dispersion. Subsequently, the ink is stirred in an ice-water bath for 12 hours. After ultrasonic dispersion in an ultrasonic cleaner for 30 minutes, the ink is set aside for spray-coating on the cathode side for the HER. Similarly, for the anode side of the OER in single-cell tests, an analogous strategy is employed for preparing and dispersing the ink of Ru oxides.

The Catalyst-Coated Membrane (CCM) method is used, where the proton exchange membrane Nafion 115 is laid flat on the operation platform of an ultrasonic spray coater at a temperature of 100 °C. The ink is injected into the coater via a syringe and sprayed onto the membrane in a criss-cross pattern using ultrasonics. The flow rate of the spray is adjusted by regulating the carrier nitrogen gas pressure. The quantity of liquid is controlled by a motor while the distribution of the liquid is further managed by varying the coordinates, motor running speed, nozzle height, and nozzle angle. After coating, the final product is dried for 20 minutes. The same ultrasonic spray-coating process is then applied to the other side with the same parameters. The catalyst loadings are 0.4 mgPt cm $^{-2}$ for the cathode and 1 mg cm$^{-2}$ for the anode, with the overall MEA size of 2.5cm*2.5cm, resulting in a total area of 6.25 cm$^2$. It is important to note that in both half-cell and single-cell configurations, the Ru content is not directly calculated but estimated based on the weight of the final product. According to ICP results, the Ru loadings for the final products A~D in the half-cell are estimated to be 0.221, 0.181, 0.2015, 0.163 mgRu cm $^{-2}$, respectively. For the single cell, these values are doubled: 0.442, 0.362, 0.403, 0.326 mgRu cm$^{-2}$. This use of precious metals is relatively low compared to other works on PEM electrolyzer MEAs[15] in the literature.

For the anode and cathode, the Ti fiber felt used previously for the half-cell stability test and carbon fiber paper (Freudenberg) with a thickness of 250 μm are used as porous transport layers (PTLs), respectively. The PTLs are hot-pressed with the MEA at 140 °C under a pressure of 6.0 MPa. The hot-pressed samples are then placed between two serpentine flow field plates and sealed with PTFE gaskets.

Before each test of the single cell, 80 °C deionized (DI) water (conductivity $\sigma \leq 1.0$ μS cm$^{-1}$) is supplied to the anode of the electrolyzer at a flow rate of 20.0 mL min$^{-1}$ for eight hours. Subsequently, the electrolyzer is operated at a constant electrolysis voltage of 1.6 V until the current fluctuation is less than 1.0 mA per minute, ensuring proper hydration and stability of the MEA and activation procedure. Then, polarization curves are recorded in constant current mode, with steps of 0.1 A cm$^{-2}$ when the current density exceeds 0.1 A cm$^{-2}$. More details, such as the instrument models, the MEA stability test method, and the electrochemical impedance spectroscopy (EIS) measurement can be found in our recent work on the poisoning effect of Ti ions in porous transport layers on PEM electrolyzer MEAs[16].

The chemicals used and corresponding suppliers are listed below:

$Zn(NO_3)_2 \cdot 6H_2O$ Sinopharm Chemical Reagent Co., Ltd.

$Fe(NO_3)_3 \cdot 9H_2O$ Sinopharm Chemical Reagent Co., Ltd.

$Co(NO_3)_2 \cdot 6H_2O$ Sinopharm Chemical Reagent Co., Ltd.

$Ni(NO_3)_2 \cdot 6H_2O$ Sinopharm Chemical Reagent Co., Ltd.

$Sc(NO_3)_3 \cdot 6H_2O$ Shandong Desheng New Material Co.

$Cu(NO_3)_2 \cdot 3H_2O$ Macklin

$Ga(NO_3)_3 \cdot 9H_2O$ Sinopharm Chemical Reagent Co., Ltd.

$Y(NO_3)_3 \cdot 6H_2O$ Shandong Desheng New Material Co.

$Zr(NO_3)_4 \cdot 6H_2O$ Macklin

$NbCl_5$ Macklin

$VCl_3$ jkchemical

$MoCl_3$ jkchemical

$Cr(NO_3)_3 \cdot 9H_2O$ Sinopharm Chemical Reagent Co., Ltd.

$Mn(NO_3)_2 \cdot 2H_2O$ Macklin

$RuCl_3.xH_2O$ Macklin

$CdCl_2 \cdot 5H_2O$ Macklin

$In(NO_3)_3$ Macklin

$RhCl_3 \cdot 3H_2O$ Macklin

$LaCl_3$ Macklin

$PrCl3 \cdot 6H_2O$ Shandong Desheng New Material Co.

$NdCl_3$ Shandong Desheng New Material Co.

$PmCl_3$ Shandong Desheng New Material Co.

$SmCl_3$ Shandong Desheng New Material Co.

$EuCl_3$ Shandong Desheng New Material Co.

$GdCl_3$ Shandong Desheng New Material Co.

$ErCl_3$ Shandong Desheng New Material Co.

$TmCl_3$ Shandong Desheng New Material Co.

$YbCl_3 \cdot H_2O$ Shandong Desheng New Material Co.

$WCl_6$ Aladdin

$ReCl_3$ Aladdin

H2$IrCl_6 \cdot 6H_2O$ Aladdin

$AuCl_3$ Aladdin

$Pb(NO_3)_2$ Sinopharm Chemical Reagent Co., Ltd.

$BaCO_3$ Macklin

$SrCO_3$ Sigma Aldrich

$Na_2CO_3$ Sinopharm Chemical Reagent Co., Ltd.

$K_2CO_3$ Sinopharm Chemical Reagent Co., Ltd.

$CaCO_3$ Sinopharm Chemical Reagent Co., Ltd.

MgCl2 Sinopharm Chemical Reagent Co., Ltd.

$Li_2CO_3$ Sinopharm Chemical Reagent Co., Ltd.

$SeCl_2$ jkchemical

$Al(NO_3)_3$ Sinopharm Chemical Reagent Co., Ltd.

Ethanol Sinopharm Chemical Reagent Co., Ltd.

$H_2SO_4$ Aladdin

Ultrapure water Nanjing Peiyin Instrument Co.

Isopropanol Sinopharm Chemical Reagent Co., Ltd.

Nafion perfluorinated resin solution (5 wt. % in isopropanol) Dupont

**Table S3** Variable ranges and constant settings during GA search process.

| Feature Name (Unit) | Variable Range (constant settings) |
|---|---|
| Metal_Dopant_1 （1~4 represent the proportion in precursor from high to low） | 71 different metal elements |
| Metal_Dopant_2 | 71 different metal elements or none |
| Metal_Dopant_3 | 71 different metal elements or none |
| Metal_Dopant_4 | 71 different metal elements or none |
| Metal_Dopant_1 Proportion in Precursor （at. %; refers to that in total four types of metal） | 50~100 |
| Metal_Dopant_2 Proportion in Precursor （at. %） | 0~50 |
| Metal_Dopant_3 Proportion in Precursor （at. %） | 0~33.33 |
| Metal_Dopant_4 Proportion in Precursor （at. %） | 0~25 |
| Hydrothermal Temperature (°C) (or precursor mixing) | 25~60 |
| Hydrothermal Time (min) (or precursor mixing) | 360~1,440 |
| Hydrothermal Still/Stirring (0/1) (or precursor mixing) | 0: still incubation; 1: stirring or sonication |
| Hydrothermal Strong Reductant in Liquid (0/1) (or precursor mixing) | 0: False; 1: True |
| Hydrothermal Weak Reductant in Liquid (0/1) (or precursor mixing) | Constant; 0: False |
| Mixed in Solid or Liquid (0/1) | 0: False (liquid mixing or hydrothermal); 1: True (ball milling) |
| Annealing Temperature (°C) | Constant; set to 400 and 500 manually |
| Annealing Time (min) | Constant; 360 |
| Annealing Still/Stirring (0/1) | Constant; 0: False |
| Annealing Atmosphere Inert (0/1) | Constant; 0: False |
| Annealing Atmosphere Reducing (0/1) | Constant; 0: False |
| Post-processing Acid Wash, etc. (after annealing; 0/1) | 0: False; 1: True |
| Catalyst Loading (mg cm$^{-2}$) | Constant; 0.5 |
| Support Material Loading (mg cm$^{-2}$) | Constant; 0 |
| Support is not Carbon (support material, TiOx, etc.; 0/1) | Constant; 0: False |
| Electrode Type_Glassy Carbon/Carbon Paper or Ti Mesh (0/1) | Constant; 0: Glassy Carbon |
| LSV Scanning Speed (mV s$^{-1}$) | Constant; 10 |
| Electrolyte Proton Concentration (M) | Constant; 1 |

Note:

It is noteworthy that in the GA search process, each of the two (database-based committees) ×2 (400 °C or 500 °C) ×2 (Weighted: Maximum Uncertainty or Lowest Predicted Overpotential) =eight types of suggestions which are randomly obtained 48 times per iteration. From these 48 suggestions, one-sixth, i.e., eight, are randomly selected further for experimental synthesis and testing. This means each iteration should have 8*8=64 data points. However, in the first to third iterations, some samples synthesized according to the ML committee exhibited poor OER performance and almost no catalytic activity, and the polarization curves could not reach 10 mA cm$^{-2}$. This is why in the online repository's experimental data records, as readers can see in **Figure 4g**, only 34, 39, and 52 data points are available for the first, second, and third iterations, respectively, because the remaining data points among the 64 samples were unusable (due to poor performance). By the fourth iteration, all 64 ML-recommended synthesized samples could at least measure $\eta_{10}$. From this perspective, the proportion of effective experimental engineering parameter suggestions from the ML committee increased from 53% to 61%, 81%, and finally 100% in the fourth iteration, reflecting the ML committee's increasing ability to effectively guide experiments in the active learning loop. For the fifth iteration, it was no longer necessary to continue the DASH iteration; therefore, we only used the ML committee to search for the lowest overpotential, resulting in 32 outcomes (excluding the prediction of maximum uncertainty from the first four iterations). We found that all 32 ML-recommended synthesis formulations could successfully test the $\eta_{10}$ data.

Moreover, as mentioned in main text, while we have confirmed through data mining that Ru is the promising element, we have also set an additional constraint in the GA searching process that Ru must be included as one of the four elements in the selection group. To be noted, Ru is not forced to be the primary metal, but is free to be placed from the first to the fourth place in the dopant order. This would significantly reduce the searching space, and the recipes given by our workflow are likely to be more meaningful and show OER activity in experimental synthesis and evaluations. For more details and parameter settings, please refer to the script in this section, available in the “Active Learning Loop&GA Prediction” directory of our GitHub repository.

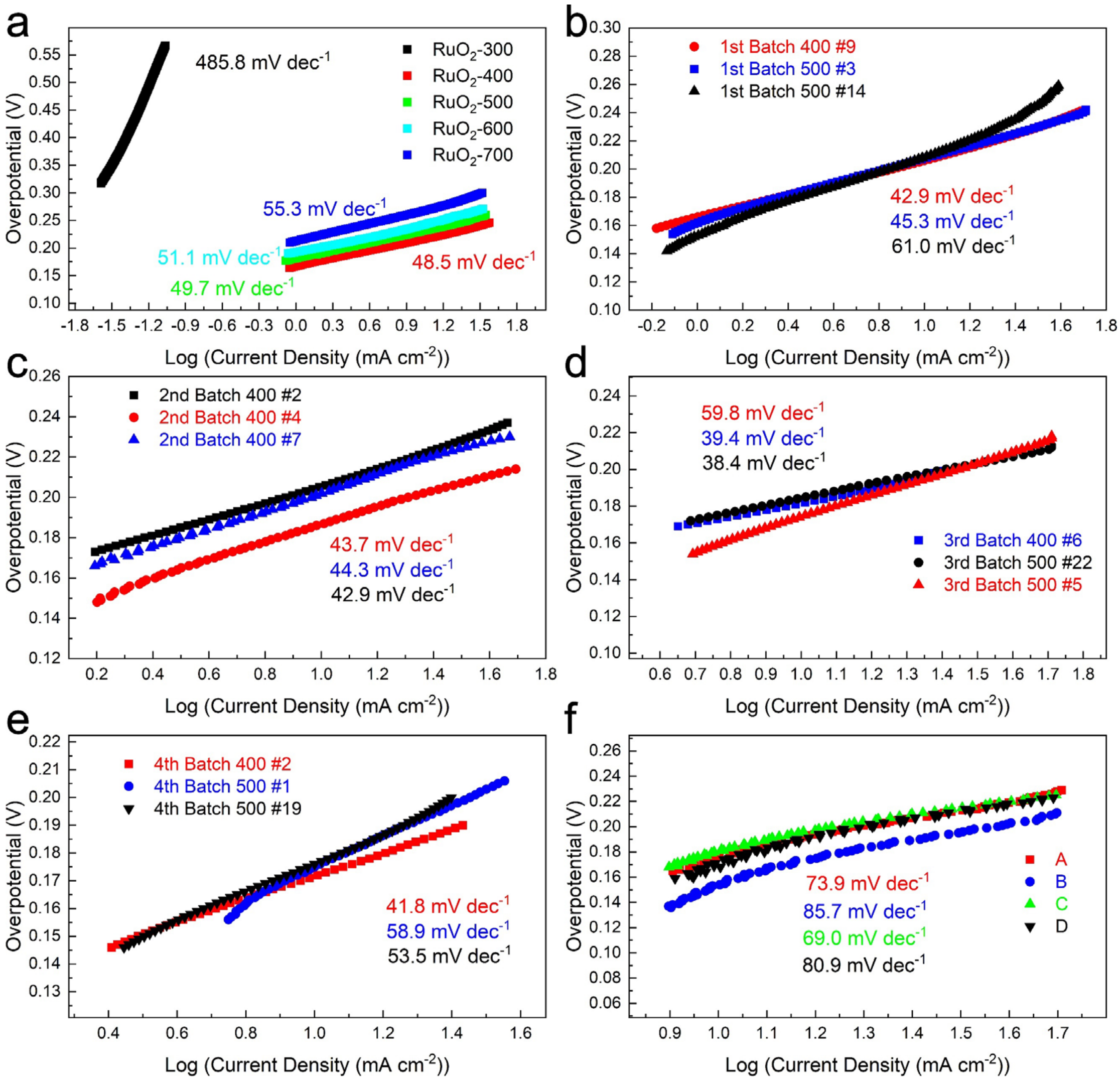

**Figure S14** a)-f) are Tafel plots (around 10 mA cm$^{-2}$) corresponding to the samples in **Figure 4**. It could be found that in batches 2, 4, and 5, the Tafel slopes of the samples are not in consistent order of their $\eta_{10}$ values. The discrepancy between the Tafel slopes in the OER catalysts arises from various factors. Different catalytic mechanisms can lead to a catalyst with a low overpotential while exhibiting a higher Tafel slope, reflecting diverse reaction kinetics. The Tafel slope, indicative of surface-reaction kinetics, contrasts with the overpotential, which relates to the number of active sites. A catalyst with many active sites might not have optimal reaction rates at these sites, resulting in a higher Tafel slope. Additionally, the electrode surface's microstructure and physical properties, along with experimental conditions like pH and temperature, can affect these measurements. This highlights the need for a multi-dimensional approach in evaluating and designing electrocatalysts that considers both the quantity and quality of active sites and their performance under varying conditions. Based on further results in **Figure S15** and **Figure S16**, we could derive that although A, B, and D have advantages on the surface area and the exposure of active sites, which lead to their better overpotential, their kinetics is inferior compared to C. C has a better Tafel slope and specific current density per ECSA, indicating that it has better intrinsic OER activity.

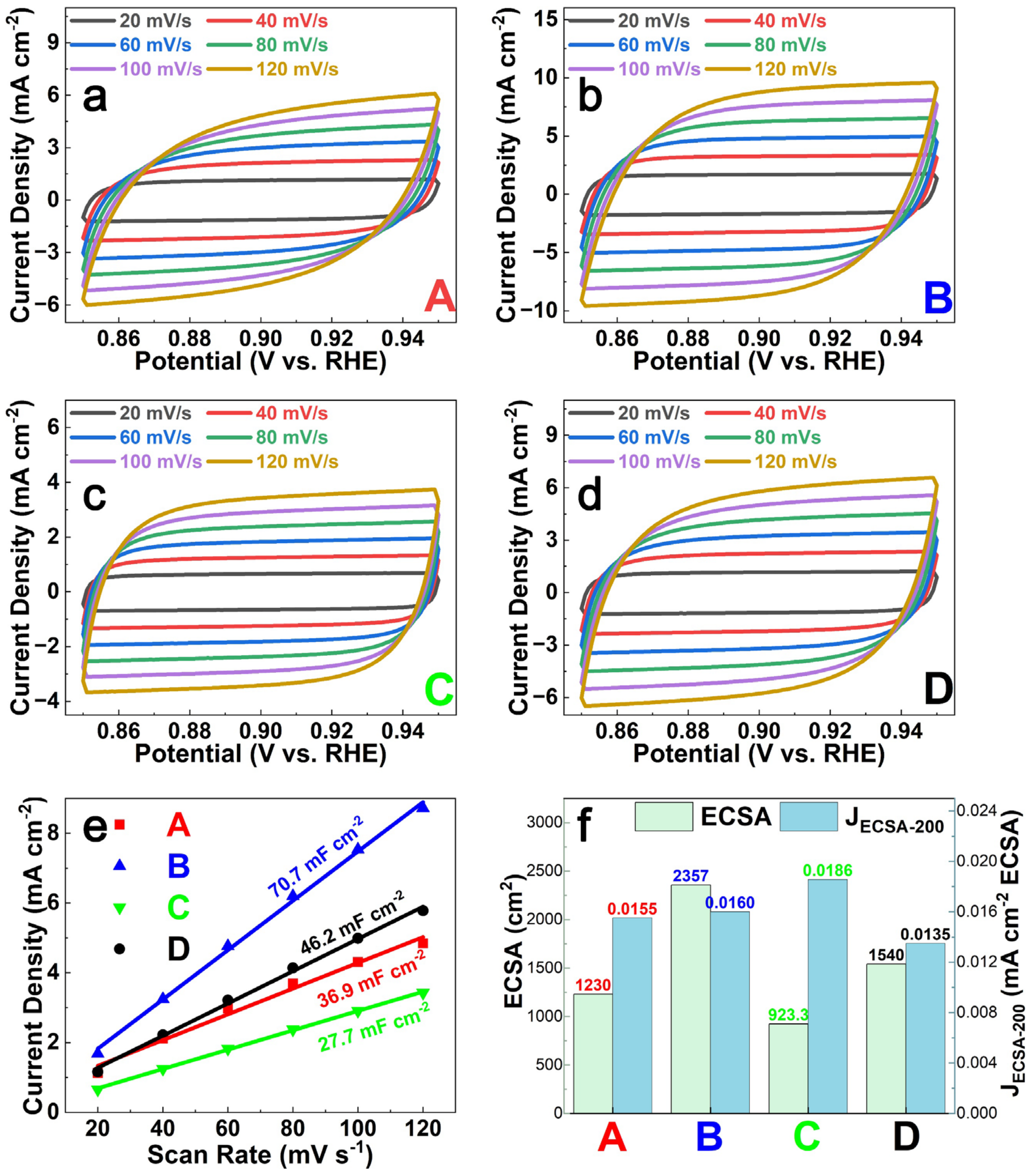

**Figure S15** a)-d) Electrochemical cyclic voltammetry scans recorded for A, B, C, and D. Scan rates are 20, 40, 60, 80, 100, and 120 mV/s, respectively. e) Linear fitting of the capacitive currents versus cyclic voltammetry scans for these catalysts. f) The calculated ECSA and specific current density per ECSA at η=200 mV ($J_{ECSA-200}$) values for A, B, C, and D.

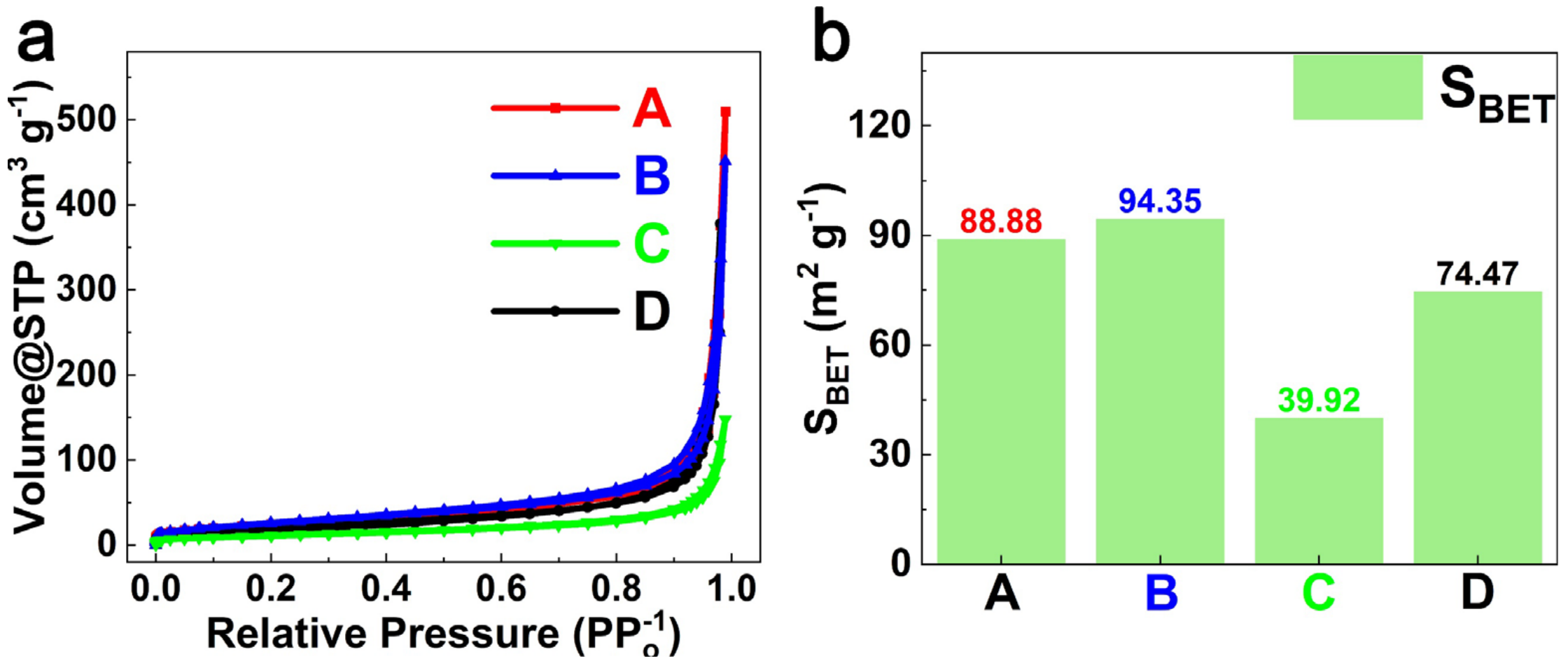

**Figure S16** a) Typical $N_2$ adsorption–desorption isotherms for A, B, C, and D. (b) Summary of the BET surface area for A, B, C, and D.

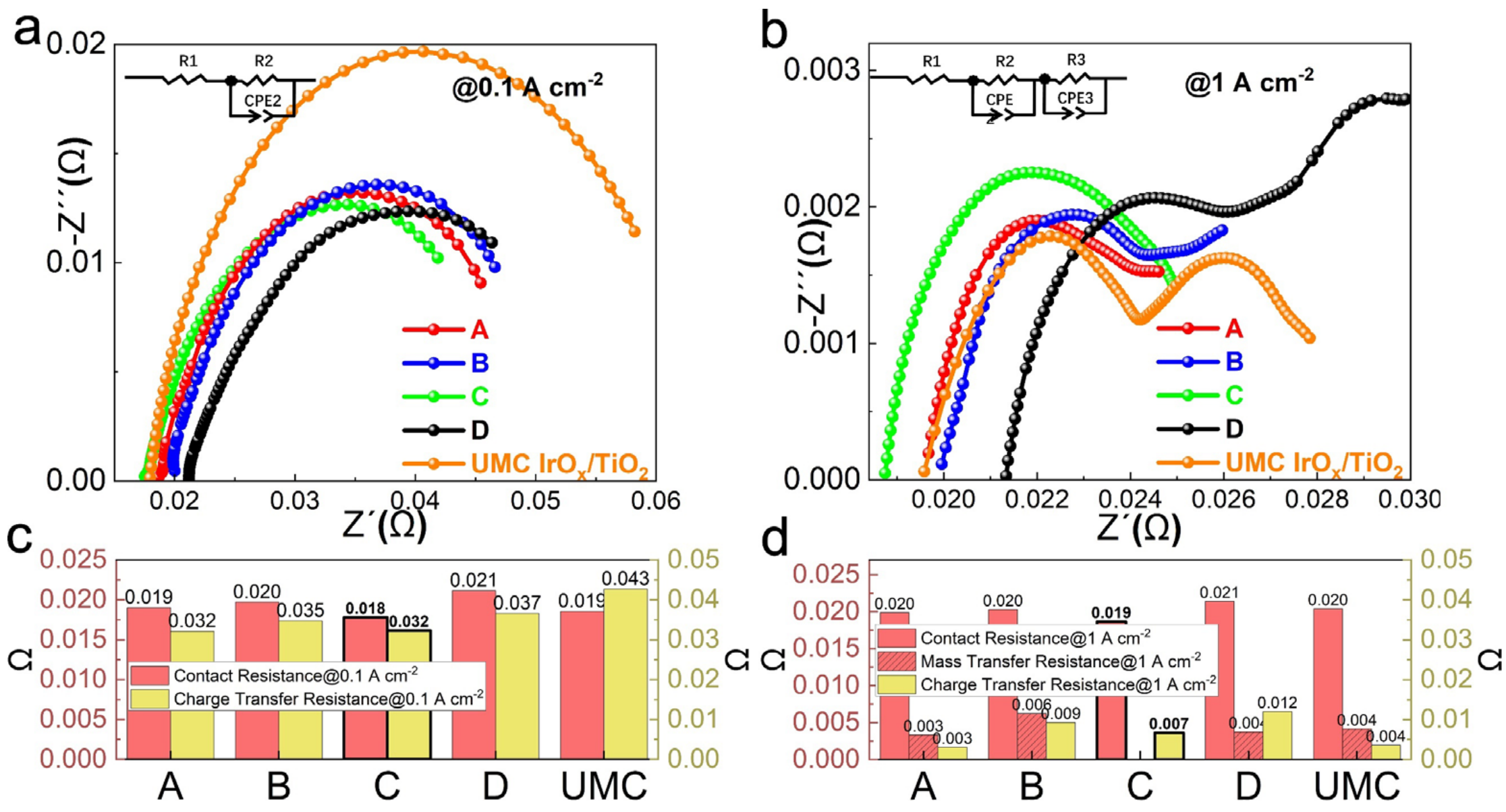

**Figure S17** a) and b) EIS spectra of the MEA samples measured in **Figure 4h**, at 0.1 and 1 A $cm^{-2}$ current densities, respectively, with their schematic equivalent circuits. c) and d) are the analyzed resistance components corresponding to a) and b).

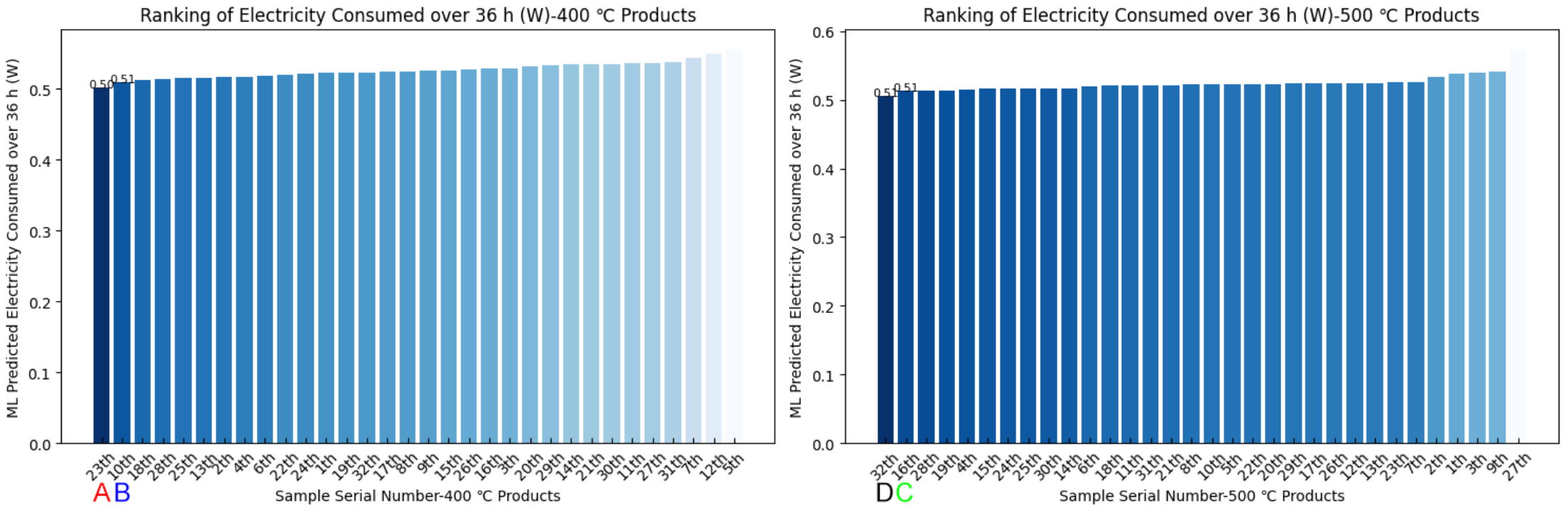

**Figure S18** Results from the direct use of the high-quality domain knowledge-based ML committee to predict the long-term performances of the electrocatalysts. Here, we enter the final samples information in the fifth batch as the input for the stability prediction models to predict the total needed electricity in 36 h while electrolyzing at a constant current density of 10 mA $cm^{-2}$. We also use a very simple hypothesis: the average voltage decay rate is constant; namely, the voltage needed to maintain the electrolysis current density will only increase linearly with time (overpotential becomes larger over time). As a final result, we found that samples A~D, which had the lowest overpotential, were still predicted to decay at the lowest rate, thus consuming the least amount of electricity in a practical situation. Corresponding script is in directory online: “Half-Cell Stability Quick Predict”.

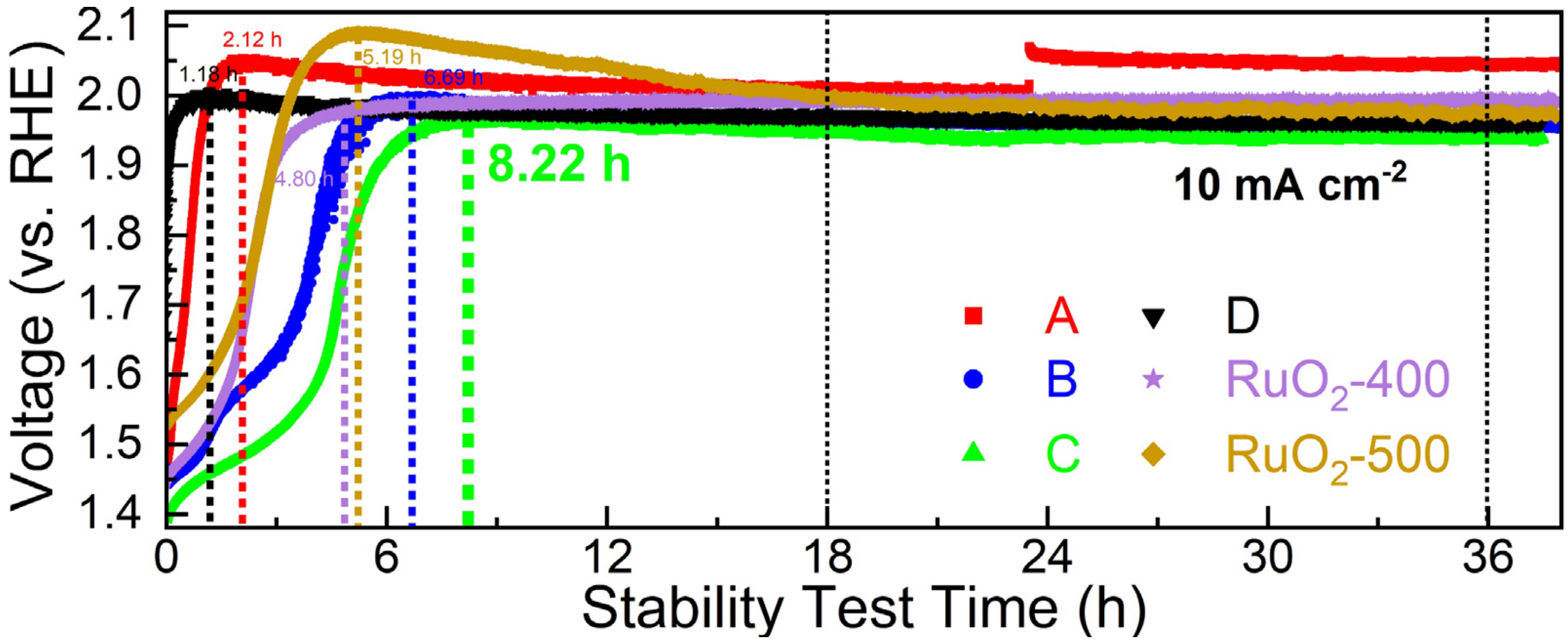

**Figure S19** The stability test in the half-cell of different samples: the dashed line indicates when the voltage is recorded to reach the maximum values. Sample C took 8.22 hours, while other samples generally failed quickly. Moreover, the final voltage of sample C is also lower.

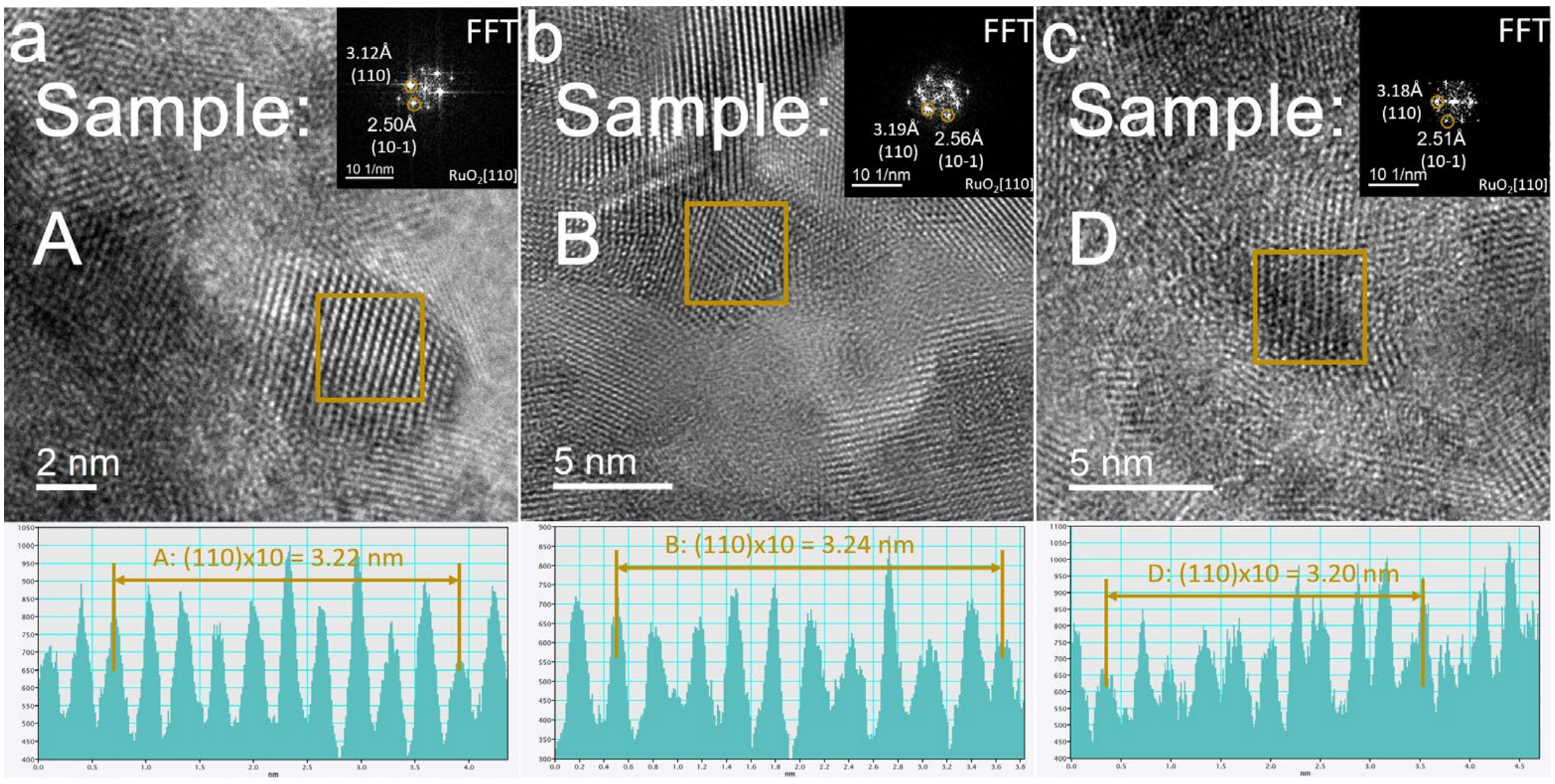

**Figure S20** Similar to **Figure 5a-b**, here we present the high-resolution TEM images, selected area FFT diffraction patterns, and corresponding statistical analysis of the average interplanar spacing. a) for sample A; b) sample B; c) sample D.

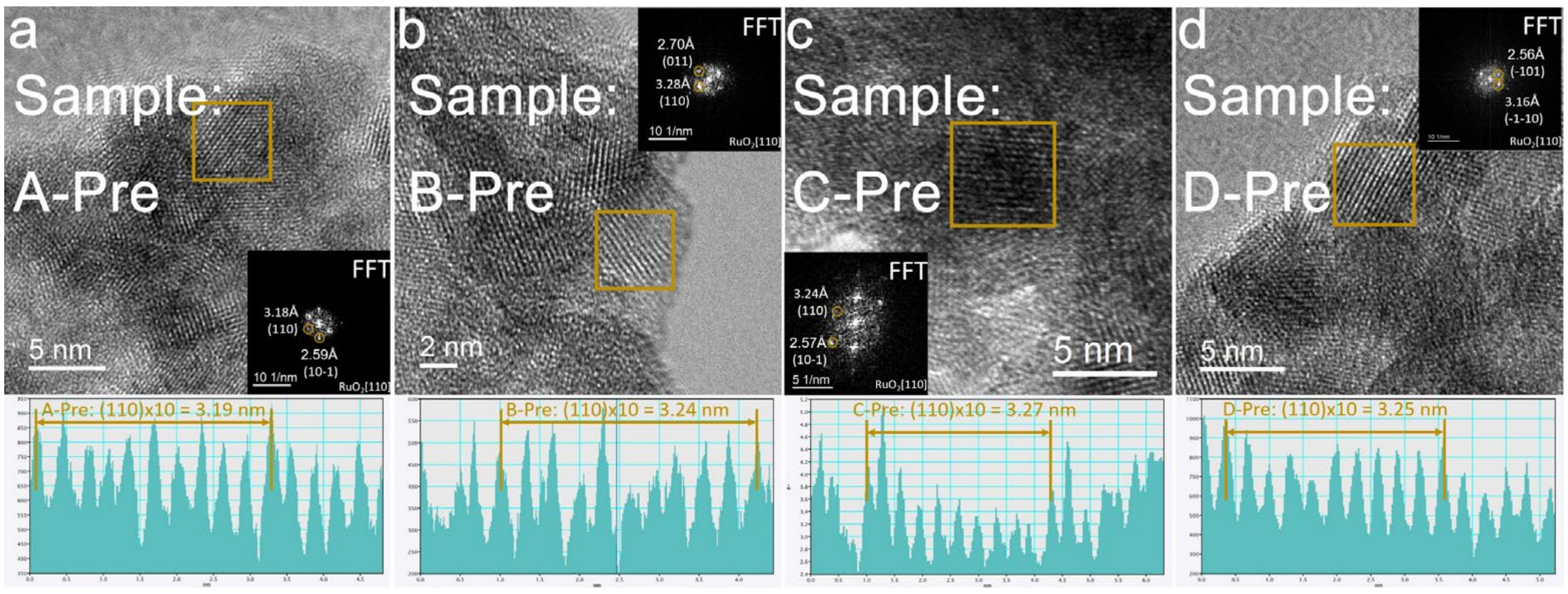

**Figure S21** High-resolution TEM images, selected area FFT diffraction patterns, and corresponding statistical analysis of the average interplanar spacing. a) for sample A-Pre; b): sample B-Pre; c) sample C-Pre; d): sample D-Pre.

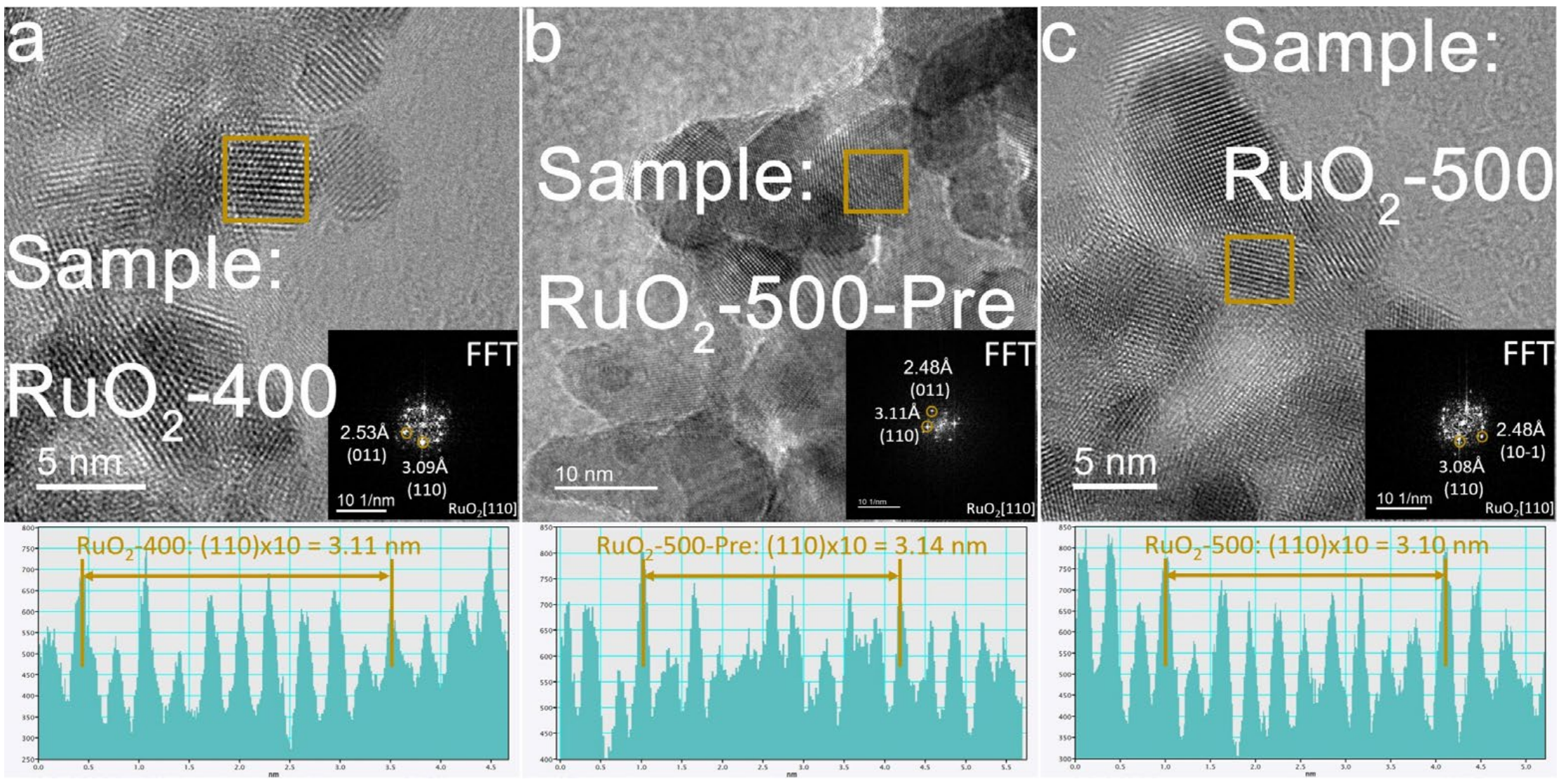

**Figure S22** High-resolution TEM images, selected area FFT diffraction patterns, and corresponding statistical analysis of the average interplanar spacing. a) for $RuO_2$-400; b) for $RuO_2$-500-Pre; c) for $RuO_2$-500.

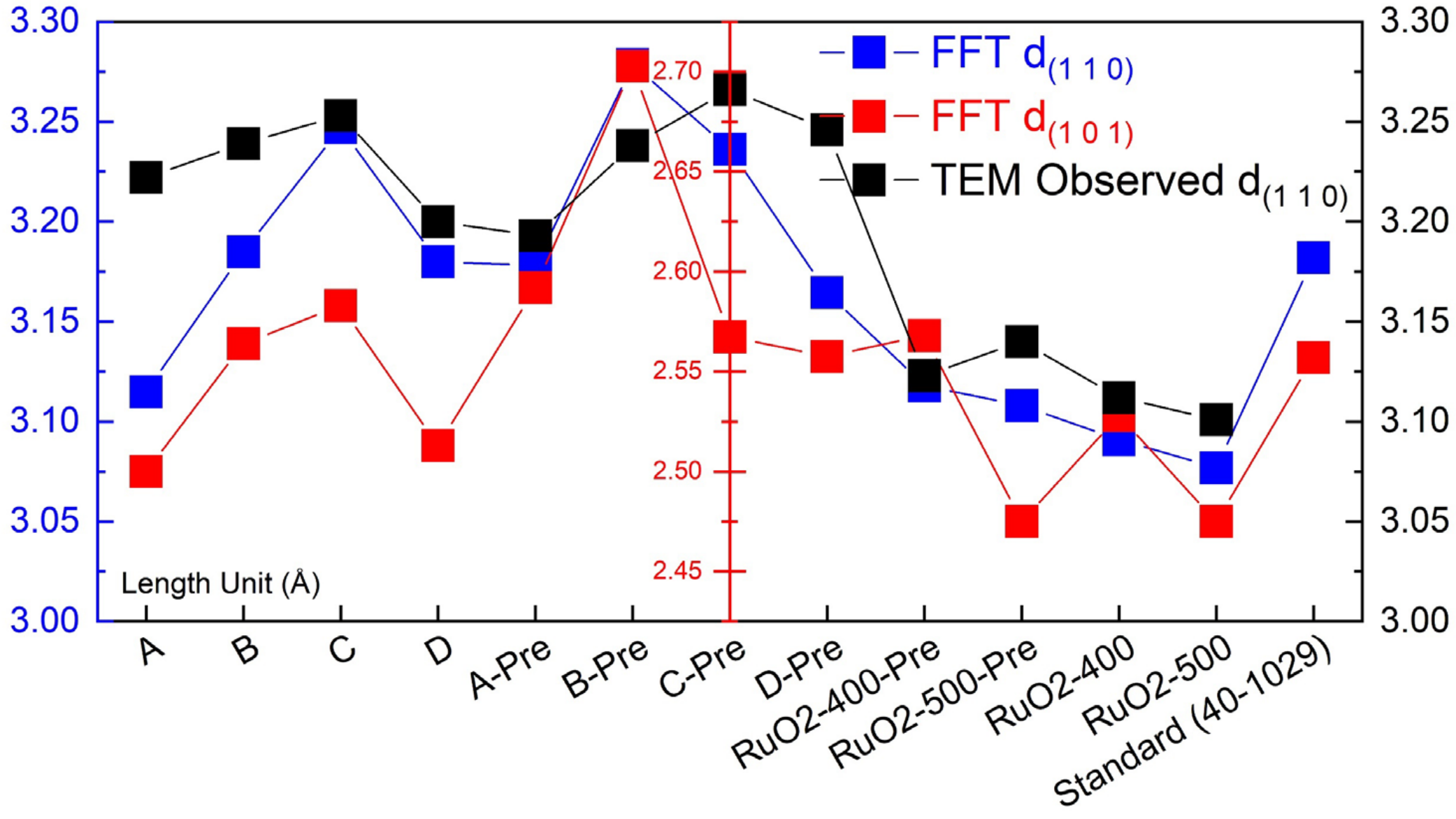

**Figure S23** The statistical line graph of the interplanar spacings for the (110) and (101) planes calculated from the inverse space diffraction patterns obtained through selected area FFT transformation, and the average (110) plane spacing directly measured in the selected areas of the TEM images for different samples of $RuO_2$ crystalline particles.

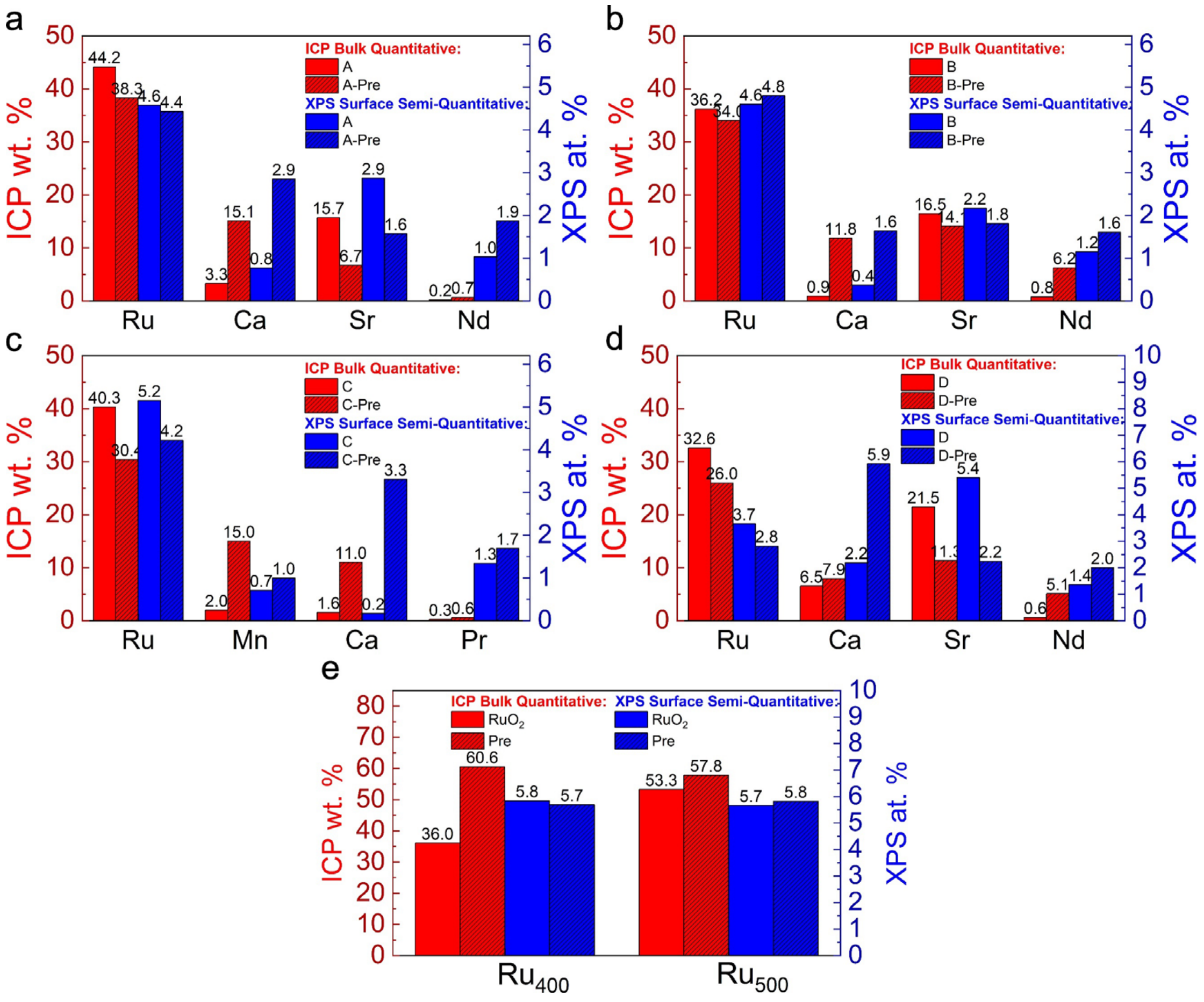

**Figure S24** a)-e) The bar chart showing the quantitative bulk mass fractions of various metal elements in 12 different samples obtained through ICP testing, and the semi-quantitative surface atomic fractions derived from XPS spectra.

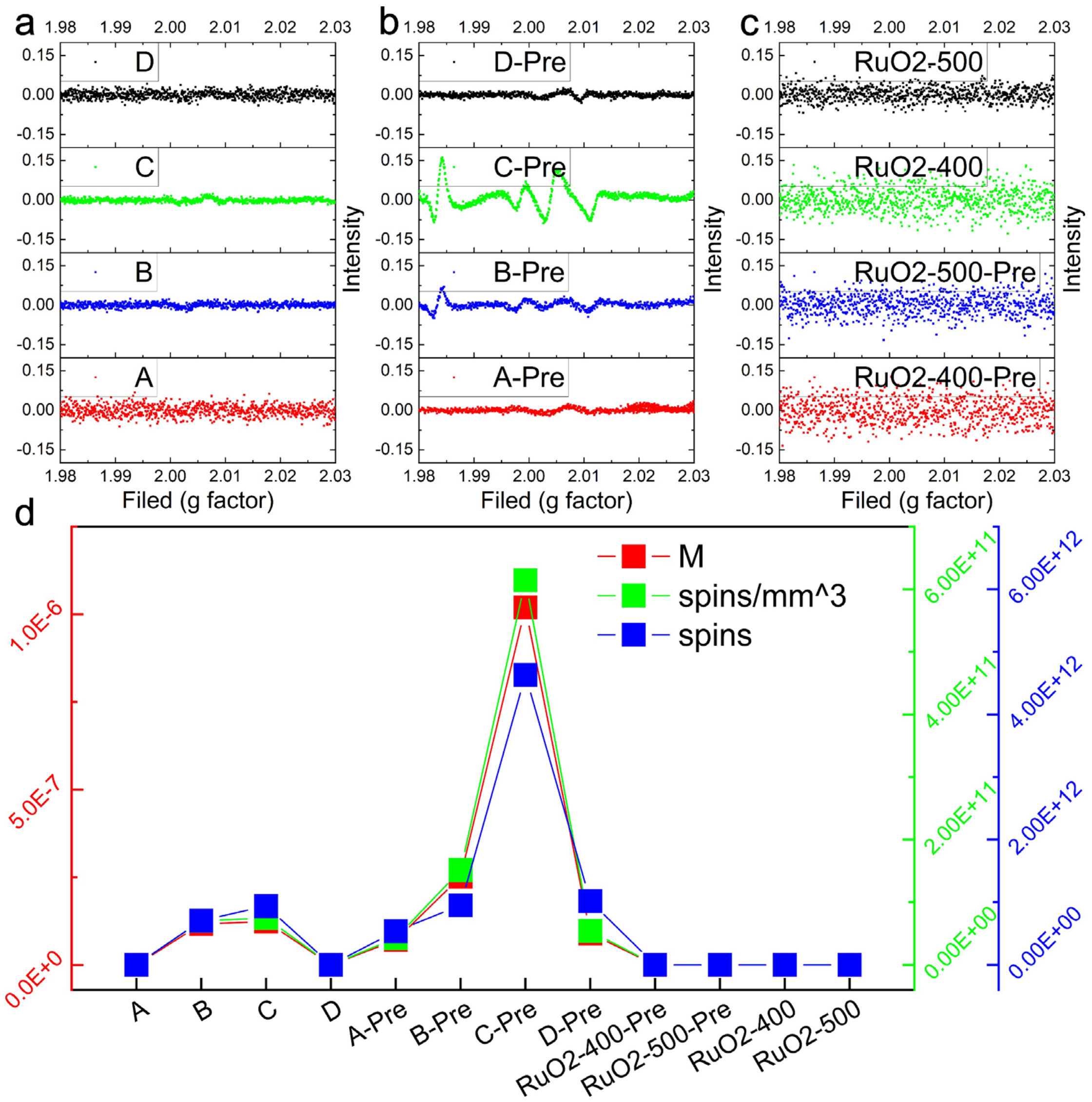

**Figure S25** a)-c) EPR spectra of 12 different samples, with the g-factor as the horizontal axis. d) Statistical line graph of magnetic susceptibility (M), in terms of total spins and spins per cubic millimeter (spins/$mm^3$) for different samples.

The data clearly shows that the C-Pre sample exhibits a high intensity of M, spins per cubic millimeter, and total spins, whereas all other samples exhibit significantly lower signal levels. This observation suggests that C-Pre contains manganese oxides, contributing to its strong magnetism due to the element composition of Ru, Mn, Ca, and Pr. However, upon observing C, the signal significantly diminishes, indicating the successful removal of manganese oxides during the acid wash process. Consequently, the remaining Mn element in C is inferred to be doped into the lattice of the primary $RuO_2$ matrix.

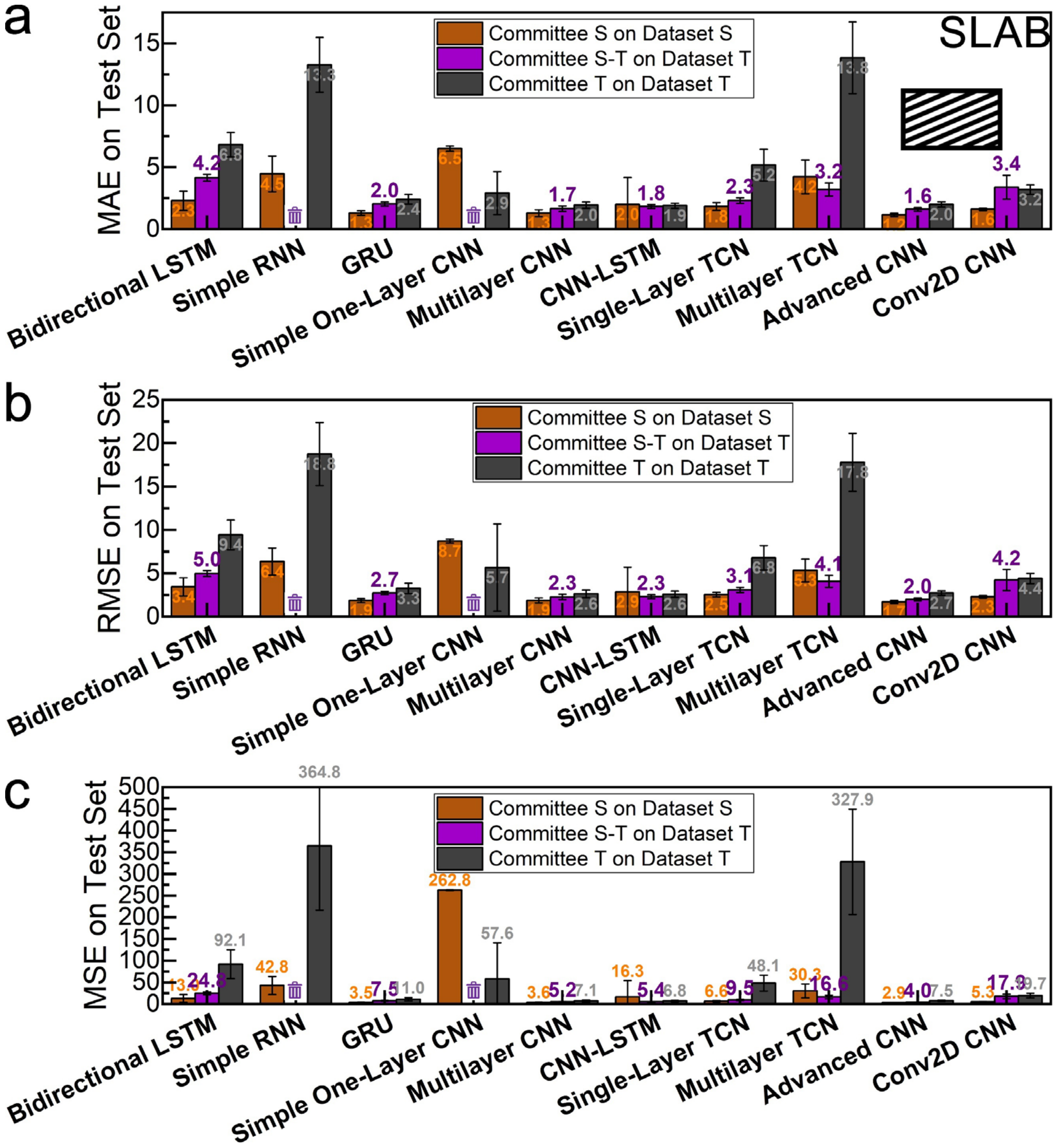

**Figure S26** Summary of the regression performance metrics of the committee predicting the energy of the slab.

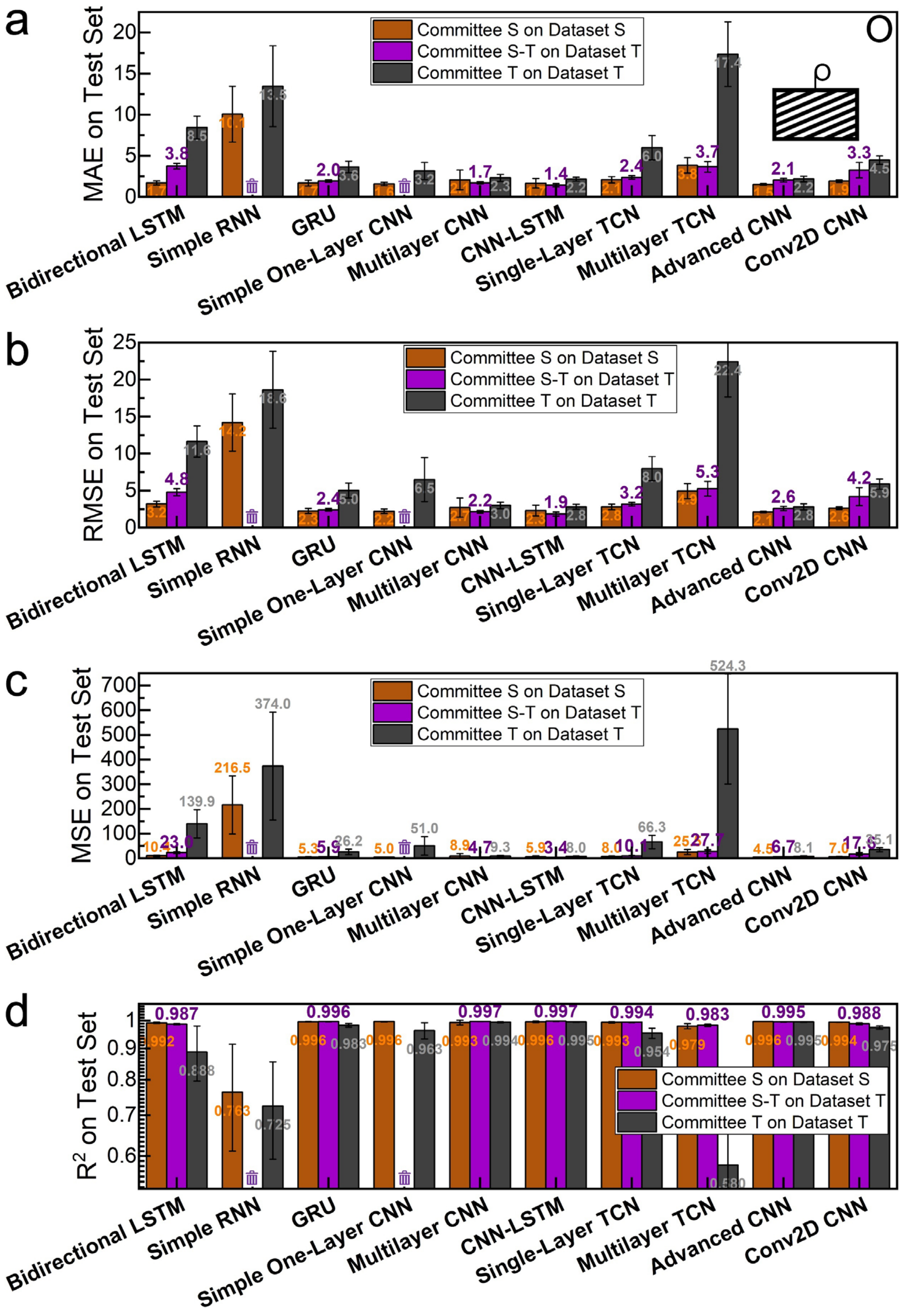

**Figure S27** Summary of the regression performance metrics of the committee predicting the energy of the slab with O species adsorbed.

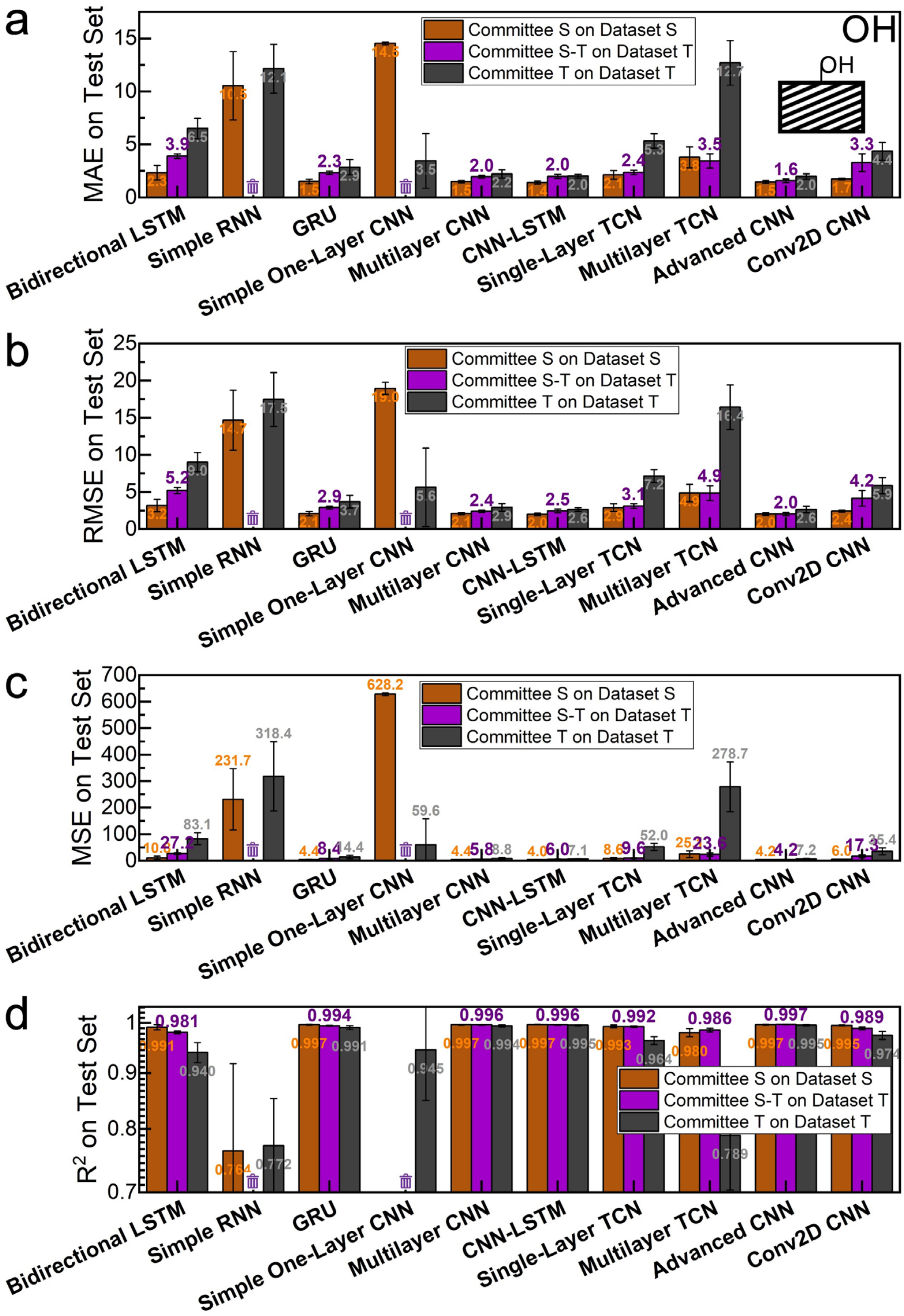

**Figure S28** Summary of the regression performance metrics of the committee predicting the energy of the slab with OH species adsorbed.

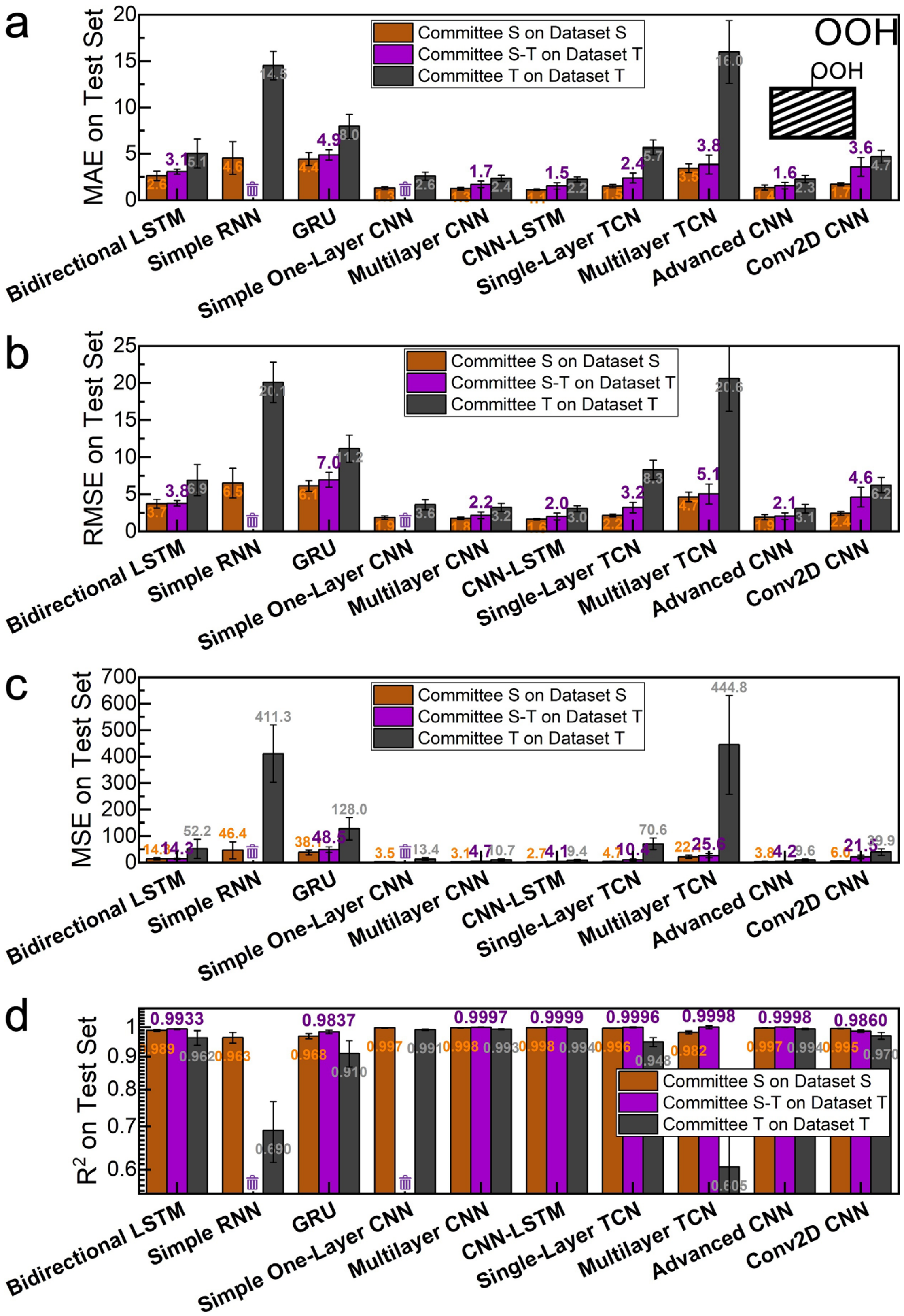

**Figure S29** Summary of the regression performance metrics of the committee predicting the energy of the slab with OOH species adsorbed.

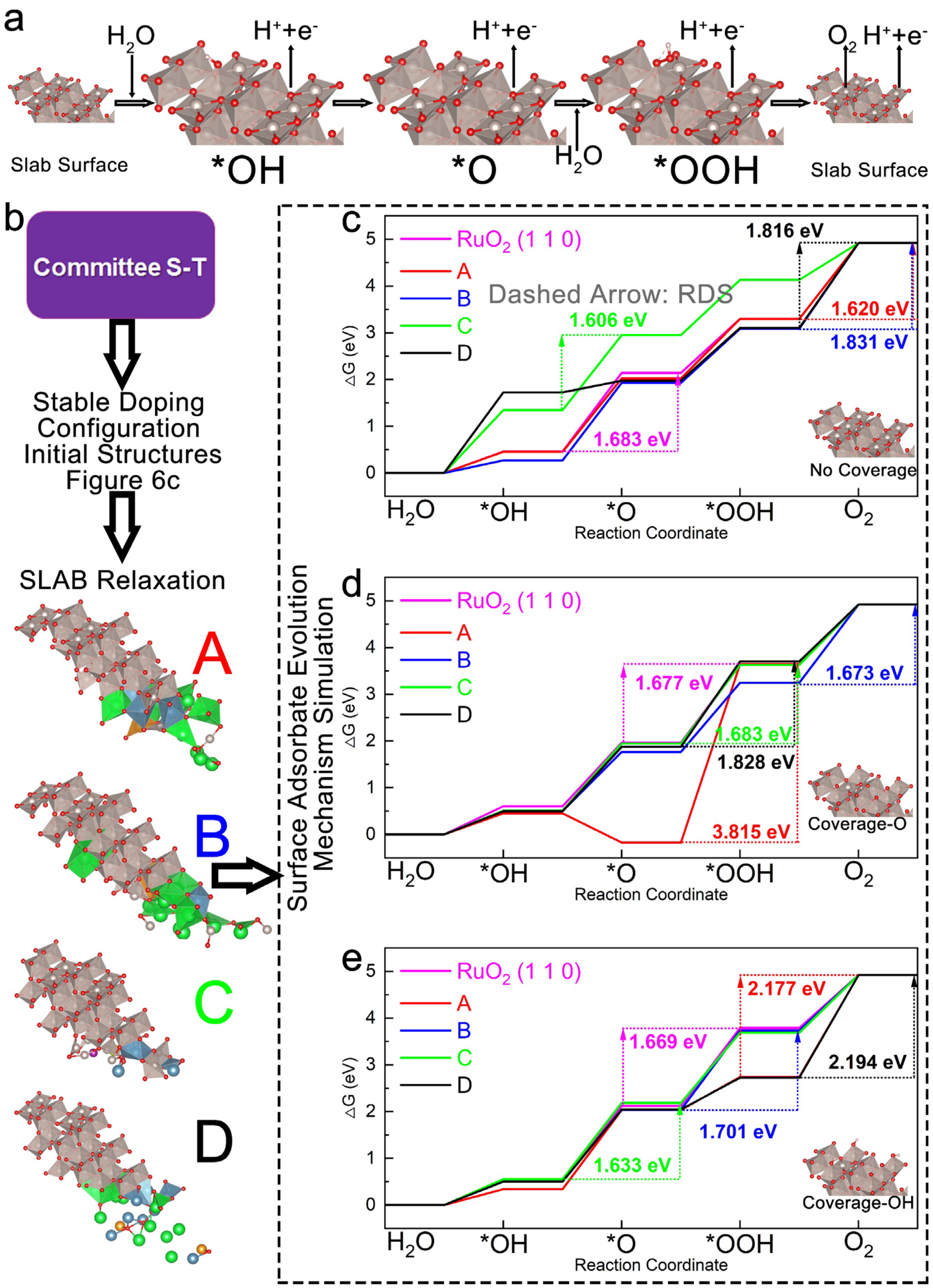

**Figure S30** a) Schematic illustration of the OER reaction pathway based on the adsorption evolution mechanism. b) Relaxed crystal structures of sample A~D corresponding to the doping structures obtained by GA search presented in **Figure 6c**. c) The comparison of OER and reaction pathways on the surface corresponding to the configurations of structures with a clean surface, namely, no additional species covered. Similarly, d) and e) are the corresponding results when the O and OH species have covered the nearing unsaturated coordinated Ru sites, respectively, as shown in the schematics. Dashed lines are used to highlight the rate determining step (RDS) in the reactions.

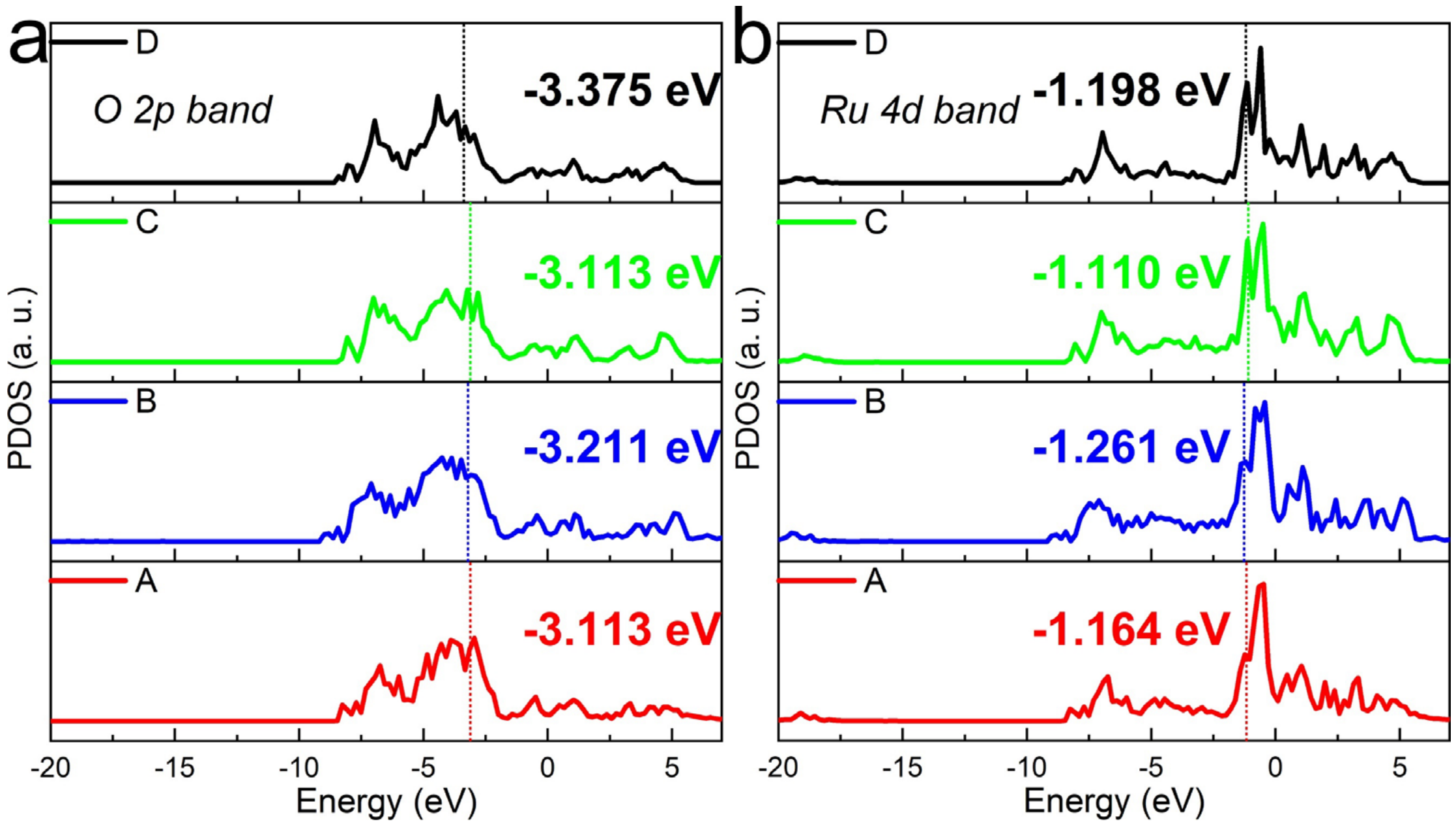

**Figure S31** a) Density of states plots of the O 2p band of different samples, b) Density of states plots of the Ru 4d band of different samples.

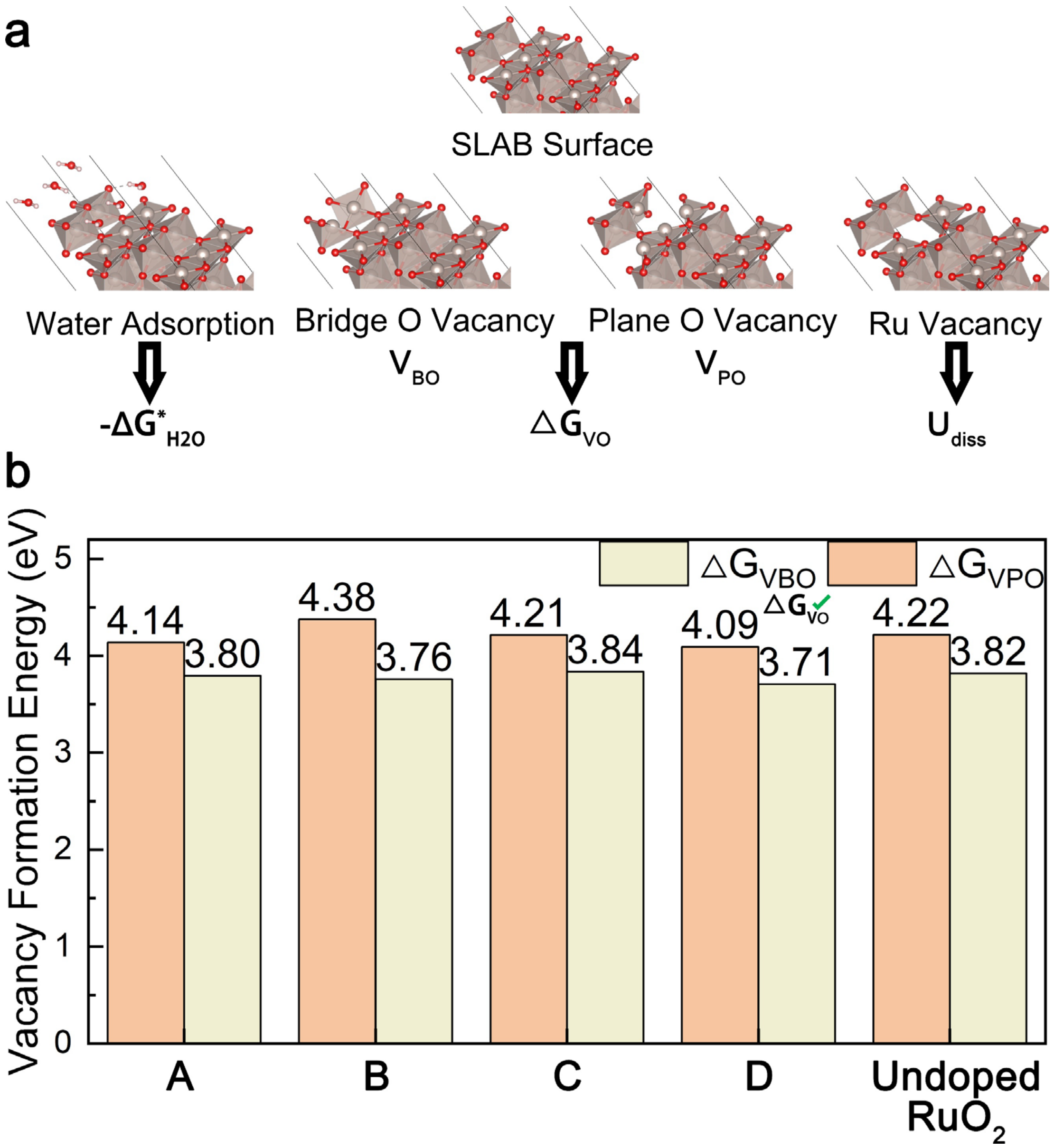

**Figure S32** a) Schematic of how different descriptors for theoretical stability was computed by surface manipulation, such as removing O and Ru atoms or the adsorption of water molecules. b) Summary bar plots of the vacancy formation energy of different samples.

Note: As we can observe from b), forming the plane O vacancy will always require more energy compared to forming the bridge O vacancy. Hence, we could determine that the O vacancy formation energy in our case is on bridge O, namely, $V_O=V_{BO}$. This is also consistent with the previous study presented by Hao et al.[17]

**Supplementary Materials References**